\documentclass[12pt]{article}
\usepackage{amscd}
\usepackage{curves}
\usepackage{latexsym,amsfonts,amssymb,amsmath,lscape,graphics,setspace}
\usepackage{graphicx}        
\usepackage{longtable}
\usepackage{youngtab}
\usepackage{float}
\usepackage{multirow}
\usepackage{color}

\usepackage{newtxtext,newtxmath}
\usepackage{anyfontsize}

\usepackage{slashed,epsfig}
\usepackage{slashed,epsfig}
\usepackage{cite}
\usepackage{hyperref}
\newcommand{\mathsym}[1]{{}}
\newcommand{\unicode}[1]{{}}
\newtheorem{definizione}{Definition}[section]

\newcommand{\bd}{\begin{definizione}}
\newcommand{\ed}{\end{definizione}}

\def\IC{\relax\,\hbox{$\inbar\kern-.3em{\rm C}$}}
\def\IG{\relax\,\hbox{$\inbar\kern-.3em{\rm G}$}}
\def\IB{\relax{\rm I\kern-.18em B}}
\def\ID{\relax{\rm I\kern-.18em D}}
\def\IL{\relax{\rm I\kern-.18em L}}
\def\IF{\relax{\rm I\kern-.18em F}}
\def\IH{\relax{\rm I\kern-.18em H}}
\def\II{\relax{\rm I\kern-.17em I}}
\def\IN{\relax{\rm I\kern-.18em N}}
\def\IP{\relax{\rm I\kern-.18em P}}
\def\IQ{\relax\,\hbox{$\inbar\kern-.3em{\rm Q}$}}
\def\bfzero{\relax\,\hbox{$\inbar\kern-.3em{\rm 0}$}}
\def\IK{\relax{\rm I\kern-.18em K}}
\def\IG{\relax\,\hbox{$\inbar\kern-.3em{\rm G}$}}
 \font\cmss=cmss10 \font\cmsss=cmss10 at 7pt
\def\IR{\relax{\rm I\kern-.18em R}}
\def\ZZ{\relax\ifmmode\mathchoice
{\hbox{\cmss Z\kern-.4em Z}}{\hbox{\cmss Z\kern-.4em Z}} {\lower.9pt\hbox{\cmsss Z\kern-.4em Z}}
{\lower1.2pt\hbox{\cmsss Z\kern-.4em Z}}\else{\cmss Z\kern-.4em Z}\fi}
\def\bfone{\relax{\rm 1\kern-.35em 1}}

\def\n010{N^{0,1,0}}
\def\inbar{\vrule height1.5ex width.4pt depth0pt}
\def\bfzero{\relax{\rm I\kern-.18em 0}}
\def\bfone{\relax{\rm 1\kern-.35em 1}}

\DeclareFontFamily{U}{rsf}{} \DeclareFontShape{U}{rsf}{m}{n}{
  <5> <6> rsfs5 <7> <8> <9> rsfs7 <10-> rsfs10}{}
\DeclareMathAlphabet\Scr{U}{rsf}{m}{n}

\newcommand{\spl}{\mathop{\rm sl}}

\newcommand{\ft}[2]{{\textstyle\frac{#1}{#2}}}
\def\tilde{\widetilde}

\def\1bar{1\hskip -.275cm -}
\def\2bar{2\hskip -.275cm -}
\def\3bar{3\hskip -.275cm -}

\newsavebox{\uuunit}
\sbox{\uuunit}
                 {\setlength{\unitlength}{0.825em}
                      \begin{picture}(0.6,0.7)
                                      \thinlines
                                      \put(0,0){\line(1,0){0.5}}
                                      \put(0.15,0){\line(0,1){0.7}}
                                      \put(0.35,0){\line(0,1){0.8}}
                                     \multiput(0.3,0.8)(-0.04,-0.02){10}{\rule{0.5pt}{0.5pt}}
                      \end {picture}}

\makeatletter \@addtoreset{equation}{section} \makeatother

\def\bfone{\relax{\rm 1\kern-.35em 1}}

\def\bfone{\relax{\rm 1\kern-.35em 1}}
\font\cmss=cmss10 \font\cmsss=cmss10 at 7pt

\newcommand{\slal}{\mathfrak{sl}}

\def\bfone{\relax{\rm 1\kern-.35em 1}}
\def\inbar{\vrule height1.5ex width.4pt depth0pt}
\def\IC{\relax\,\hbox{$\inbar\kern-.3em{\rm C}$}}
\def\ID{\relax{\rm I\kern-.18em D}}
\def\IF{\relax{\rm I\kern-.18em F}}
\def\IH{\relax{\rm I\kern-.18em H}}
\def\II{\relax{\rm I\kern-.17em I}}
\def\IN{\relax{\rm I\kern-.18em N}}
\def\IP{\relax{\rm I\kern-.18em P}}
\def\IQ{\relax\,\hbox{$\inbar\kern-.3em{\rm Q}$}}
\def\IR{\relax{\rm I\kern-.18em R}}
\font\cmss=cmss10 \font\cmsss=cmss10 at 7pt
\def\ZZ{\relax\ifmmode\mathchoice
{\hbox{\cmss Z\kern-.4em Z}}{\hbox{\cmss Z\kern-.4em Z}} {\lower.9pt\hbox{\cmsss Z\kern-.4em Z}}
{\lower1.2pt\hbox{\cmsss Z\kern-.4em Z}}\else{\cmss Z\kern-.4em Z}\fi}

\def\tilde{\widetilde}
\def\bar{\overline}

\def\hat{\widehat}

\def\Coe#1.#2.{\frac{#1}{#2}}

\def\coe#1.#2.{\relax{\textstyle \frac{#1}{#2}}\displaystyle}

\def\to{\rightarrow}
\def\notin{\hbox{{$\in$}\kern-.51em\hbox{/}}}

\def\IE{\relax{{\rm I\kern-.18em E}}}

\def\IGam{\relax{{\rm I}\kern-.18em \Gamma}}

\def\IA{\relax{\hbox{{\rm A}\kern-.82em {\rm A}}}}

\def\ben{\begin{equation}}
\def\een{\end{equation}}
\def\bea{\begin{eqnarray}}
\def\eea{\end{eqnarray}}
\input amssym.def
\input amssym.tex
\def\bea{\begin{eqnarray}}
\def\eea{\end{eqnarray}}
\input amssym.def
\input amssym.tex
\makeatletter
\g@addto@macro\bfseries{\boldmath}
\makeatother
\begin{document}
\hypersetup{pageanchor=false}
\begin{titlepage}
\begin{center}
\begin{flushright}
ARC-18-19
\end{flushright}
\vskip20pt
{\Large\sc The role of $\mathrm{PSL(2,7)}$ in M-theory:  \\
\vskip 0.2cm M2-branes, Englert equation
and the septuples }  \\[1cm]
{Bianca Letizia~Cerchiai${}^{\; a,c,d,e}$,   Pietro~Fr\'e${}^{\;
b,d,e}$ and Mario Trigiante${}^{\; a,d,e}$\\[10pt]
{${}^a${{\sl\small  Dipartimento di Fisica Politecnico di Torino,}}\\
{\em C.so Duca degli Abruzzi, 24, I-10129 Torino, Italy}~\quad\\
\emph{e-mail:}\quad  {\small {\tt bianca.cerchiai@polito.it}},
\quad {\small {\tt mario.trigiante@polito.it}}\\
\vspace{5pt} {${}^b$\sl \small Dipartimento di Fisica, Universit\'a di Torino\\
P. Giuria 1,  10125 Torino, \
Italy\\}
\emph{e-mail:} \quad {\small {\tt pietro.fre@unito.it}}\\
\vspace{5pt}
{{\em $^{c}$\sl\small Centro Fermi - Museo Storico della Fisica e Centro Studi e Ricerche ``Enrico Fermi'',}}\\
{\em Piazza del Viminale, 1, Roma 00184, Italy}~\quad\\
\vspace{5pt}
{${}^d$\sl \small INFN, Sezione di Torino\\
P. Giuria 1,  10125 Torino, \ Italy\\}
\vspace{5pt} {{\em ${}^{e}$
\sl\small Arnold-Regge Center,}}
\\{\em via P. Giuria 1,  10125 Torino, Italy} }}
\vspace{15pt}
\begin{abstract}
Reconsidering the M2-brane solutions of $d=11$ supergravity with a
transverse Englert flux introduced by one of us in 2016, we present
a new purely group theoretical algorithm to solve Englert equation
based on a specific embedding of the $\mathrm{PSL(2,7)}$ group into
$\mathrm{Weyl[\mathfrak{e}_7]}$.  The aforementioned embedding is
singled out by the identification of $\mathrm{PSL(2,7)}$ with the
automorphism group of the Fano plane. Relying on the revealed
intrinsic $\mathrm{PSL(2,7)}$ symmetry of Englert equation and on
the new algorithm we present an exhaustive classification of Englert
fluxes. The residual supersymmetries of the corresponding M2-brane
solutions associated with the first of the 8 classes into which we
have partitioned Englert fluxes are exhaustively analyzed and we
show that all residual $d=3$ supersymmetries with $\mathcal{N} \in
\left\{1,2,3,4,5,6\right\}$ are available. Our constructions
correspond to a particular case in the category of M2-brane
solutions with transverse self-dual fluxes.
\end{abstract}
\end{center}
\end{titlepage}
\hypersetup{pageanchor=false}
\vspace*{-6.8em}
\tableofcontents \noindent {}
\section{Introduction}
\label{introibo} The scenario underlying the \textit{gauge/gravity
correspondence}
\cite{Maldacena:1997re,Kallosh:1998ph,Ferrara:1998jm,Ferrara:1998ej,sergiotorino,witkleb,Fabbri:1999hw,Fabbri:1999ay,Aharony:2008ug,Gaiotto:2007qi,Gaiotto:2009tk}
is multi-faceted and involves many different
geometrical aspects. In particular there are two main paradigms:
\begin{description}
  \item[a)] The case of M2-branes solutions of $d=11$ supergravity,
  where the eight-dimensional space $\mathcal{M}_8$ transverse to the brane world
  volume is taken to be the metric cone over a five dimensional compact Einstein manifold
  $\mathcal{M}_7$ characterized by the metric:
  \begin{equation}
    ds^2_{(8)}\, = \, dr^2 \,+\,r^2 \, ds^2_{\mathcal{M}_7} \quad ;
    \quad r\in \mathbb{R}_+\, .
  \end{equation}
  \item[b)] The case of D3-brane solutions of type IIB supergravity,
   where the six-dimensional space $\mathcal{M}_6$ transverse to the brane world
  volume is taken to be the metric cone over a five-dimensional compact Einstein manifold
  $\mathcal{M}_5$ characterized by the metric:
  \begin{equation}
    ds^2_{(6)}\, = \, dr^2 \,+\,r^2 \, ds^2_{\mathcal{M}_5} \quad ;
    \quad r\in \mathbb{R}_+ \, .
  \end{equation}
\end{description}
In the case of the M2-branes, variants of the above solution included the introduction of a self-dual 4-form flux in the transverse 8-dimensional space and were extensively studied in the literature (see for instance \cite{Duff:1997xja,Hawking:1997bg,Cvetic:2000mh,Cvetic:2000db}). The properties of these solutions, such as supersymmetry, strongly depend on the topology of the transverse space as well as on the structure of the internal flux and only specific examples where analyzed.\par A new class of M2-brane solutions with self-dual transverse flux was recently introduced in \cite{miol168}. Inspired by  previous results in $d=7$
\cite{pietroantoniosorin,Fre:2015wvs},  the $11$-dimensional
manifold at the base of the M2--branes was chosen with  the
following topology:
\begin{equation}\label{rabolatus}
    \mathcal{M}_{11} \, = \, \mathrm{Mink}_{1,2} \times \mathbb{R}_+\times \mathrm{T^7},
\end{equation}
where $\mathrm{Mink}_{1,2}$ is Minkowski space in $1+2$ dimensions
and represents the world-volume of the $M2$-brane, while
$\mathrm{T^7}$ is a flat compact seven-torus. $\mathbb{R}_+\times
\mathrm{T^7}$ is the eight-dimensional space transverse to the
brane. It was shown that one can obtain exact solutions of $d=11$
supergravity where the metric is of the form:
\begin{equation}\label{colabrodone}
    ds_{11}^2 \, = \, H(y)^{-\ft{2}{3}}\left(d\xi^\mu \otimes d\xi^\nu\,
    \eta_{\mu\nu}\right)\, - \,H(y)^{\ft{1}{3}}\left(dy^I \otimes dy^J\,
    \delta_{IJ}\right),
\end{equation}
the function $H(y)$ over the transverse eight-dimensional space
being defined by an inhomogeneous Laplace equation whose source is
provided by the norm of an Englert flux. By this we mean a solution
of the following linear equation for a three-form $\mathbf{Y}^{[3]}$
living on the $\mathrm{T^7}$ torus:\footnote{The relation between equation (\ref{engleterro}) and the self-duality condition on the 4-for field-strength in the Euclidean 8-dimensional transverse space is illustrated in Appendix \ref{travertinus}. We shall refer to Eq.~(\ref{engleterro}), somewhat improperly, as the \emph{Englert equation}, since it describes the internal flux in the original Englert solution \cite{englert}, though on  a space with a different topology.}
\begin{equation}\label{engleterro}
    \star_{\,\mathrm{T^7}} \,\mathrm{d} \mathbf{Y}^{[3]}
\,=\,-\,\frac{\mu}{4} \, \mathbf{Y}^{[3]},
\end{equation}
which is the natural generalization of Beltrami equation for a
$1$-form on a $\mathrm{T^3}$-torus:
\begin{equation}\label{francoterro}
   \star_{\,\mathrm{T^3}}  \mathrm{d} \mathbf{Y}^{[1]}
\,=\,-\,\nu \, \mathbf{Y}^{[1]}.
\end{equation}
Just as in \cite{Fre:2015mla,pietroantoniosorin,Fre:2015wvs}, the
torus $\mathrm{T^3}$ was chosen to be:
\begin{equation}
    \mathrm{T^3} \simeq \frac{\mathbb{R}^3}{\Lambda_{cubic}} \, ,
\end{equation}
where $\Lambda_{cubic}$ is the cubic lattice, which endowed Beltrami
equation with the discrete symmetry provided by the point group of
such a lattice, namely the octahedral group $\mathrm{O_{24}}$, in
the same way in \cite{miol168} the torus $\mathrm{T^7}$ was chosen
to be:
\begin{equation}
    \mathrm{T^7} \simeq \frac{\mathbb{R}^7}{\Lambda^{root}} \, ,
\end{equation}
where $\Lambda^{root}$ is a root lattice of a suitable Lie algebra, prescribed to admit a
point group isomorphic to the simple group $\mathrm{PSL(2,7)}$ of
order 168. The main motivation for such an a priori choice performed
in \cite{miol168} was the embedding
$\mathrm{PSL(2,7)}\hookrightarrow \mathrm{G_{2(-14)}} \subset
\mathrm{SO(7)}$ which appeared to be promising in view of the
possible existence of Killing spinors for the corresponding M2-brane
solution. In any case just as it happens that Beltrami equation is
covariant with respect to the $\mathrm{O_{24}}$ group, the adopted
point group endows Englert equation with a
$\mathrm{PSL(2,7)}$-symmetry.
\par
In this paper we adopt a substantially new approach to the problem of constructing solutions of this kind.
It is based on a deeper understanding of the significance of the group $\mathrm{PSL(2,7)}$, entering in a different role as the \textit{automorphism group of the Fano plane} and, as such, as a subgroup of the Weyl group of $E_{7(7)}$~\cite{Cerchiai:2010tk}. This allows for the construction of novel M2-brane solutions with non constant fluxes.
The main point of the present work is  the realization that the $\mathrm{PSL(2,7)}$-symmetry of the Englert
equation is much less a matter of choice than it appeared to
be in the approach of \cite{miol168}. Indeed, as we explain in
section \ref{L168teoria}, which is a full fledged revisitation of
the theory of the $\mathrm{PSL(2,7)}$ group, this latter, in its
role as automorphism group of the Fano plane, provides
a systematic group-theoretical construction of the solutions to the Englert equation on
a flat space. These can be written in terms of elementary solutions, each defined by seven (a \emph{septuple} of) triples of integers 
$\{n_1,n_2,n_3\}$, $n_i \in \{1,2,3,4,5,6,7\}$, $n_1<n_2<n_3$, corresponding to the vertices of a Fano plane (which define a so-called \emph{Steiner triple system}), combined with a suitably defined \emph{complementary} septuple. The elementary solutions are characterized by the property that the non-vanishing internal components ${\bf Y}^{[3]}$ of the 3-form field ${\bf A}^{[3]}$ are only defined by the triplets of integers in the two septuples. The automorphism group $\mathrm{PSL(2,7)}$ of the Fano plane used for this construction is chosen to be the point group of the torus-lattice as well as the underlying symmetry group of the final solutions. Its action on the $d=11$ fields (and in particular on the internal components of the 3-form) can be inferred as follows.
The 35 components $Y_{ijk}$, $i,j,k=1,\dots, 7$, of ${\bf Y}^{[3]}$  on the seven-torus are in one-to-one correspondence with weights of the $35$-dimensional representation of
$\mathrm{SO(7)}\subset \mathrm{SL(7,\mathbb{R})}$ according to:
\begin{equation}\label{caballinus}
  dx^i\wedge dx^j \wedge dx^k  \Leftrightarrow \begin{array}{c} \\ \young(i,j,k)\\ \end{array} \Leftrightarrow  \mathbf{w}_{{\bf 35}}
  \in\Lambda^{weight}_{\mathfrak{a}_6} \, ,
\end{equation}
where $x^i$ are the torus coordinates, $\Lambda^{weight}_{\mathfrak{a}_6}$ denotes the weight lattice
of the $\mathfrak{a}_6$ Lie algebra and $\mathbf{w}_{{\bf 35}}$ a weight 
 of the $35$-dimensional representation. The automorphism group $\mathrm{PSL(2,7)}$ of the chosen Fano plane, being a subgroup of the Weyl group of $\mathfrak{a}_6$, acts in terms of permutations on the seven values of the internal indices, with respect to which the 35 weights $\{\mathbf{w}_{{\bf 35}}\}$ split
into an orbit of length 7 plus another of length 28. This embedding of $\mathrm{PSL(2,7)}$ into $\mathrm{SL(7,\mathbb{R})}$ is different from the
crystallographic embedding considered in \cite{miol168}. Such
observation provides an intrinsic group theoretical algorithm to
construct solutions of Englert equation.
\par
In \cite{miol168} some solutions of the Englert equation were
constructed using the obvious uplifting to 7-dimensions of the
technique utilized in
\cite{Fre:2015mla,pietroantoniosorin,Fre:2015wvs} to construct
solutions to Beltrami equation, namely the Fourier expansion of the
field $\mathbf{Y}^{[3]}$ and the restriction of the considered
momenta to  orbits of the $\mathrm{PSL(2,7)}$ in the weight lattice
of the $\mathfrak{a}_7$ Lie algebra. Such constructions were
particularly cumbersome since they produced rather large parameter
spaces that had to be organized a posteriori into irreducible
representations of $\mathrm{PSL(2,7)}$ and of its subgroups.
Furthermore there was no clear cut strategy for an exhaustive
classification.
\par In this paper, utilizing this new viewpoint and in particular the different
inequivalent embedding mentioned above, we have been able to classify
all solutions according to 424 generating schemes grouped into 8
classes, each class labeled by an invariant signature. This
classification is displayed in Table \ref{finisterrae}. We have also
provided an exhaustive analysis of the residual supersymmetries for
the solutions of the first class in which both the original septuple and the complementary one are of Steiner type (they both have signature $(0,21,0)$ and define two distinct Fano planes). The
result of this analysis is summarized in Table (\ref{barimondo}). It
shows that M2-branes with all possible number of supercharges can be
obtained from our construction. The analysis of the remaining seven
classes of solutions is postponed to a future publication.
Similarly, as we discuss in the conclusive section \ref{zumildo}, we
postpone to a future publication of the possible interpretation of
our M2-solutions in various classical contexts of the gauge/gravity
correspondence or of the Kaluza-Klein expansion.

Although the approach followed in the present paper and the results are substantially different from those of \cite{miol168},
for the sake of completeness we shall recall some general properties of the group  $\mathrm{PSL(2,7)}$ which are illustrated in the same reference.
\par
The paper is organized as follows:
\begin{itemize}
  \item In section \ref{genM2Engle} we review the structure of the
  Ansatz of M2-branes with Englert fluxes.
  \item In section \ref{sembranormale} we study the normal form of
  Englert three-forms and we introduce the role played by the group
  $\mathrm{PSL(2,7)}$.
  \item In sections \ref{L168teoria},\ref{zuccamarcia} we revisit the entire theory of
  the group $\mathrm{PSL(2,7)}$ and of its crystallographic
  irreducible representations. In particular we illustrate the
  difference between the crystallographic irreducible representation of
  dimension $7$ utilized in \cite{miol168} and a new
  crystallographic irreducible representation of dimension $6$ which
  is the key weapon for our algorithm to construct solutions of
  Englert equation.
  \item In section \ref{baciogallone} we present the intrinsic group theoretical
  algorithm to solve Englert equation and we arrive at the
  classification of table \ref{finisterrae}.
  \item In section \ref{kilspisection} we review the criterion,
  found in \cite{miol168}, for the preservation of $\mathcal{N} =2,\dots,6$
  residual supersymmetries in $d=3$.
  \item In section \ref{0210susy} we derive the classification of
  residual supersymmetries for the solutions of type $(0,21,0)$.
  \item In section \ref{zumildo} we draw our conclusions and we
  illustrate the perspectives for the interpretation of our M2-brane
  solutions.
  \item In appendix \ref{travertinus} we consider the more general
  case of M2-brane solutions of $d=11$ supergravity with a transverse
  internal flux and we show that Englert fluxes are a particular
  subclass in this class. 
\end{itemize}
The reader who is only interested in the construction and study of the new solutions and their supersymmetry, can skip the more mathematical sections \ref{L168teoria}, \ref{zuccamarcia}.
\section{M2-branes with Englert fluxes}
\label{genM2Engle} In this section we shortly review the structure
of M2-brane solutions of $d=11$ supergravity with Englert fluxes that
were introduced in \cite{miol168} and constitute the object of 
study, from a new viewpoint, of the present paper.
\par
In order to describe  the general form of these  solutions with
Englert fluxes  we need to consider the effective low energy
lagrangian of $M$-theory, namely $d=11$ supergravity for which we
utilize the geometric rheonomic formulation of
\cite{FreDauriaHidden,FDAgauge}\footnote{For a recent review in
modernized notations see \cite{maiobuk}, Volume II, Chapter 6.}.
Appendix \ref{peritoniteA} provides a dictionary between the
normalization used in the first paper on $d=11$ supergravity
\cite{Cremmer:1978km} and those of \cite{FreDauriaHidden,FDAgauge}.
\subsection{Summary of \texorpdfstring{$d=11$}{d=11} supergravity in the rheonomy framework}
\label{summario} The complete set of curvatures defining the
relevant Free Differential Algebra is given below
(\cite{FreDauriaHidden,FDAgauge}):
\begin{eqnarray}
\mathfrak{T}^{a} & = & \mathcal{D}V^a - {\rm i} \ft 12 \, \overline{\psi} \, \wedge \,
\Gamma^a \, \psi \nonumber\\
\mathfrak{R}^{ab} & = & d\omega^{ab} - \omega^{ac} \, \wedge \,
\omega^{cb}
\nonumber\\
\rho & = & \mathcal{D}\psi \equiv d \psi - \ft 14 \, \omega^{ab} \, \wedge \, \Gamma_{ab} \, \psi\nonumber\\
\mathbf{F^{[4]}} & = & d\mathbf{A^{[3]}} - \ft 12\, \overline{\psi}
\, \wedge \, \Gamma_{ab} \, \psi \,
\wedge \, V^a \wedge V^b \nonumber\\
\mathbf{F^{[7]}} & = & d\mathbf{A^{[6]}} -15 \, \mathbf{F^{[4]}} \,
\wedge \, \mathbf{A^{[3]}} - \ft {15}{2} \, \, V^{a}\wedge V^{b} \,
\wedge \, {\bar \psi} \wedge \, \Gamma_{ab} \, \psi
\, \wedge \, \mathbf{A^{[3]}} \nonumber\\
\null & \null & - {\rm i}\, \ft {1}{2} \, \overline{\psi} \, \wedge
\, \Gamma_{a_1 \dots a_5} \, \psi \, \wedge \, V^{a_1} \wedge \dots
\wedge V^{a_5} \label{FDAcompleta}
\end{eqnarray}
There is a unique rheonomic parametrization of the curvatures
(\ref{FDAcompleta}) which solves the Bianchi identities and it  is
the following one:
\begin{eqnarray}
\mathfrak{T}^a & = & 0 \nonumber\\
\mathbf{F^{[4]}} & = & F_{a_1\dots a_4} \, V^{a_1} \, \wedge \dots \wedge \, V^{a_4} \nonumber\\
\mathbf{F^{[7]}} & = & \ft {1}{84} F^{a_1\dots a_4} \, V^{b_1} \,
\wedge \dots \wedge \,
V^{b_7} \, \epsilon_{a_1 \dots a_4 b_1 \dots b_7} \nonumber\\
\rho & = & \rho_{a_1a_2} \,V^{a_1} \, \wedge \, V^{a_2} - {\rm i}
\ft 13 \, \left(\Gamma^{a_1a_2 a_3} \psi \, \wedge \, V^{a_4} + \ft
1 8 \Gamma^{a_1\dots a_4 m}\, \psi \, \wedge \, V^m
\right) \, F^{a_1 \dots a_4} \nonumber\\
\mathfrak{R}^{ab} & = & R^{ab}_{\phantom{ab}cd} \, V^c \, \wedge \,
V^d + {\rm i} \, \rho_{mn} \, \left( \ft 12 \Gamma^{abmn} - \ft 2 9
\Gamma^{mn[a}\, \delta^{b]c} + 2 \,
\Gamma^{ab[m} \, \delta^{n]c}\right) \, \psi \wedge V^c\nonumber\\
 & &+\overline{\psi} \wedge \, \Gamma^{mn} \, \psi \, F^{mnab} + \ft 1{24} \overline{\psi} \wedge \,
 \Gamma^{abc_1 \dots c_4} \, \psi \, F^{c_1 \dots c_4}
\label{rheoFDA}
\end{eqnarray}
The expressions (\ref{rheoFDA}) satisfy the Bianchi.s provided the
space--time components of the curvatures satisfy the following
constraints
\begin{eqnarray}
0 & = & \mathcal{D}_m F^{mc_1 c_2 c_3} \, + \, \ft 1{96} \,
\epsilon^{c_1c_2c_3 a_1 a_8} \, F_{a_1 \dots a_4}
\, F_{a_5 \dots a_8}  \label{maxwell}\\
0 & = & \Gamma^{abc} \, \rho_{bc} \label{gravitino}\\
R^{am}_{\phantom{bm}cm} & = & 6 \, F^{ac_1c_2c_3} \,F^{bc_1c_2c_3} -
\, \ft 12 \, \delta^a_b \, F^{c_1 \dots c_4} \,F^{c_1 \dots
c_4}\label{fieldeque}
\end{eqnarray}
which are the  space--time field equations. 
\subsection{M2-brane solutions with \texorpdfstring{$\mathbb{R}_+\times \mathrm{T^7}$}{R+ x T7} in the transverse dimensions }
\label{EngleM2T7} Among all the possible solutions to the field
equations (\ref{maxwell}-\ref{fieldeque}) we are interested in those
that describe $M2$-branes of the form described below.
\par
According to the general rules of brane-chemistry (see for
instance \cite{maiobuk}, page 288 and following ones), we introduce
the following $d=11$ metric:
\begin{equation}\label{colabrodo}
    ds_{11}^2 \, = \, H(y)^{-\ft{4 \tilde{d}}{9\Delta}}\left(d\xi^\mu \otimes d\xi^\nu\,
    \eta_{\mu\nu}\right)\, - \,H(y)^{\ft{4 {d}}{9\Delta}}\left(dy^I \otimes dy^J\,
    \delta_{IJ}\right)
\end{equation}
where:
\begin{equation}\label{worldindex}
    \xi^\mu \quad ; \quad \mu \, = \, \underline{0},\underline{1},\underline{2}
\end{equation}
are the coordinates on $\mathrm{Mink}_{1,2}$, while:
\begin{equation}\label{trasversale}
    y^I \quad ; \quad I \, = \, 1,2,\,\dots ,8
\end{equation}
are the coordinates of the 8-dimensional transverse space. Since in
$d=11$ there is no dilaton  we have
\begin{equation}\label{bamboccione}
    \Delta \, = \, 2 \frac{\tilde{d} \, d}{9} \, = \, 2 \, \frac{6 \times 3}{9} \, = \, 4 \quad ; \quad d=3 \, ;
    \quad \tilde{d} =6
\end{equation}
and the appropriate $M2$ Ansatz for the metric becomes (\ref{colabrodone}):
\begin{equation}
    ds_{11}^2 \, = \, H(y)^{-\ft{2}{3}}\left(d\xi^\mu \otimes d\xi^\nu\,
    \eta_{\mu\nu}\right)\, - \,H(y)^{\ft{1}{3}}\left(dy^I \otimes dy^J\,
    \delta_{IJ}\right)
\end{equation}
Because of the chosen topology of the transverse space, see Eq. (\ref{rabolatus}), it is
convenient to set:
\begin{equation}\label{faccialei}
    y^8 \, = \, U \in \mathbb{R}_+ \quad ; \quad y^i \, = \, x^i \in \mathrm{T^7} \quad (i=1,\dots,7)
\end{equation}
The next point is to choose an appropriate Ansatz for the three-form
$\mathbf{A^{[3]}}$. We set:
\begin{equation}\label{coffilasio}
    \mathbf{A^{[3]}}\, = \, \frac{2}{H(y)} \, \Omega^{[3]} \, + \, e^{-\mu \,U} \mathbf{Y}^{[3]}
\end{equation}
where:
\begin{eqnarray}
  \Omega^{[3]} &=& \ft{1}{6} \epsilon_{\mu\nu\rho} \, d\xi^\mu \wedge d\xi^\nu \wedge d\xi^\rho \label{wvol}\\
 \mathbf{Y}^{[3]} &=& Y_{ijk}(x)\, dx^i \wedge dx^j \wedge dx^k
\end{eqnarray}
The essential point in the above formula is that the antisymmetric
tri-tensor $Y_{ijk}(x)$ depends only on the coordinates $x$ of the
seven-torus $\mathrm{T^7}$. The geometry of $\mathrm{T^7}$ is defined by a lattice $\Lambda$ whose point group is the ${\rm PSL(2,7)}$ group to be introduced in the next Sections. 
\par
As shown in \cite{miol168}, with the Ansatz (\ref{coffilasio}), the
non-vanishing components of the $4$-form $\mathbf{F}^{[4]}$ are the
following ones:
\begin{eqnarray}
  F_{\underline{abc}I} &=& \frac{1}{12 }\,H(y)^{-\ft 76}\,\partial_I H(y) \\
 F_{8ijk} &=& - \frac{\mu}{4} e^{-\mu \, U}\, H(y)^{-\ft 23}\, Y_{ijk}  \\
F_{ijk\ell} &=& H(y)^{-\ft 23} \, e^{-\mu \,U} \partial_i Y_{jk\ell}
\end{eqnarray}
Then we can easily verify that the Maxwell field equation
(\ref{maxwell}) is satisfied provided the following two differential
constraints hold:
\begin{eqnarray}\label{harmoniusca}
    \Box_{\mathbb{R}_+\times \mathrm{T^7}} H(y)& = & \frac{\mu}{4}\, e^{-2\,\mu \,
    U}\epsilon^{ijk\ell mnr} \,\partial_i Y_{jk\ell} \,Y_{mnr}\label{2harmoniusca2}\\
\frac{1}{4!}\epsilon^{pqrijk\ell}\,\partial_iY_{jk\ell} &=&
-\,\frac{\mu}{4} \, Y_{pqr}\label{englerta}
\end{eqnarray}
The two equations admit the following index-free rewriting:
\begin{eqnarray}\label{FharmoniuscaF}
    \Box_{\mathbb{R}_+\times \mathrm{T^7}} H(y)& = & - \frac{3\,\mu^2}{2}\,
    e^{-2\,\mu \,U}\, \parallel\mathbf{Y}\parallel^2 \, \equiv \, J(y)\label{3harmoniusca3}\\
\star_{\,\mathrm{T^7}} \,\mathrm{d} \mathbf{Y}^{[3]}
&=& -\,\frac{\mu}{4} \, \mathbf{Y}^{[3]}\label{englertaF}
\end{eqnarray}
As we see Eq. (\ref{englertaF}) is the generalization to a
$7$-dimensional torus of Beltrami equation on the three-dimensional
one. It is just Englert equation and in the present work we pursue a new systematic group theoretical approach to the construction of its solutions and the study of their supersymmetries. 
\par
As shown in \cite{miol168},  Einstein equations are also satisfied
once Eq.s (\ref{3harmoniusca3}-\ref{englertaF}) are satisfied.\par
As mentioned earlier and shown in detail in Appendix \ref{travertinus}, these solutions fall in the general class of M2-branes with self-dual transverse flux.
\section{Normal form and the role of \texorpdfstring{$\mathrm{PSL(2,7)}$}{PSL(2,7)}}
\label{sembranormale}
Our approach to a systematic study of the solutions  to the Englert equation is to construct elementary solutions in which the 
only non-vanishing components $Y_{ijk}$ (the internal
part of the $3$-form ${\bf A}^{[3]}$) are defined by the \emph{normal form} of the representation ${\bf 35}$ with respect to the action of ${\rm SO(7)}$. The normal form is defined by the subspace of the representation space $V_{{\bf 35}}=\{Y_{ijk}\}$ of least dimension, in which a generic vector in $V_{{\bf 35}}$ can be rotated by means of an ${\rm SO}(7)$ transformation.
This subspace has 14 parameters since a
generic vector in $V_{{\bf 35}}$ has a trivial little group in
${\rm SO(7)}$, so that the number of parameters of a generic element
modulo the action of ${\rm SO(7)}$ is just $14=35-21$. The normal
form of $Y_{ijk}$ can be chosen in various ways, some of which have
a special geometric interpretation. It is important to stress that in our solution $Y_{ijk}(x)$ are not constant and thus in general one cannot recover the most general tensor $Y_{ijk}(x)$ from its restriction to the normal form through an ${\rm SO(7)}$ transformation. Nevertheless
the normal form will define elementary tensors satisfying the Englert equation which are the building blocks for our systematic study of its solutions.
\par
As mentioned in the Introduction, the components of $Y_{ijk}$  can be put
into one-to-one correspondence with a the weights of the ${\bf 35}$ representation of ${\rm SL}(7,\mathbb{R})$ group acting linearly on $x^i$.
 Formally these weights can be thought of as part of the $63$
positive roots of an $\mathfrak{e}_{7(7)}$ Lie algebra. The latter has a special role in $d=11$ supergravity since it generates the global symmetry group ${\rm E_{7(7)}}$ of the $d=4$ supergravity obtained from the eleven-dimensional one through toroidal reduction \cite{Cremmer:1979up}. However, it must be emphasized at this point that our solutions in general do not admit an effective $d=4$ description and that they are covariant only with respect to the ${\rm SL}(7,\mathbb{R})$ subgroup of ${\rm E_{7(7)}}$. The action of ${\rm SL}(7,\mathbb{R})$ will change the metric on the torus into a different constant one. The action of the ${\rm SO}(7)$ subgroup of ${\rm SL}(7,\mathbb{R})$, leaves the metric $\delta_{ij}$ on $\mathrm{T^7}$ invariant but transforms the lattice $\Lambda$ defining it. The latter is left invariant only by its point group which is a subgroup of ${\rm SO}(7)$ and which will be chosen to be ${\rm PSL(2,7)}$.

Let us denote by $\alpha_{ijk}$ the $\mathfrak{e}_{7(7)}$ positive roots corresponding to $Y_{ijk}$. The action of ${\rm SO(7)}$ on $Y_{ijk}$ can be
fixed by choosing seven non-vanishing components to correspond to a
maximal subset of mutually orthogonal roots $\alpha^{(I)}$,
$I=1,\ldots,7$ among the the 35 that we named $\alpha_{ijk}$.\footnote{With an abuse of notation we use the same letters to label $\alpha^{(I)}$ and the eight transverse directions to the M2-brane. The different interpretation of these letters will be clear from the context.} The
normal form is then obtained by complementing this set of components
with an other set of seven parameters, so that the total number of
independent components amounts to 14.
\par
Adopting this viewpoint the normal form is defined by two septuples
of parameters, the first of which is, defined,  as we have said by
the roots $\alpha^{(I)}$.
\par
Let us recall the main properties of this particular set of seven
roots. They define, together with their negative $-\alpha^{(I)}$, an $\mathfrak{sl}(2)^7$ subalgebra of $\mathfrak{e}_7$. Moreover it can be shown that the seven triplets $(i,j,k)$ of indices defining the $\alpha^{(I)}$ among the $\alpha_{ijk}$ form a so-called Steiner triple system and are in one-to-one correspondence with the vertices of a Fano plane (see Figure \ref{Fanoex} for a particular choice of this septuple).
It is at this level that the group $\mathrm{PSL(2,7)}$ enters the
game. As we show in the next section \ref{L168teoria} entirely
devoted to an in depth discussion of $\mathrm{PSL(2,7)}$, of his
subgroups and of its representations, this simple group has a
crystallographic action on the $\mathfrak{e}_7$ root lattice and
actually maps the $\mathfrak{e}_{7}$ root system
$\Delta_{\mathfrak{e}_7}$ into itself, so that it happens to be  a
subgroup of the $\mathrm{Weyl}[\mathfrak{e}_7]$ group.
\par
We anticipate that we can have two distinct conjugacy classes of
embeddings:
\begin{equation}\label{sposalizio}
    \mathrm{PSL(2,7)} \,\hookrightarrow \,\mathrm{Weyl}[\mathfrak{e}_7]
\end{equation}
one based on the $7$-dimensional irreducible representation of
$\mathrm{PSL(2,7)}$, the other on its $6$-dimensional one. With
respect to the former embedding there are no orbits of length 7 in
the $\mathfrak{e}_7$ root lattice and in particular in the root
system $\Delta_{\mathfrak{e}_7}$. With respect to the latter
embedding, as discussed below, there are instead orbits of length 7 and the unique such
one that is contained in the subset of 35 positive roots 
$\alpha_{ijk}$ precisely consists of the septuple $\alpha^{(I)}$ of mutually
orthogonal roots that define the embedding of the
$\mathfrak{sl}(2)^7 $ subalgebra into the $\mathfrak{e}_{7(7)}$. This embedding of $\mathrm{PSL(2,7)}$ indeed acts as the automorphism group of the 
Fano plane associated with this septuple since, as a subgroup of ${\rm Weyl[\mathfrak{a}_6]}=S_7$, its effect is of permuting the $\alpha^{(I)}$s.
\par
Hence, as firstly shown in \cite{Cerchiai:2010tk}, there are 135
inequivalent choices of the septuple of commuting roots which is the
ratio between the order of  $\mathrm{Weyl}[\mathfrak{e}_7]$ and the
order of the product of $(\mathbb{Z}_2)^7$ (that reverses the sign of
each $\alpha^{(I)}$), times the order of $\mathrm{PSL(2,7)}$ that
permutes the $\alpha^{(I)}$s in the septuple:
\begin{equation}
135=\frac{|\mathrm{Weyl}[\mathfrak{e}_7]|}{2^7\times
|\mathrm{PSL(2,7)}|}=\frac{ 2903040}{2^7\times 168}\,.
\end{equation}
Let us refer to the two conjugacy classes of $\mathrm{PSL(2,7)}$
subgroups within $\mathrm{Weyl}[\mathfrak{e}_7]$ as:
\begin{equation}
    \mathrm{PSL(2,7)_7} \, \subset \,
    \mathrm{Weyl}[\mathfrak{e}_7] \quad ; \quad \mathrm{PSL(2,7)_{1+6}} \, \subset \,
    \mathrm{Weyl}[\mathfrak{e}_7]\, .
\end{equation}
The reason for this naming, thoroughly explained in section
\ref{L168teoria}, is that the embedding into the Weyl group occurs
via the crystallographic embedding into the point group
$\mathrm{SO\left(7,\mathbb{Z}\right)_{\mathfrak{e}_7}}$ of the root
lattice $\Lambda^{\mathbf{r}}_{\mathfrak{e}_7}$:
\begin{equation}\label{subiudice}
    \mathrm{PSL(2,7)_7} \, \hookrightarrow \,
    \mathrm{SO\left(7,\mathbb{Z}\right)_{\mathfrak{e}_7}}
    \quad ; \quad \mathrm{PSL(2,7)_{1+6}} \, \hookrightarrow \,
    \mathrm{SO\left(7,\mathbb{Z}\right)_{\mathfrak{e}_7}}\, .
\end{equation}
By $\mathrm{SO\left(7,\mathbb{R}\right)_{\mathfrak{e}_7}}$ we denote
the standard $\mathrm{SO(7)}$ Lie group presented in the basis where
the invariant metric $\eta=\mathfrak{C}_{\mathfrak{e}_7}$ is the
Cartan matrix of the $\mathfrak{e}_7$ Lie algebra:
\begin{equation}\label{sapoco}
    L \, \in \,\mathrm{SO\left(7,\mathbb{R}\right)_{\mathfrak{e}_7}}
    \, \Leftrightarrow \, L^T \, \mathfrak{C}_{\mathfrak{e}_7} \, L
    \, = \, \mathfrak{C}_{\mathfrak{e}_7} \, .
\end{equation}
The point group of the root lattice
$\mathrm{SO\left(7,\mathbb{Z}\right)_{\mathfrak{e}_7}} \subset
\mathrm{SO\left(7,\mathbb{R}\right)_{\mathfrak{e}_7}}$ is the
discrete subgroup made by those $7 \times 7$ matrices $L$ that
satisfy (\ref{sapoco}) and have integer valued entries. The two
embeddings (\ref{subiudice}) are distinguished by the fact that the
character of the $7$ dimensional representation realized by the
embedding is the irreducible character $\pmb{\chi_7^{irr}}$ of the
$7$ dimensional representation of $\mathrm{PSL(2,7)}$ in the first
case, while it is the sum of the irreducible characters
$\pmb{\chi_6^{irr}}\oplus \pmb{\chi_1^{irr}}$ in the second:
\begin{equation}\label{sparuto}
    \pmb{\chi}\left[\mathrm{PSL(2,7)_7}\right] \, =
    \,\pmb{\chi_7^{irr}} \quad ; \quad \pmb{\chi}\left[\mathrm{PSL(2,7)_{1+6}}\right] \, =
    \,\pmb{\chi_6^{irr}}\oplus \pmb{\chi_1^{irr}}.
\end{equation}
Choosing the embedding $\mathrm{PSL(2,7)_{1+6}}$ we obtain that the
root lattice of the $\mathfrak{a}_6$ subalgebra of $\mathfrak{e}_7$
is left invariant by the action of $\mathrm{PSL(2,7)_{1+6}}$. This
obviously extends to the weight lattice of the same algebra. It
follows that the set of positive roots of $\mathfrak{e}_7$ splits
into subsets corresponding to irreducible representations of
$\mathfrak{a}_6\sim \slal(7,\mathbb{R})$. In particular a group of
$\mathbf{35}$ positive roots corresponds to the weights of the $35$
irreducible representation of $\slal(7,\mathbb{R})$, the three times
antisymmetric, which means the tensor $Y_{ijk}$. This is the
rigorous definition of the roots $\alpha_{ijk}$ mentioned above.
\par
The $\mathbf{35}$ dimensional set is invariant under the action of
$\mathrm{PSL(2,7)_{1+6}}$ and splits in two orbits:
\begin{equation}\label{carisma}
    \mathbf{35} \, \stackrel{\mathrm{PSL(2,7)_{1+6}}}{\Longrightarrow} \, \mathbf{7}_A \oplus \mathbf{28}.
\end{equation}
The orbit $\mathbf{7}_A$, group theoretically defined in a unique
way, provides, as mentioned above, the set of $7$ mutually commuting roots $\alpha^{(I)}$ and, in the
correspondence between $\mathfrak{a}_7$ weights and the triples of
indices $\{ijk\}$ (see table \ref{pesotti}) a first  septuple of
\textit{Steiner triples}. Such system of triples can be characterized by the property that any two of the
seven triplets of indices $\{ijk\}$ must have only one index in
common. 
\par
We can easily count the possible number of the septuples
$\mathbf{7}_A$ measuring the number of conjugate copies of the group
$\mathrm{PSL(2,7)_{1+6}}$ inside $\mathrm{Weyl}[\mathfrak{a}_6]
\subset \mathrm{Weyl}[\mathfrak{e}_7]$:
\begin{equation}
\#\text{  of septuples $\mathbf{7}_A$}\, =\,
\frac{|\mathrm{Weyl}[\mathfrak{a}_6] |}{2^7\times
|\mathrm{PSL(2,7)_{1+6} }|} \, = \, \frac{7!}{2^7\times 168} \, = \,
30.
\end{equation}

\par
The second step in order to obtain the $14$ parameters of the normal
form is to adjoin to septuple $\mathbf{7}_A$ of Steiner triples a
second septuple $\mathbf{7}_B$ which is complementary to the first.
\par
The concept of complementarity is briefly described in the lines
below.
\par
\par Let us denote a set of seven
triples of indices $\{ijk\}$, $1\le i<j<k\le 7$, by $\vec{\sigma}$:
\begin{equation}
\vec{\sigma}=\{\vec{\sigma}_I\}_{I=1,\dots,7}\,\,\,,\,\,\,\,\vec{\sigma}_I
=(\sigma^1_I,\sigma^2_I,\sigma^3_I)\,\,,\,\,\,\,
1\le\sigma^1_I<\sigma^2_I<\sigma^3_I\le 7\,.
\end{equation}
If $P$ is a permutation of the seven values of the index $I$
labeling the triplets in $\vec{\sigma}$, we shall denote the
permuted set of triplets by $\vec{\sigma}\cdot P$:
\begin{equation}
\vec{\sigma}\cdot P=\{\vec{\sigma}_{P(I)}\}_{I=1,\dots,7}\,.
\end{equation}
Two septuples $\vec{\sigma}$ and $\vec{\gamma}$ are \emph{complementary} or \emph{mutually non-local}
if there exist two permutations $P,\,P'\in S_7$ such that:
\begin{equation}
\forall
I=1,\dots,7\,\,:\,\,\,\,\,\,\epsilon^{i_I\,\sigma^1_{P(I)}\sigma^2_{P(I)}\sigma^3_{P(I)}
\gamma^1_{P'(I)}\gamma^2_{P'(I)}\gamma^3_{P'(I)}}\neq 0\,,\label{epsilonPP}
\end{equation}
where $I\,\rightarrow\,i_I$ is a mapping of the set $\{1,2,3,4,5,6,7\}$ into itself which need not be onto. For $I=1,\dots, 7$, the numbers $i_I$ are uniquely defined by the condition (\ref{epsilonPP}). 

The selection of a septuple $\mathbf{7}_B$ complementary to the
septuple $\mathbf{7}_A$ can be derived automatically in a group
theoretical way considering the maximal subgroup of order $21$ of
$\mathrm{PSL(2,7)_{1+6}}$, denoted $\mathrm{G_{21}}$ (see section
\ref{g21gruppo}). Under the action of $\mathrm{G_{21}}$ we have:
\begin{equation}\label{marasma}
    \mathbf{35} \, \stackrel{\mathrm{PSL(2,7)_{1+6}}}{\Longrightarrow} \,
    \mathbf{7}_A \oplus \mathbf{28}\, \stackrel{\mathrm{G_{21}}}{\Longrightarrow} \,
    \mathbf{7}_A \oplus \mathbf{7}_B \oplus \mathbf{21}
\end{equation}
and the septuple $\mathbf{7}_B$ is automatically complementary to
septuple $\mathbf{7}_B$. How many are the possible choices of $\mathbf{7}_B$
for fixed $\mathbf{7}_A$? There is an easy answer: they are as many
as the different subgroups
$\mathrm{G_{21}}\subset\mathrm{PSL(2,7)_{1+6}}$ in the unique
conjugacy class, namely:
\begin{equation}\label{pallinus}
    \#\text{  of septuples $\mathbf{7}_B$}\, =\,
\frac{|\mathrm{PSL(2,7)_{1+6} }|}{|\mathrm{G_{21} }|} \, = \,
\frac{168}{21} \, = \, 8.
\end{equation}
With the above preliminary arguments and anticipations we have
illustrated the crucial role played by the group $\mathrm{PSL(2,7)}$
in deriving a normal $14$-parameter form of the solution to Englert
equation. In particular in Section \ref{minsol} we shall illustrate how to construct a solution from a couple of complementary septuples $\vec{\sigma},\,\vec{\gamma}$, see Eq. (\ref{seed}).\par
 In the next long section we present the theory of
$\mathrm{PSL(2,7)}$ in a systematic way, providing a great deal of relevant
constructive details about representations, subgroups,
crystallographic action on root lattices and orbits that, up to our
knowledge, are not available in the mathematical literature. After
such a preparation we will return to the explicit construction of
the normal form of the solution to Englert equation in section
\ref{baciogallone}.\par
 The reader who is only interested in the construction of the solutions to the Englert equation and the study of their supersymmetry can skip the next two mathematical Sections and move directly to Section \ref{baciogallone}.
\section{Theory of the simple group \texorpdfstring{$\mathrm{PSL(2,7)}$}{PSL(2,7)}}
\label{L168teoria} Since the finite simple group $\mathrm{PSL(2,7)}$
plays a fundamental role in the derivation of the normal form of
solutions to the Englert equation (\ref{engleterro}) we devote the
present section and its subsections to the structural theory of this
remarkable group. One of its most relevant  properties, which turns
out to be quite momentous for M--theory and was not duely observed
in the mathematical literature, is that it is crystallographic in
$7$-dimensions. It is also crystallographic in $6$ dimensions. In
both cases the crystallographic representation corresponds to the
irreducible representation of the same dimension predicted by
general group theory; furthermore the lattice that is left invariant
by the action of the $\mathrm{PSL(2,7)}$ group is, respectively, the
root lattice $\Lambda_{\mathfrak{a}_7}^r$ and the root lattice
$\Lambda_{\mathfrak{a}_6}^r$, having denoted by $\mathfrak{a}_\ell$
the simple complex Lie algebra whose maximal split real form is
$\spl(\ell+1,\mathbb{R})$. Because of duality it follows that also
the corresponding weight lattices $\Lambda_{\mathfrak{a}_7}^w$ and
$\Lambda_{\mathfrak{a}_6}^w$ are equally preserved by the action of
$\mathrm{PSL(2,7)}$ that is provided by integer valued matrices both
in the root and in the weight basis. Since the symmetric Cartan
matrices $\mathfrak{C}_{\mathfrak{a}_7}$ and
$\mathfrak{C}_{\mathfrak{a}_6}$ are left invariant by
$\mathrm{PSL(2,7})$ it follows that this latter has a natural
irreducible embedding both in $\mathrm{SO(7)}$ and in
$\mathrm{SO(6)}$. Last but not least, since the root lattice
$\Lambda_{\mathfrak{a}_7}^r$ is a sublattice of the $\mathfrak{e}_7$
root lattice it follows that $\mathrm{PSL(2,7)}$ is crystallographic
with respect also to this latter and is actually a subgroup of the
Weyl group $\mathrm{Weyl}[\mathfrak{e}_7]$. It is just this property
what provides the link of $\mathrm{PSL(2,7)}$ with exceptional field
theory and with the solutions of Englert equation.
\subsection{Definition of the group \texorpdfstring{$\mathrm{PSL(2,7)}$}{PSL(2,7)}}
The finite group:
\begin{equation}\label{L168}
  \mathrm{PSL(2,7)}\, \equiv \, \mathrm{PSL(2,\mathbb{Z}_7)}
\end{equation}
is the second smallest simple group after the alternating group
$A_5$ which has 60 elements and coincides with the symmetry group of
the regular icosahedron or dodecahedron.
$\mathrm{PSL(2,7)}$  has 168 elements: they can be
identified with all the possible $2\times2$ matrices with
determinant one whose entries belong to the finite field
$\mathbb{Z}_7$, counting them up to an overall sign. In projective
geometry, $\mathrm{PSL(2,7)}$ is classified as a
\textit{Hurwitz group} since it is the automorphism group of a
Hurwitz Riemann surface, namely a surface of genus $g$ with the
maximal number $84\,(g-1)$ of conformal
automorphisms\footnote{Hurwitz's automorphisms theorem proved in
1893  states that the order $|\mathcal{G}|$ of the group
$\mathcal{G}$ of orientation-preserving conformal automorphisms, of
a compact Riemann surface of genus $g > 1$ admits the following
upper bound $|\mathcal{G}| \le 84(g - 1)$}. The Hurwitz surface
pertaining to the Hurwitz group $\mathrm{PSL(2,7)}$ is the
Klein quartic, namely the locus $\mathcal{K}_4$ in
$\mathbb{P}_2(\mathbb{C})$ cut out by the following quartic
polynomial constraint on the homogeneous coordinates $\{x,y,z\}$:
\begin{equation}\label{kleinusquart}
   x^3 \,y + y^3 \, z \, + z^3 \, x \, = \, 0
\end{equation}
Indeed $\mathcal{K}_4$ is a genus $g=3$ compact Riemann surface and
it can be realized as the quotient of the hyperbolic Poincar\'e
plane $\mathbb{H}_2$ by a certain  group $\Gamma$ that acts freely
on $\mathbb{H}_2$ by isometries.
\par
The $\mathrm{PSL(2,7)}$ group, which is also isomorphic to
$\mathrm{GL(3,\mathbb{Z}_2)}$, has received a lot of attention in
Mathematics and it has  important applications in algebra, geometry,
and number theory: for instance, besides being associated with the
Klein quartic, $\mathrm{PSL(2,7)}$ is the automorphism
group of the Fano plane.
\par
The reason why we consider $\mathrm{PSL(2,7)}$ in this
section is associated with another property of this finite simple
group which was proved almost twenty years ago in \cite{kingus},
namely:
\begin{equation}\label{caralino}
    \mathrm{PSL(2,7)} \, \subset \, \mathrm{G_{2(-14)}}
\end{equation}
This means that $\mathrm{PSL(2,7)}$ is a finite subgroup of
the compact form of the exceptional Lie group $\mathrm{G_2}$ and the
$7$-dimensional fundamental representation of the latter is
irreducible upon restriction to $\mathrm{PSL(2,7)}$.
\par
As we already mentioned the group $\mathrm{PSL(2,7)}$ is
crystallographic in $d=7$, and in $d=6$.
\subsection{Structure of \texorpdfstring{$\mathrm{PSL(2,7)}$}{PSL(2,7)}}
For the reasons outlined above we consider the simple group
(\ref{L168}) and its crystallographic action in $d=7$. The Hurwitz
simple group  $\mathrm{PSL(2,7)}$ is abstractly presented
as follows\footnote{In the rest of this section we follow closely
the results obtained by the present author in a recent paper
\cite{miol168}}:
\begin{equation}\label{abstroL168}
\mathrm{PSL(2,7)} \, = \, \left(R,S,T \,\parallel \, R^2 \,
= \, S^3 \, = \, T^7 \, = \, RST \, = \, \left(TSR\right)^4 \, = \,
\mathbf{e}\right)
\end{equation}
and it has order 168:
\begin{equation}\label{order168}
  \mid \mathrm{PSL(2,7)} \mid \, = \, 168
\end{equation}
For practical convenience we distinguish the abstract description of
the group, from its concrete realization in terms of matrices, by
rewriting Eq. (\ref{abstroL168}) in terms of abstract generators
denoted by the corresponding greek letters:
\begin{equation}\label{abstroL168Ab}
\mathrm{PSL(2,7)} \, = \, \left(\rho,\sigma,\tau
\,\parallel \, \rho^2 \, = \, \sigma^3 \, = \, \tau^7 \, = \,
\rho.\sigma.\tau \, = \, \left(\tau.\sigma.\rho\right)^4 \, = \,
\epsilon\right)
\end{equation}
In this way we can give an exhaustive enumeration of all the group
elements as words in the three symbols $\rho$,$\sigma$,$\tau$.
\par
 The elements of this simple group are organized in six conjugacy classes
according to the scheme displayed below:
\begin{equation}\label{coniugini}
\mbox{
\begin{tabular}{||c|c|c|c|c|c|c||}
\hline \hline
Conjugacy class &$\mathcal{C}_1$&$\mathcal{C}_2$&$\mathcal{C}_3$&$\mathcal{C}_4$
&$\mathcal{C}_5$&$\mathcal{C}_6$\\
  \hline
  \hline
  representative of the class  & $\mathbf{e}$ & $R$ & $S$ &$TSR$ & $T$ & $SR$ \\
  \hline
  order of the elements in the class & 1 & 2 & 3 & 4 & 7 & 7 \\
  \hline
  number of elements in the class & 1 & 21 & 56 & 42 & 24 & 24  \\
  \hline
  \hline
\end{tabular}}
\end{equation}
As one sees from the above table (\ref{coniugini}) the group
contains elements of order $2$, $3$, $4$ and $7$ and there are two
inequivalent conjugacy classes of elements of the highest order.
According to the general theory of finite groups, there are $6$
different irreducible representations of dimensions $1,6,7,8,3,3$,
respectively. The character table of the group
$\mathrm{PSL(2,7)}$ can be found in the mathematical
literature. It reads as follows:
\begin{center}
\begin{tabular}{||c|c|c|c|c|c|c||}
\hline \hline
Representation &$\mathcal{C}_1$&$\mathcal{C}_2$&$\mathcal{C}_3$&$\mathcal{C}_4$
&$\mathcal{C}_5$&$\mathcal{C}_6$\\
  \hline
 $\mathrm{D_1}\left[\mathrm{PSL(2,7)}\right]$  & $1$ & $1$ & $1$ &$1$ & $1$ & $1$ \\
\hline
$\mathrm{D_6}\left[\mathrm{PSL(2,7)}\right]$  & $6$ & $2$ & $0$ &$0$ & $-1$ & $-1$ \\
\hline
$\mathrm{D_7}\left[\mathrm{PSL(2,7)}\right]$  & $7$ & $-1$ & $1$ &$-1$ & $0$ & $0$ \\
\hline
$\mathrm{D_8}\left[\mathrm{PSL(2,7)}\right]$  & $8$ & $0$ & $-1$ &$0$ & $1$ & $1$ \\
\hline $\mathrm{DA_{3}}\left[\mathrm{PSL(2,7)}\right]$  &
$3$ & $-1$ & $0$ &$1$ & $\ft 12
\left(-1+{\rm i}\sqrt{7}\right)$ & $\ft 12 \left(-1-{\rm i}\sqrt{7}\right)$ \\
\hline $\mathrm{DB_{3}}\left[\mathrm{PSL(2,7)}\right]$  &
$3$ & $-1$ & $0$ &$1$ & $\ft 12
\left(-1-{\rm i}\sqrt{7}\right)$& $\ft 12 \left(-1+{\rm i}\sqrt{7}\right)$ \\
  \hline
  \hline
\end{tabular}
\begin{equation}\label{caratterini}
\null
\end{equation}
\end{center}
Soon we will retrieve it  by constructing explicitly all the
irreducible representations
\subsection{The 7-dimensional irreducible representation}
The two representations most relevant for our purposes are the $7$
and the $6$-dimensional ones. We begin with the former.
\par
The following three statements are true:
\begin{enumerate}
  \item The $7$-dimensional irreducible representation  is crystallographic
  since all elements $\gamma\in \mathrm{PSL(2,7)}$ are represented by integer valued
  matrices $D_7\left(\gamma\right)$ in a basis of vectors that span a lattice,
  namely the root lattice $\Lambda^{\mathbf{r}}_{\mathfrak{a}_7}$ of the
  $\mathfrak{a}_7$ simple Lie algebra.
  \item The $7$-dimensional irreducible representation provides an immersion
  $\mathrm{PSL(2,7)} \hookrightarrow \mathrm{SO(7)}$ since its elements preserve
  the symmetric Cartan matrix of $A_7$:
  \begin{eqnarray}
  \forall \gamma \in \mathrm{PSL(2,7)}\quad:\quad D_7^T\left(\gamma\right) \,
  \mathfrak{C}_{\mathfrak{a}_7} \, D_7\left(\gamma\right)& = & \mathfrak{C}_{\mathfrak{a}_7}\nonumber \\
  \mathfrak{C}_{i,j} & = & \alpha_i \cdot \alpha_j \quad\quad \quad  (i,j \, =\,1 \,\dots ,7) \label{gomorito}
  \end{eqnarray}
  defined in terms of the simple roots $\alpha_i$ whose standard construction
  in terms of the unit vectors $\epsilon_i$ of $\mathbb{R}^8$ is recalled below:
      \begin{equation}\label{simplerutte}
      \begin{array}{cccccccccccc}
        \alpha_1 & = & \epsilon_1 -\epsilon_2 & ; & \alpha_2 & = &
        \epsilon_2-\epsilon_3 & = & ; &\alpha_3 & = & \epsilon_3 - \epsilon_4 \\
       \alpha_4 & = & \epsilon_4 -\epsilon_5 & ;
       & \alpha_5 & = & \epsilon_5-\epsilon_6 & = & ; &\alpha_6 & = & \epsilon_6 - \epsilon_7 \\
        \alpha_7 & = & \epsilon_7 -\epsilon_8 &\null & \null&\null & \null &\null &\null&\null &\null &\null \\
      \end{array}
      \end{equation}
  \item Actually the $7$-dimensional representation defines an
  embedding $\mathrm{PSL(2,7)} \hookrightarrow \mathrm{G_2} \subset \mathrm{SO(7)}$ since there exists
  a three-index antisymmetric tensor $\phi_{ijk}$ satisfying the relations of
  octonionic structure constants that is preserved by all the matrices $D_7(\gamma)$:
      \begin{equation}\label{cantonus}
      \forall \gamma \in \mathrm{PSL(2,7)} \quad : \quad
      D_7(\gamma)_{ii^\prime}\,D_7(\gamma)_{jj^\prime}\
      \,D_7(\gamma)_{kk^\prime}\,\phi_{i^\prime j^\prime k^\prime} \, = \, \phi_{ijk}
      \end{equation}
\end{enumerate}
\par
Let us prove the above statements. It suffices to write the explicit
form of the generators $R$, $S$ and $T$ in the crystallographic
basis of the considered root lattice:
\begin{equation}
 \mathbf{ v} \, \in \, \Lambda^{\mathbf{r}}_{\mathfrak{a}_7}  \quad \Leftrightarrow
 \quad  \mathbf{ v} \, = \, n_i \, \alpha_i \quad n_i \in \mathbb{Z}
\end{equation}
Explicitly if  we set:
\begin{eqnarray}\label{rgen}
R_7 \, = \,\mathcal{R} & \equiv &  \left(
\begin{array}{ccccccc}
 0 & 0 & 0 & 0 & 0 & 0 & -1 \\
 0 & 0 & 0 & 0 & 0 & -1 & 0 \\
 0 & 0 & -1 & 1 & 0 & -1 & 0 \\
 0 & -1 & 0 & 1 & 0 & -1 & 0 \\
 0 & -1 & 0 & 1 & -1 & 0 & 0 \\
 0 & -1 & 0 & 0 & 0 & 0 & 0 \\
 -1 & 0 & 0 & 0 & 0 & 0 & 0 \\
\end{array}
\right) \nonumber\\
 S_7 \, = \,\mathcal{S} & \equiv & \left(
\begin{array}{ccccccc}
 0 & 0 & 0 & 0 & 0 & 0 & -1 \\
 1 & 0 & 0 & 0 & 0 & 0 & -1 \\
 1 & 0 & 0 & -1 & 1 & 0 & -1 \\
 1 & 0 & -1 & 0 & 1 & 0 & -1 \\
 1 & 0 & -1 & 0 & 1 & -1 & 0 \\
 1 & 0 & -1 & 0 & 0 & 0 & 0 \\
 1 & -1 & 0 & 0 & 0 & 0 & 0 \\
\end{array}
\right) \nonumber\\
T_7 \, = \,\mathcal{T} & \equiv & \left(
\begin{array}{ccccccc}
 0 & 0 & 0 & 0 & 0 & -1 & 1 \\
 1 & 0 & 0 & 0 & 0 & -1 & 1 \\
 0 & 1 & 0 & 0 & 0 & -1 & 1 \\
 0 & 0 & 1 & 0 & 0 & -1 & 1 \\
 0 & 0 & 0 & 1 & 0 & -1 & 1 \\
 0 & 0 & 0 & 0 & 1 & -1 & 1 \\
 0 & 0 & 0 & 0 & 0 & 0 & 1 \\
\end{array}
\right)
\end{eqnarray}
we find that the defining relations of $\mathrm{PSL(2,7)}$
are satisfied:
\begin{equation}\label{relationibus}
 \mathcal{R}^2 \, = \, \mathcal{S}^3 \, = \, \mathcal{T}^7 \,=\,
 \mathcal{RST} \, = \, (\mathcal{TSR})^4 \, = \, \mathbf{1}_{7\times7}
\end{equation}
and furthermore we have:
\begin{equation}
  \mathcal{R}^T \,\mathfrak{C}_{\mathfrak{a}_7}\, \mathcal{R} \,=\, \mathcal{S}^T
  \,\mathfrak{C}_{\mathfrak{a}_7}\,\mathcal{ S} \, = \,
  \mathcal{T}^T \,\mathfrak{C}_{\mathfrak{a}_7}\, \mathcal{T} \, = \, \,\mathfrak{C}_{\mathfrak{a}_7}\,
\end{equation}
where the explicit form of the $\mathfrak{a}_7$ Cartan matrix is
recalled below:
\begin{equation}\label{cartanschula}
\,\mathfrak{C}_{\mathfrak{a}_7}\, = \,  \left(
\begin{array}{ccccccc}
 2 & -1 & 0 & 0 & 0 & 0 & 0 \\
 -1 & 2 & -1 & 0 & 0 & 0 & 0 \\
 0 & -1 & 2 & -1 & 0 & 0 & 0 \\
 0 & 0 & -1 & 2 & -1 & 0 & 0 \\
 0 & 0 & 0 & -1 & 2 & -1 & 0 \\
 0 & 0 & 0 & 0 & -1 & 2 & -1 \\
 0 & 0 & 0 & 0 & 0 & -1 & 2 \\
\end{array}
\right)
\end{equation}
This proves statements 1) and 2).
\par
In order to prove statement 3) we proceed as follows. In
$\mathbb{R}^7$ we consider the antisymmetric three-index tensor
$\phi_{ABC}$ that is required to satisfy the algebraic relations of
the octonionic structure constants, namely\footnote{In this equation
the indices of the $G_2$-invariant tensor are denoted with capital
letter of the Latin alphabet, as it was the case in the quoted
literature on weak $G_2$-structures. In the following we will use
lower case latin letters, the upper Latin letters being reserved for
$d=8$}:
\begin{eqnarray}\label{2colibri2}
  \phi_{ABM} \, \phi_{CDM} & = & \frac{1}{18} \delta^{AB}_{CD} \, +\, \frac{2}{3} \Phi_{ABCD} \\
  \phi_{ABC} & = &- \frac{1}{6} \epsilon_{ABCPQRS} \, \Phi_{ABCD}
\end{eqnarray}
The subgroup of $\mathrm{SO(7)}$ which leaves $\phi_{ABC} $
invariant is, by definition, the compact section
$\mathrm{G_{(2,-14)}}$ of the complex $\mathrm{G_2}$ Lie group. We
mention here two different realizations of the
$\mathrm{G_2}$-tensor, $\phi_{ABC}$ and $\varphi_{ABC}$, that we
utilize in the sequel in relation with two different irreducible
representations of $\mathrm{PSL(2,7)}$:
\begin{eqnarray}\label{gorlandus}
  \begin{array}{ccc|ccc}
 \phi_{1,2,7} &=&\frac{1}{6}& \varphi_{1,2,6} &=&\frac{1}{6} \\
 \phi_{1,3,5} &=& \frac{1}{6}& \varphi_{1,3,4} &=& - \frac{1}{6} \\
 \phi_{1,4,6} &=& \frac{1}{6}& \varphi_{1,5,7} &=& - \frac{1}{6} \\
  \phi_{2,3,6} &=& \frac{1}{6}& \varphi_{2,3,7} &=& \frac{1}{6} \\
 \phi_{2,4,5} &=& -\frac{1}{6}& \varphi_{2,4,5} &=& \frac{1}{6}  \\
  \phi_{3,4,7} &=& \frac{1}{6}& \varphi_{3,5,6} &=& -\frac{1}{6} \\
  \phi_{5,6,7} &=& -\frac{1}{6}& \varphi_{4,6,7} &=& -\frac{1}{6}\\
\end{array} &; & \mbox{all other components vanish}
\end{eqnarray}
A particular matrix that transforms the standard orthonormal basis
of $\mathbb{R}^7$ into the basis of simple roots $\alpha_i$ is the
following one:
\begin{equation}\label{mgothica}
  \mathfrak{M} \, = \, \left(
\begin{array}{ccccccc}
 \sqrt{2} & -\frac{1}{\sqrt{2}} & 0 & 0 & 0 & 0 & 0 \\
 0 & -\frac{1}{\sqrt{2}} & \sqrt{2} & -\frac{1}{\sqrt{2}} & 0 & 0 &
   0 \\
 0 & 0 & 0 & -\frac{1}{\sqrt{2}} & \sqrt{2} & -\frac{1}{\sqrt{2}} &
   0 \\
 0 & 0 & 0 & 0 & 0 & -\frac{1}{\sqrt{2}} & \sqrt{2} \\
 0 & -\frac{1}{\sqrt{2}} & 0 & \frac{1}{\sqrt{2}} & 0 &
   -\frac{1}{\sqrt{2}} & 0 \\
 0 & 0 & 0 & -\frac{1}{\sqrt{2}} & 0 & 0 & 0 \\
 0 & \frac{1}{\sqrt{2}} & 0 & 0 & 0 & -\frac{1}{\sqrt{2}} & 0 \\
\end{array}
\right)
\end{equation}
since:
\begin{equation}\label{rattus}
  \mathfrak{M}^T \, \mathfrak{M} \, = \, \mathfrak{C}_{\mathfrak{a}_7}\,
\end{equation}
Defining the transformed tensor:
\begin{equation}\label{perenospero}
  \hat{\varphi}_{ijk} \, \equiv \, \left(\mathfrak{M}^{-1}\right)_i^{\phantom{i}I} \,
  \left(\mathfrak{M}^{-1}\right)_j^{\phantom{j}J}
   \,  \left(\mathfrak{M}^{-1}\right)_k^{\phantom{k}K} \, \varphi_{IJK}
\end{equation}
we can explicitly verify that:
\begin{eqnarray}
\hat{\varphi}_{ijk} &=& \left(\mathcal{R}\right)_i^{\phantom{i}p} \,
\left(\mathcal{R}\right)_j^{\phantom{i}q}
   \,  \left(\mathcal{R}\right)_k^{\phantom{i}r} \, \hat{\varphi}_{pqr} \nonumber\\
 \hat{\varphi}_{ijk} &=& \left(\mathcal{S}\right)_i^{\phantom{i}p} \, \left(\mathcal{S}\right)_j^{\phantom{i}q}
   \,  \left(\mathcal{S}\right)_k^{\phantom{i}r} \, \hat{\varphi}_{pqr}\nonumber\\
\hat{\varphi}_{ijk} &=& \left(\mathcal{T}\right)_i^{\phantom{i}p} \,
\left(\mathcal{T}\right)_j^{\phantom{i}q}
   \,  \left(\mathcal{T}\right)_k^{\phantom{i}r} \, \hat{\varphi}_{pqr} \label{gargiulo}
\end{eqnarray}
Hence, being preserved by the three-generators $\mathcal{R}$,
$\mathcal{S}$ and $\mathcal{T}$, the antisymmetric tensor
$\varphi_{ijk}$ is preserved by the entire discrete group
$\mathrm{PSL(2,7)}$ which, henceforth, is  a subgroup of
$\mathrm{G_{(2,-14)}}\subset \mathrm{SO(7)}$, as it was shown by
intrinsic group theoretical arguments in \cite{kingus}. The other
representations of the group $\mathrm{PSL(2,7)}$ were
explicitly constructed about ten years ago by Pierre Ramond and his
younger collaborators in \cite{ramonus}. They are completely
specified by giving the matrix form of the three generators
$\rho,\sigma ,\tau $ satisfying the defining relations
\ref{abstroL168Ab}. For the $6$-dimensional representation we will
instead use the crystallographic basis provided by the
$\mathfrak{a}_6$ root lattice.
\subsection{The 6-dimensional representation} Introducing the following short-hand notation:
\begin{eqnarray}
  c_n &=& \cos \left[ \frac{2\pi}{7} \, n \right] \nonumber\\
 s_n &=& \sin \left[ \frac{2\pi}{7} \, n \right] \label{crisanus}
\end{eqnarray}
in \cite{ramonus} the generators of the group
$\mathrm{PSL(2,7)}$ in the $6$-dimensional irreducible
representation were
explicitly written as it is displayed below: {
\begin{eqnarray*}
  D[\rho]_6 &=& \left(
\begin{array}{|l@{\hspace{3pt}}|l@{\hspace{3pt}}|l@{\hspace{3pt}}|l@{\hspace{3pt}}|
l@{\hspace{3pt}}|l@{\hspace{3pt}}}
 \frac{c_3-1}{\sqrt{2}} & \frac{c_2-1}{\sqrt{2}} &
   \frac{c_1-1}{\sqrt{2}} & c_3-c_1 & c_1-c_2 &
   c_2-c_3 \\
   \hline
 \frac{c_2-1}{\sqrt{2}} & \frac{c_1-1}{\sqrt{2}} &
   \frac{c_3-1}{\sqrt{2}} & c_2-c_3 & c_3-c_1 &
   c_1-c_2 \\
   \hline
 \frac{c_1-1}{\sqrt{2}} & \frac{c_3-1}{\sqrt{2}} &
   \frac{c_2-1}{\sqrt{2}} & c_1-c_2 & c_2-c_3 &
   c_3-c_1 \\
   \hline
 c_3-c_1 & c_2-c_3 & c_1-c_2 &
   \frac{c_1-1}{\sqrt{2}} & \frac{c_2-1}{\sqrt{2}} &
   \frac{c_3-1}{\sqrt{2}} \\
   \hline
 c_1-c_2 & c_3-c_1 & c_2-c_3 &
   \frac{c_2-1}{\sqrt{2}} & \frac{c_3-1}{\sqrt{2}} &
   \frac{c_1-1}{\sqrt{2}} \\
   \hline
 c_2-c_3 & c_1-c_2 & c_3-c_1 &
   \frac{c_3-1}{\sqrt{2}} & \frac{c_1-1}{\sqrt{2}} &
   \frac{c_2-1}{\sqrt{2}} \\
\end{array}
\right)
\end{eqnarray*}}
{
\begin{eqnarray*}
&D[\sigma]_6 \, =\,&\nonumber\\
  &\left(
\begin{array}{l@{\hspace{2pt}}|l@{\hspace{1pt}}|l@{\hspace{1pt}}|l@{\hspace{1pt}}|l@{\hspace{1pt}}|
l@{\hspace{1pt}}}
 \frac{(c_3-1) \rho ^2}{\sqrt{2}} & \frac{(c_2-1)
   \rho ^4}{\sqrt{2}} & \frac{(c_1-1) \rho }{\sqrt{2}}
   & (c_3-c_1) \rho ^3 & (c_1-c_2) \rho ^5 &
   (c_2-c_3) \rho ^6 \\
   \hline
 \frac{(c_2-1) \rho ^2}{\sqrt{2}} & \frac{(c_1-1)
   \rho ^4}{\sqrt{2}} & \frac{(c_3-1) \rho }{\sqrt{2}}
   & (c_2-c_3) \rho ^3 & (c_3-c_1) \rho ^5 &
   (c_1-c_2) \rho ^6 \\
   \hline
 \frac{(c_1-1) \rho ^2}{\sqrt{2}} & \frac{(c_3-1)
   \rho ^4}{\sqrt{2}} & \frac{(c_2-1) \rho }{\sqrt{2}}
   & (c_1-c_2) \rho ^3 & (c_2-c_3) \rho ^5 &
   (c_3-c_1) \rho ^6 \\
   \hline
 (c_3-c_1) \rho ^2 & (c_2-c_3) \rho ^4 &
   (c_1-c_2) \rho  & \frac{(c_1-1) \rho
   ^3}{\sqrt{2}} & \frac{(c_2-1) \rho ^5}{\sqrt{2}} &
   \frac{(c_3-1) \rho ^6}{\sqrt{2}} \\
   \hline
 (c_1-c_2) \rho ^2 & (c_3-c_1) \rho ^4 &
   (c_2-c_3) \rho  & \frac{(c_2-1) \rho
   ^3}{\sqrt{2}} & \frac{(c_3-1) \rho ^5}{\sqrt{2}} &
   \frac{(c_1-1) \rho ^6}{\sqrt{2}} \\
   \hline
 (c_2-c_3) \rho ^2 & (c_1-c_2) \rho ^4 &
   (c_3-c_1) \rho  & \frac{(c_3-1) \rho
   ^3}{\sqrt{2}} & \frac{(c_1-1) \rho ^5}{\sqrt{2}} &
   \frac{(c_2-1) \rho ^6}{\sqrt{2}} \\
\end{array}
\right)& \nonumber\\
\end{eqnarray*}}
\begin{eqnarray}
  D[\tau]_6 & = &\left( D[\rho]_6 \cdot D[\sigma]_6\right)^{-1}\label{parapatto}
\end{eqnarray}
and where shown to satisfy the required relations
(\ref{abstroL168Ab}).
\par
We rather introduce the crystallographic basis in a completely
analogous way to the case of the $7$-dimensional irreducible
representation.
\par
 The following two statements are true:
\begin{enumerate}
  \item The $6$-dimensional irreducible representation  is crystallographic
  since all elements $\gamma\in \mathrm{PSL(2,7)}$ are represented by integer valued
  matrices $D_6\left(\gamma\right)$ in a basis of vectors that span a lattice,
  namely the root lattice $\Lambda^{\mathbf{r}}_{\mathfrak{a}_6}$ of the
  $\mathfrak{a}_6$ simple Lie algebra.
  \item The $6$-dimensional irreducible representation provides an immersion
  $\mathrm{PSL(2,7)} \hookrightarrow \mathrm{SO(6)}$ since its elements preserve
  the symmetric Cartan matrix of $\mathfrak{a}_6$:
      \begin{eqnarray}
        \forall \gamma \in \mathrm{PSL(2,7)}\quad:\quad D_6^T\left(\gamma\right) \,
        \mathfrak{C}_{\mathfrak{a}_6} \, D_6\left(\gamma\right)& = &
        \mathfrak{C}_{\mathfrak{a}_6}\nonumber \\
         \mathfrak{C}_{i,j} & = & \alpha_i \cdot \alpha_j \quad\quad \quad  (i,j \, =\,1 \,\dots ,6)
         \label{gomorito6}
      \end{eqnarray}
     defined in terms of the simple roots $\alpha_i$ whose standard construction
     in terms of the unit vectors $\epsilon_i$ of $\mathbb{R}^7$ is recalled below:
      \begin{equation}\label{simplerutte6}
      \begin{array}{cccccccccccc}
        \alpha_1 & = & \epsilon_1 -\epsilon_2 & ; & \alpha_2 & = &
        \epsilon_2-\epsilon_3 & = & ; &\alpha_3 & = & \epsilon_3 - \epsilon_4 \\
       \alpha_4 & = & \epsilon_4 -\epsilon_5 & ;
       & \alpha_5 & = & \epsilon_5-\epsilon_6 & = & ; &\alpha_6 & = & \epsilon_6 - \epsilon_7 \\
      \end{array}
      \end{equation}
\end{enumerate}
\par
Let us prove the above statements. It suffices to write the explicit
form of the generators $\rho$, $\sigma$ and $\tau$ in the
crystallographic basis of the considered root lattice:
\begin{equation}
 \mathbf{ v} \, \in \, \Lambda^{\mathbf{r}}_{\mathfrak{a}_6}  \quad \Leftrightarrow
 \quad  \mathbf{ v} \, = \, n_i \, \alpha_i \quad n_i \in \mathbb{Z}
\end{equation}
Explicitly if  we set:
\begin{eqnarray}\label{rgen6}
{R}_6 & = &  \left(
\begin{array}{cccccc}
 0 & -1 & 1 & 0 & 0 & 0 \\
 -1 & 0 & 1 & 0 & 0 & 0 \\
 0 & 0 & 1 & 0 & 0 & 0 \\
 0 & 0 & 0 & 1 & 0 & 0 \\
 0 & 0 & 0 & 0 & 1 & 0 \\
 0 & 0 & 0 & 0 & 1 & -1 \\
\end{array}
\right) \quad ; \quad {S}_6 \, = \, \left(
\begin{array}{cccccc}
 -1 & 1 & 0 & 0 & 0 & 0 \\
 -1 & 1 & 0 & 0 & 0 & -1 \\
 -1 & 1 & 0 & -1 & 1 & -1 \\
 -1 & 0 & 1 & -1 & 1 & -1 \\
 -1 & 0 & 0 & 0 & 1 & -1 \\
 -1 & 0 & 0 & 0 & 0 & 0 \\
\end{array}
\right) \nonumber\\
{T}_6 & = & \left(
\begin{array}{cccccc}
 0 & 0 & 0 & 0 & -1 & 1 \\
 0 & -1 & 1 & 0 & -1 & 1 \\
 0 & -1 & 0 & 1 & -1 & 1 \\
 0 & -1 & 0 & 0 & 0 & 1 \\
 1 & -1 & 0 & 0 & 0 & 1 \\
 1 & -1 & 0 & 0 & 0 & 0 \\
\end{array}
\right)
\end{eqnarray}
we find that the defining relations of $\mathrm{PSL(2,7)}$
are satisfied:
\begin{equation}\label{relationibus6}
 {R}_6^2 \, = \, {S}_6^3 \, = \, T_6^7 \,=\,
 \left({T}_6\, {S}_6\, {R}_6\right)^4 \, = \, \mathbf{1}_{6\times 6}
\end{equation}
and furthermore we have:
\begin{equation}
   {R}^T_6 \,\mathfrak{C}_{\mathfrak{a}_6}\,  {R}_6 \,=\,  {S}^T_6
  \,\mathfrak{C}_{\mathfrak{a}_6}\, { S}_6 \, = \,
   {T}^T_6 \,\mathfrak{C}_{\mathfrak{a}_6}\,  {T}_6 \, = \, \,\mathfrak{C}_{\mathfrak{a}_6}\,
\end{equation}
where the explicit form of the $\mathfrak{a}_6$ Cartan matrix is
recalled below:
\begin{equation}\label{cartanschula6}
\,\mathfrak{C}_{\mathfrak{a}_6}\, = \,  \left(
\begin{array}{cccccc}
 2 & -1 & 0 & 0 & 0 & 0 \\
 -1 & 2 & -1 & 0 & 0 & 0 \\
 0 & -1 & 2 & -1 & 0 & 0 \\
 0 & 0 & -1 & 2 & -1 & 0 \\
 0 & 0 & 0 & -1 & 2 & -1 \\
 0 & 0 & 0 & 0 & -1 & 2 \\
\end{array}
\right)
\end{equation}
\subsection{The 8-dimensional representation}
Utilizing the same notations as before in \cite{ramonus}   the
matrix form of the generators pertaining to the irreducible
$8$-dimensional representation was given as follows:
{
\begin{eqnarray*}
  &D[\sigma]_8 = &\nonumber\\
  &\left(
\begin{array}{cccccccc}
 c_1 & s_1 & 0 & 0 & 0 & 0 & 0 & 0 \\
 -s_1 & c_1 & 0 & 0 & 0 & 0 & 0 & 0 \\
 0 & 0 & 1 & 0 & 0 & 0 & 0 & 0 \\
 0 & 0 & 0 & c_3 & s_3 & 0 & 0 & 0 \\
 0 & 0 & 0 & -s_3 & c_3 & 0 & 0 & 0 \\
 0 & 0 & 0 & 0 & 0 & c_2 & s_2 & 0 \\
 0 & 0 & 0 & 0 & 0 & -s_2 & c_2 & 0 \\
 0 & 0 & 0 & 0 & 0 & 0 & 0 & 1 \\
\end{array}
\right) \end{eqnarray*} }
\begin{flushleft}
{\fontsize{5.8}{6.2} {\hskip -4cm \begin{eqnarray*}
 & D[\rho]_8 \, = \, &\nonumber\\
 & \left(
\begin{array}{c@{\hspace{2pt}}c@{\hspace{1pt}}c@{\hspace{1pt}}c@{\hspace{1pt}}c@{\hspace{1pt}}
c@{\hspace{1pt}}c@{\hspace{1pt}}c@{\hspace{1pt}}}
 2-2 c_1 & 0 & 2 c_1+2 c_2-4 c_3 & 2-2 c_2 & 0 &
   2-2 c_3 & 0 & 2 \sqrt{3} c_1-2 \sqrt{3} c_2 \\
 0 & -2 c_1+4 c_2-2 & 0 & 0 & 2 c_2-4 c_3+2 & 0 & 4
   c_1-2 c_3-2 & 0 \\
 2 c_1+2 c_2-4 c_3 & 0 & -c_1+2 c_2-c_3 & -4
   c_1+2 c_2+2 c_3 & 0 & 2 c_1-4 c_2+2 c_3 & 0 &
   \sqrt{3} c_1-\sqrt{3} c_3 \\
 2-2 c_2 & 0 & -4 c_1+2 c_2+2 c_3 & 2-2 c_3 & 0 &
   2-2 c_1 & 0 & 2 \sqrt{3} c_2-2 \sqrt{3} c_3 \\
 0 & 2 c_2-4 c_3+2 & 0 & 0 & 4 c_1-2 c_3-2 & 0 & 2
   c_1-4 c_2+2 & 0 \\
 2-2 c_3 & 0 & 2 c_1-4 c_2+2 c_3 & 2-2 c_1 & 0 &
   2-2 c_2 & 0 & 2 \sqrt{3} c_3-2 \sqrt{3} c_1 \\
 0 & 4 c_1-2 c_3-2 & 0 & 0 & 2 c_1-4 c_2+2 & 0 & -2
   c_2+4 c_3-2 & 0 \\
 2 \sqrt{3} c_1-2 \sqrt{3} c_2 & 0 & \sqrt{3}
   c_1-\sqrt{3} c_3 & 2 \sqrt{3} c_2-2 \sqrt{3} c_3
   & 0 & 2 \sqrt{3} c_3-2 \sqrt{3} c_1 & 0 & c_1-2
   c_2+c_3 \\
\end{array}
\right)&
\end{eqnarray*}}}
\end{flushleft}

\begin{eqnarray}
  D[\tau]_8 & = &\left( D[\rho]_8 \cdot D[\sigma]_8\right)^{-1}
\end{eqnarray}
It remains to be seen whether there exists a crystallographic basis
also for this irreducible representation. We have not explored the
matter but we conjecture that if such a basis exists it is that of
the simple roots of $\mathfrak{a}_8$ leading to an embedding into
the $\mathfrak{e}_8$ Weyl group.
\subsection{The 3-dimensional complex representations}
\label{tritioreppo} Before passing to other items of
$\mathrm{PSL(2,7)}$ theory we mention the last two irreducible
representations of this simple group. They are very important in the
context of the resolution of $\mathbb{C}^3/\Gamma$ singularities and
its relation with the AdS/CFT correspondence (see
\cite{Bruzzo:2017fwj} and \cite{marcovaldo}). Indeed the two three
dimensional irreducible representations are complex and they are
conjugate to each other. They define an embedding:
\begin{equation}\label{su3vaialetto}
    \mathrm{PSL(2,7)} \hookrightarrow \mathrm{SU(3)}
\end{equation}
so that the resolution of $\mathbb{C}^3/\mathrm{PSL(2,7)}$ is
crepant and defines a Ricci flat K\"ahler manifold of Calabi Yau
type (non-compact).
\par
To define these two representations it suffices to give the form of
the generators for one of them. The generators of the conjugate
representation are the complex conjugates of the same matrices.
\par
Setting:
\begin{equation}\label{rholukko}
  \psi \, \equiv \, e^{\frac{2 i \pi }{7}}
\end{equation}
we have the following form for the representation $\mathbf{3}$:
\begin{eqnarray}
 D[\rho]_3 &=& \left(
\begin{array}{ccc}
 \frac{i \left(\psi ^2-\psi ^5\right)}{\sqrt{7}} &
   \frac{i \left(\psi -\psi ^6\right)}{\sqrt{7}} &
   \frac{i \left(\psi ^4-\psi ^3\right)}{\sqrt{7}} \\
 \frac{i \left(\psi -\psi ^6\right)}{\sqrt{7}} &
   \frac{i \left(\psi ^4-\psi ^3\right)}{\sqrt{7}} &
   \frac{i \left(\psi ^2-\psi ^5\right)}{\sqrt{7}} \\
 \frac{i \left(\psi ^4-\psi ^3\right)}{\sqrt{7}} &
   \frac{i \left(\psi ^2-\psi ^5\right)}{\sqrt{7}} &
   \frac{i \left(\psi -\psi ^6\right)}{\sqrt{7}} \\
\end{array}
\right) \nonumber\\
 D[\sigma]_3 &=& \left(
\begin{array}{ccc}
 \frac{i \left(\psi ^3-\psi ^6\right)}{\sqrt{7}} &
   \frac{i \left(\psi ^3-\psi \right)}{\sqrt{7}} &
   \frac{i (\psi -1)}{\sqrt{7}}\nonumber \\
 \frac{i \left(\psi ^2-1\right)}{\sqrt{7}} & \frac{i
   \left(\psi ^6-\psi ^5\right)}{\sqrt{7}} & \frac{i
   \left(\psi ^6-\psi ^2\right)}{\sqrt{7}} \\
 \frac{i \left(\psi ^5-\psi ^4\right)}{\sqrt{7}} &
   \frac{i \left(\psi ^4-1\right)}{\sqrt{7}} & \frac{i
   \left(\psi ^5-\psi ^3\right)}{\sqrt{7}} \\
\end{array}
\right) \nonumber\\
D[\tau]_3 &=& \left(
\begin{array}{ccc}
 -i e^{\frac{3 i \pi }{14}} & 0 & 0 \\
 0 & -i e^{-\frac{i \pi }{14}} & 0 \\
 0 & 0 & -e^{-\frac{i \pi }{7}} \\
\end{array}
\right) \label{tridimirep}
\end{eqnarray}
\subsection{The proper subgroups of \texorpdfstring{$\mathrm{PSL(2,7)}$}{PSL(2,7)}}
The crystallographic nature of the group in $d=7$ has already been
stressed. We  introduce the $\mathfrak{a}_7$ weight lattice which,
by definition, is just the dual of the root lattice. According with
\begin{equation}\label{roncisvaldo}
  \pmb{\pi}\, \in \,  {\Lambda}^{\mathbf{w}}_{\mathfrak{a}_7} \, \Leftrightarrow \,
  \pmb{\pi}\, = \, n_i \, \lambda^i \quad : \quad n^i \in \mathbb{Z}
\end{equation}
the root lattice is spanned by the simple weights that are
implicitly defined by the relations:
\begin{equation}\label{dualabasata}
    \lambda^i \cdot \alpha_j \, = \, \delta^i_j \quad \Rightarrow \quad \lambda^i \, = \,
    \left(\mathfrak{C}^{-1}_{\mathfrak{a}_7}\right)^{ij} \,
    \alpha_j
\end{equation}
Since the group $\mathrm{PSL(2,7)}$ is crystallographic on the root
lattice, by necessity it is crystallographic also on the weight
lattice. Given the generators of the group
$\mathrm{PSL(2,7)}$ in the basis of simple roots we obtain
the same in the basis of simple weights through the following
transformation:
\begin{equation}\label{trasforlinzo}
    \mathcal{R}_w \, = \, \mathfrak{C}_{\mathfrak{a}_7} \, \mathcal{R} \,
    \mathfrak{C}_{\mathfrak{a}_7}^{-1} \quad ; \quad
    \mathcal{S}_w \, = \, \mathfrak{C}_{\mathfrak{a}_7} \, \mathcal{S} \,
    \mathfrak{C}_{\mathfrak{a}_7}^{-1}\quad ; \quad
    \mathcal{T}_w \, = \, \mathfrak{C}_{\mathfrak{a}_7} \, \mathcal{T} \,
    \mathfrak{C}_{\mathfrak{a}_7}^{-1}
\end{equation}
Explicitly we find:
\begin{eqnarray*}
  \mathcal{R}_w &=& \left(
\begin{array}{ccccccc}
 0 & 0 & 0 & 0 & 0 & 0 & -1 \\
 0 & 0 & 0 & -1 & -1 & -1 & 0 \\
 0 & 0 & -1 & 0 & 0 & 0 & 0 \\
 0 & 0 & 1 & 1 & 1 & 0 & 0 \\
 0 & 0 & 0 & 0 & -1 & 0 & 0 \\
 0 & -1 & -1 & -1 & 0 & 0 & 0 \\
 -1 & 0 & 0 & 0 & 0 & 0 & 0 \\
\end{array}
\right) \quad ; \quad   \mathcal{S}_w\, = \,\left(
\begin{array}{ccccccc}
 -1 & -1 & -1 & -1 & -1 & -1 & -1 \\
 1 & 1 & 1 & 1 & 0 & 0 & 0 \\
 0 & 0 & 0 & -1 & 0 & 0 & 0 \\
 0 & 0 & 0 & 1 & 1 & 1 & 0 \\
 0 & 0 & 0 & 0 & 0 & -1 & 0 \\
 0 & 0 & -1 & -1 & -1 & 0 & 0 \\
 0 & -1 & 0 & 0 & 0 & 0 & 0 \\
\end{array}
\right) \end{eqnarray*}
\begin{eqnarray}
  \mathcal{T}_w &=& \left(
\begin{array}{ccccccc}
 -1 & -1 & -1 & -1 & -1 & -1 & 0 \\
 1 & 0 & 0 & 0 & 0 & 0 & 0 \\
 0 & 1 & 0 & 0 & 0 & 0 & 0 \\
 0 & 0 & 1 & 0 & 0 & 0 & 0 \\
 0 & 0 & 0 & 1 & 0 & 0 & 0 \\
 0 & 0 & 0 & 0 & 1 & 0 & 0 \\
 0 & 0 & 0 & 0 & 0 & 1 & 1 \\
\end{array}
\right)
\end{eqnarray}
Given the weight basis, which is useful in several constructions,
let us continue our survey of the remarkable simple group
$\mathrm{PSL(2,7)}$ by a discussion of its subgroups, none
of which, obviously, is normal.
\par
$\mathrm{PSL(2,7)}$ contains maximal subgroups only of
index 8 and 7, namely of order 21 and 24. The order 21 subgroup
$\mathrm{G_{21}}$ is the unique non-abelian group of that order and
abstractly it has the structure of the semidirect product
$\mathbb{Z}_3 \propto \mathbb{Z}_7$. Up to conjugation there is only
one subgroup $\mathrm{G_{21}}$ as we have explicitly verified with
the computer. On the other hand, up to conjugation, there are two
different groups of order 24 that are both isomorphic to the
octahedral group $\mathrm{O_{24}}\sim S_4$.
\subsubsection{The maximal subgroup \texorpdfstring{$\mathrm{G_{21}}$}{G21}}
\label{g21gruppo} The group $\mathrm{G_{21}}$ has two generators
$\mathcal{X}$ and $\mathcal{Y}$ that satisfy the following
relations:
\begin{equation}
\mathcal{X}^3 \, = \, \mathcal{Y}^7 \, = \,\mathbf{ 1} \quad ; \quad
\mathcal{X}\mathcal{Y}  = \mathcal{Y}^2 \mathcal{X} \label{faloluna}
\end{equation}
The organization of the 21 group elements into conjugacy classes is
displayed below:
\begin{equation}\label{corsaro}
\begin{array}{|c|c|c|c|c|c|}
\hline \mbox{Conjugacy} \mbox{Class} & C_1 & C_2 & C_3 & C_4 & C_5
\\ \hline \mbox{representative of the class} & e & \mathcal{Y} &
\mathcal{X}^2 \mathcal{Y}\mathcal{X}\mathcal{Y}^2 &
\mathcal{Y}\mathcal{X}^2 & \mathcal{X}
\\ \hline \mbox{order of the elements in the class} & 1 & 7 & 7 & 3
& 3 \\ \hline \mbox{number of elements in the class} & 1 & 3 & 3 & 7 & 7 \\
\hline
\end{array}
\end{equation}
As we see there are five conjugacy classes which implies that there
should be five irreducible representations the square of whose
dimensions should sum up to the group order 21. The solution of this
problem is:
\begin{equation}\label{cavacchioli}
    21 \, = \, 1^2 + 1^2 + 1^2 + 3^2 + 3^2
\end{equation}
and the corresponding character table is mentioned below:
\begin{equation}\label{bruttocarattere21}
\begin{array}{|c|c|c|c|c|c|}
\hline 0 & e & \mathcal{Y} & \mathcal{X}^2
\mathcal{Y}\mathcal{X}\mathcal{Y}^2 & \mathcal{Y}\mathcal{X}^2 &
\mathcal{X}
\\ \hline
\mathrm{D_1}\left[\mathrm{G_{21}}\right] & 1 & 1 & 1 &
1 & 1 \\
\hline \mbox{DX}_1\left[\mathrm{G_{21}}\right] & 1 & 1 & 1 &
-(-1)^{1/3} &
(-1)^{2/3} \\
\hline \mbox{DY}_1\left[\mathrm{G_{21}}\right] & 1 & 1 & 1 &
(-1)^{2/3} &
-(-1)^{1/3} \\
\hline \mbox{DA}_3\left[\mathrm{G_{21}}\right] & 3 & \frac{1}{2} i
\left(i+\sqrt{7}\right) & -\frac{1}{2} i \left(-i+\sqrt{7}\right)
& 0 & 0 \\
\hline \mbox{DB}_3\left[\mathrm{G_{21}}\right] & 3 & -\frac{1}{2} i
\left(-i+\sqrt{7}\right) & \frac{1}{2} i \left(i+\sqrt{7}\right)
& 0 & 0 \\
\hline
\end{array}
\end{equation}
In the weight-basis the two generators of the $\mathrm{G_{21}}$
subgroup of $\mathrm{PSL(2,7)}$ can be chosen to be the
following matrices and this fixes our representative of the unique
conjugacy class:
\begin{equation}\label{XYgenerati}
    \mathcal{X}\, = \, \left(
\begin{array}{ccccccc}
 1 & 1 & 1 & 1 & 1 & 1 & 1 \\
 0 & 0 & 0 & 0 & 0 & 0 & -1 \\
 0 & -1 & -1 & -1 & -1 & -1 & 0 \\
 0 & 1 & 1 & 1 & 0 & 0 & 0 \\
 0 & 0 & -1 & -1 & 0 & 0 & 0 \\
 0 & 0 & 1 & 1 & 1 & 0 & 0 \\
 0 & 0 & 0 & -1 & -1 & 0 & 0 \\
\end{array}
\right) \quad  \mathcal{Y} \, = \,\left(
\begin{array}{ccccccc}
 0 & 1 & 1 & 0 & 0 & 0 & 0 \\
 0 & 0 & 0 & 1 & 1 & 1 & 1 \\
 0 & 0 & -1 & -1 & -1 & -1 & -1 \\
 0 & 0 & 1 & 1 & 0 & 0 & 0 \\
 -1 & -1 & -1 & -1 & 0 & 0 & 0 \\
 1 & 1 & 1 & 1 & 1 & 0 & 0 \\
 0 & 0 & 0 & 0 & 0 & 1 & 0 \\
\end{array}
\right)
\end{equation}
The embedding of $\mathrm{G_{21}}$ into $\mathrm{PSL(2,7)}$ can be
unambiguously fixed by writing the two generators of the former  as
words in the generators of the latter. We have:
\begin{equation}\label{puccio}
 \mathcal{Y} \, = \,  \rho \cdot\tau \cdot\tau \cdot\tau \cdot\sigma \cdot\rho
 \quad ; \quad \mathcal{X} \, = \,  \sigma \cdot\rho \cdot\sigma \cdot\rho \cdot\tau \cdot\tau
\end{equation}
Eq.(\ref{puccio}) allows to restrict any given representation of
$\mathrm{PSL(2,7)}$ to its maximal subgroup $\mathrm{G_{21}}$.
\subsubsection{The maximal subgroups \texorpdfstring{$\mathrm{O_{24A}}$}{O24A} and \texorpdfstring{$\mathrm{O_{24B}}$}{O24B}} The octahedral group
$\mathrm{O_{24}}$ has two generators $\mathfrak{s}$ and
$\mathfrak{t}$ that satisfy the following relations:
\begin{equation}
\mathfrak{s}^2 \, = \, \mathfrak{t}^3\,  =
\,(\mathfrak{s}\mathfrak{t})^4 \, = \, \mathbf{1}
\end{equation}
The 24 elements are organized in five conjugacy classes according to
the scheme displayed below:
\begin{equation}\label{o24coniugi}
  \begin{array}{|c|c|c|c|c|c|} \hline
 \mbox{Conjugacy Class} & C_1 & C_2 & C_3 & C_4 & C_5 \\
\hline
 \mbox{representative of the class} & e & \mathfrak{t} & \mathfrak{s}\mathfrak{t}\mathfrak{s}\mathfrak{t}  & \mathfrak{s} &  \mathfrak{s}\mathfrak{t}  \\
\hline
 \mbox{order of the elements in the class} & 1 & 3 & 2 & 2 & 4 \\
\hline
 \mbox{number of elements in the class} & 1 & 8 & 3 & 6 & 6 \\
\hline
\end{array}
\end{equation}
The  character table where we also mention a standard representative
of each conjugacy class is the following one:
\begin{equation}\label{caratteri24}
\begin{array}{|c|c|c|c|c|c|}
\hline
 0 & e & \mathfrak{t} & \mathfrak{s}\mathfrak{t}\mathfrak{s}\mathfrak{t}  & \mathfrak{s} &
 \mathfrak{s}\mathfrak{t} \\
\hline
 \mathrm{D_1}\left[\mathrm{O_{24}}\right] & 1 & 1 & 1 & 1 & 1 \\
\hline
 \mbox{D}_2\left[\mathrm{O_{24}}\right] & 1 & 1 & 1 & -1 & -1 \\
\hline
 \mathrm{D_3}\left[\mathrm{O_{24}}\right] & 2 & -1 & 2 & 0 & 0 \\
\hline
 \mathrm{D_4}\left[\mathrm{O_{24}}\right] & 3 & 0 & -1 & -1 & 1 \\
\hline
 \mathrm{D_5}\left[\mathrm{O_{24}}\right] & 3 & 0 & -1 & 1 & -1 \\
\hline
\end{array}
\end{equation}
By computer calculations we have verified that there are just two
disjoint conjugacy classes of $\mathrm{O_{24}}$ maximal subgroups in
$\mathrm{PSL(2,7)}$ that we have named A and B,
respectively. We have chosen two standard representatives, one for
each conjugacy class, that we have named $\mathrm{O_{24A}}$ and
$\mathrm{O_{24B}}$ respectively. To fix these subgroups it suffices
to mention the explicit form of their generators in the weight
basis.
\par
For the group $\mathrm{O_{24A}}$, we chose:
\begin{equation}\label{generatori24A}
    \mathfrak{t}_A \, = \, \left(
\begin{array}{ccccccc}
 1 & 1 & 1 & 1 & 1 & 1 & 1 \\
 0 & 0 & 0 & 0 & 0 & 0 & -1 \\
 0 & -1 & -1 & -1 & -1 & -1 & 0 \\
 0 & 1 & 1 & 1 & 0 & 0 & 0 \\
 0 & 0 & -1 & -1 & 0 & 0 & 0 \\
 0 & 0 & 1 & 1 & 1 & 0 & 0 \\
 0 & 0 & 0 & -1 & -1 & 0 & 0 \\
\end{array}
\right)    \quad  \mathfrak{s}_A \, = \,\left(
\begin{array}{ccccccc}
 0 & 0 & 0 & 1 & 1 & 1 & 0 \\
 0 & 0 & 0 & 0 & -1 & -1 & 0 \\
 -1 & -1 & -1 & -1 & 0 & 0 & 0 \\
 1 & 1 & 0 & 0 & 0 & 0 & 0 \\
 0 & 0 & 1 & 1 & 1 & 1 & 1 \\
 0 & -1 & -1 & -1 & -1 & -1 & -1 \\
 0 & 1 & 1 & 1 & 1 & 0 & 0 \\
\end{array}
\right)
\end{equation}
For the group $\mathrm{O_{24B}}$, we chose:
\begin{equation}\label{generatori24B}
    \mathfrak{t}_B \, = \, \left(
\begin{array}{ccccccc}
 1 & 1 & 1 & 1 & 0 & 0 & 0 \\
 0 & -1 & -1 & -1 & 0 & 0 & 0 \\
 0 & 1 & 1 & 1 & 1 & 0 & 0 \\
 0 & 0 & -1 & -1 & -1 & 0 & 0 \\
 0 & 0 & 1 & 1 & 1 & 1 & 0 \\
 0 & 0 & 0 & -1 & -1 & -1 & 0 \\
 0 & 0 & 0 & 1 & 1 & 1 & 1 \\
\end{array}
\right)    \quad  \mathfrak{s}_B \, = \,\left(
\begin{array}{ccccccc}
 0 & 0 & 1 & 1 & 1 & 0 & 0 \\
 -1 & -1 & -1 & -1 & -1 & 0 & 0 \\
 1 & 1 & 1 & 1 & 1 & 1 & 1 \\
 0 & 0 & 0 & 0 & 0 & 0 & -1 \\
 0 & -1 & -1 & -1 & -1 & -1 & 0 \\
 0 & 1 & 1 & 1 & 0 & 0 & 0 \\
 0 & 0 & 0 & -1 & 0 & 0 & 0 \\
\end{array}
\right)
\end{equation}
Just as in the case of the subgroup $\mathrm{G_{21}}$ we can
uniquely fix the embedding of the two octahedral subgroups into
$\mathrm{PSL(2,7)}$ in any given of its representations by writing
the two generators of the subgroup as words in the generators of the
bigger group. Explicitly we have:
\begin{eqnarray}
  \mathfrak{t}_A &=& \rho \cdot\sigma \cdot\rho \cdot\tau \cdot\tau \cdot\sigma \cdot\rho \cdot\tau \quad ;
  \quad \mathfrak{s}_A \, = \, \tau \cdot\tau \cdot\sigma \cdot\rho \cdot\tau \cdot\sigma \cdot\sigma\, \nonumber\\
  \mathfrak{t}_B &=& \rho \cdot\tau \cdot\sigma \cdot\rho \cdot\tau \cdot\tau \cdot\sigma \cdot\rho \cdot\tau \quad ;
  \quad \mathfrak{s}_B \, = \, \sigma \cdot\rho \cdot\tau \cdot\sigma \cdot\rho \cdot\tau\,
\end{eqnarray}
\subsubsection{The tetrahedral subgroup \texorpdfstring{$\mathrm{T_{12}} \subset \mathrm{O_{24}}$}{T12 in O24}}
Every octahedral group $\mathrm{O_{24}}$ has, up to
$\mathrm{O_{24}}$-conjugation, a unique tetrahedral subgroup
$\mathrm{T_{12}}$ whose order is 12. The abstract description of the
tetrahedral group is provided by the following presentation in terms
of two generators:
\begin{equation}\label{presentaT12}
  \mathrm{T_{12}} =\left(s,t\left|s^2\right. = t^3 = (st)^3 = 1\right)
\end{equation}
The 12 elements are organized into  four conjugacy classes as
displayed below:
\begin{equation}\label{coniugatoT12}
\begin{array}{|c|c|c|c|c|}
\hline
 \mbox{Classes} & C_1 & C_2 & C_3 & C_4 \\
\hline
 \mbox{standard representative} & 1 & s & t & t^2s \\
\hline
 \mbox{order of the elements in the conjugacy class} & 1 & 2 & 3 & 3 \\
\hline
 \mbox{number of elements in the conjugacy class} & 1 & 3 & 4 & 4 \\
\hline
\end{array}
\end{equation}
We do not display the character table since we will not use it. The
two tetrahedral subgroups $\mathrm{T_{12A}} \subset
\mathrm{O_{24A}}$ and $\mathrm{T_{12B}} \subset \mathrm{O_{24B}}$
are not conjugate under the big group $\mathrm{PSL(2,7)}$.
Hence we have two conjugacy classes of tetrahedral subgroups of
$\mathrm{PSL(2,7)}$.
\subsubsection{The dihedral subgroup \texorpdfstring{$\mathrm{Dih_{3}} \subset \mathrm{O_{24}}$}{Dih3 in O24}  }
Every octahedral group $\mathrm{O_{24}}$ has a dihedral subgroup
$\mathrm{Dih_{3}}$ whose order is 6. The abstract description of the
dihedral group $\mathrm{Dih_{3}}$ is provided by the following
presentation in terms of two generators:
\begin{equation}\label{presentaDih3}
  \mathrm{Dih_{3}} =\left(A,B\left| A^3=B^2=(BA)^2=1\right.\right)
\end{equation}
The 6 elements are organized into  three conjugacy classes as
displayed below:
\begin{equation}\label{coniugatoDih3}
\begin{array}{|c|c|c|c|}
\hline
 \mbox{Conjugacy} \mbox{Classes} & C_1 & C_2 & C_3 \\
\hline
 \mbox{standard representative of the class} & 1 & A & B \\
\hline
 \mbox{order of the elements in the class} & 1 & 3 & 2 \\
\hline
 \mbox{number of elements in the class} & 1 & 2 & 3 \\
\hline
\end{array}
\end{equation}
We do not display the character table since we will not use it.
Differently from the case of the tetrahedral subgroups the two
dihedral subgroups $\mathrm{Dih_{3A}} \subset \mathrm{O_{24A}}$ and
$\mathrm{Dih_{3B}} \subset \mathrm{O_{24B}}$ turn out to be
conjugate under the big group $\mathrm{PSL(2,7)}$. Actually
there is just one $\mathrm{PSL(2,7)}$-conjugacy class of
dihedral subgroups $\mathrm{Dih_{3}}$.
\subsection{Enumeration of
the possible subgroups and orbits in the \texorpdfstring{$\mathfrak{a}_7$}{a7} and
\texorpdfstring{$\mathfrak{a}_6$}{a6}  weight lattices}\label{subgroupsH} In $d=3$ the orbits  of the
octahedral group acting on the cubic lattice are the vertices of
regular geometrical figures. Since $\mathrm{PSL(2,7)}$ has
a crystallographic action on the mentioned $7$-dimensional and
$6$-dimensional weight lattices, its orbits $\mathcal{O}$ in
$\Lambda^{\mathbf{w}}_{\mathfrak{a}_7}$ and
$\Lambda^{\mathbf{w}}_{\mathfrak{a}_6}$ correspond to the analogue
of the regular geometrical figures in $d=7$ and in $d=6$. Every
orbit is in correspondence with a coset $\mathrm{G}/\mathrm{H}$
where $\mathrm{G}$ is the big group and $\mathrm{H}$ one of its
possible subgroups. Indeed $\mathrm{H}$ is the stability subgroup of
an element of the orbit.
\par
Since the maximal subgroups of $\mathrm{PSL(2,7)}$ are of
index 7 or 8 we can have subgroups $\mathrm{H} \subset
\mathrm{PSL(2,7)}$ that are either $\mathrm{G_{21}}$ or
$\mathrm{O_{24}}$ or subgroups thereof. Furthermore, as we know, the
order $|\mathrm{H}|$  of any subgroup $\mathrm{H}\subset \mathrm{G}$
must be a divisor of $|\mathrm{G}|$. Hence we conclude that
\begin{equation}\label{ordinatini}
 |\mathrm{H}| \, \in \,   \{1,2,3,4,6,7,8,12,21,24\}
\end{equation}
Correspondingly we might have $\mathrm{PSL(2,7)}$-orbits
$\mathcal{O}$ in the weight lattices
$\Lambda^{\mathbf{w}}_{\mathfrak{a}_{7,6}}$, whose length is one of
the following 10 numbers:
\begin{equation}\label{apprilunghi}
    \ell_{\mathcal{O}} \, \in \, \{168,84,56,42,28,24,21,14,8,7\}
\end{equation}
\par
Combining the information about the possible group orders
(\ref{ordinatini}) with the information that the maximal subgroups
are of index 8 or 7, we arrive at the following list of possible
subgroups $\mathrm{H}$ (up to conjugation) of the group
$\mathrm{PSL(2,7)}$:
\begin{description}
\item[Order 24)] Either H = $\mathrm{O_{24A}}$  or H = $\mathrm{O_{24B}}$.
\item[Order 21)] The only possibility is H = $\mathrm{G_{21}}$.
\item[Order 12)] The only possibilities are H = $\mathrm{T_{12A}}$ or H = $\mathrm{T_{12B}}$ where
$\mathrm{T_{12}}$ is the tetrahedral subgroup of the octahedral
group $\mathrm{O_{24}}$.
\item[Order 8)] Either H = $\mathbb{Z}_2\times\mathbb{Z}_2\times \mathbb{Z}_2$ or
H = $\mathbb{Z}_2\times\mathbb{Z}_4$, or  H = $\mathrm{Dih_4}$ where
$\mbox{Dih}_4$ denotes the dihedral subgroup of index 3 of the
octahedral group $\mathrm{O_{24}}$.
\item[Order 7)] The only possibility is $\mathbb{Z}_7$.
\item [Order 6)] Either H = $\mathbb{Z}_2\times\mathbb{Z}_3$ or  H = $\mbox{Dih}_3$,
where $\mbox{Dih}_3$ denotes the dihedral subgroup of index 4 of the
octahedral group $\mathrm{O_{24}}$.
\item[Order 4)] Either H = $\mathbb{Z}_2\times\mathbb{Z}_2$  or H = $\mathbb{Z}_4$.
\item[Order 3)] The only possibility is H = $\mathbb{Z}_3$
\item[Order 2)] The only possibility is H = $\mathbb{Z}_2$.
\end{description}
Quite curiously and inspiringly the various possibilities are
realized in a partially mutually exclusive pattern in $7$ and $6$
dimensions as recalled in the following two subsections and
summarized in table \ref{giuanen}.
\begin{table}[ht]
\caption{Summary of the $\mathrm{PSL(2,7)}$ orbits of vectors
existing in the $\mathfrak{a}_7$ and $\mathfrak{a}_6$ weight
lattices. All possible lengths enumerated in Eq. (\ref{apprilunghi})
are realized, except for $\ell = 24$, yet not at the same time in
$d=7$ and $d=6$. Most of the lower length orbits corresponding to
the largest stability subgroups are realized in either one of the
two crystallographic irreducible representations, $d=7$ or $d=6$} 
\vskip10pt
\centering 
\begin{tabular}{|c|c ||c| c|} 
 \hline 
 Orbit length & Subgroup & d=7 & d=6 \\ \hline
 $7$ & $\mathrm{O_{24A}}$ & No & Yes  \\ \hline
 $7$ & $\mathrm{O_{24B}}$ & No & Yes  \\ \hline
 $8$ & $\mathrm{G_{21}}$ & Yes & No \\ \hline
 $14$ & $\mathrm{T_{12A}}$ & Yes & No  \\ \hline
 $14$ & $\mathrm{T_{12B}}$ & Yes & No  \\ \hline
 $21$ & $\mathrm{Dih_4}$ & No & Yes \\ \hline
 $24$ & $\mathbb{Z}_7$ & No & No \\ \hline
 $28$ & $\mathrm{Dih_3}$ & Yes & Yes \\ \hline
 $42$ & $\mathbb{Z}_4$ & Yes & No \\ \hline
 $56$ & $\mathbb{Z}_3$ & Yes & No \\ \hline
 $84$ & $\mathbb{Z}_2$ & Yes & Yes \\ \hline
 $168$ & $\mathrm{Id}$ & Yes & Yes \\ \hline
\hline 
\end{tabular}
\label{giuanen} 
\end{table}
\subsubsection{Synopsis of the \texorpdfstring{$\mathrm{PSL(2,7)}$}{PSL(2,7)} orbits in the weight lattice \texorpdfstring{$\Lambda_{\mathfrak{a}_7}^{\mathbf{w}}$}{La7w} } In \cite{miol168}, the
author  presented the results, obtained by means of  computer
calculations, on the orbits of the considered simple group acting on
the $\mathfrak{a}_7$ weight lattice. They are briefly summarized
below:
\begin{enumerate}
\item Orbits of length 8 (one parameter\pmb{n}; stability subgroup $\mathrm{H}^s=\mathrm{G_{21}}$)
\item Orbits of length 14 (two types A $\&$ B) (one parameter\pmb{  n}; stability subgroup
$\mathrm{H}^s=\mathrm{T_{12A,B}}$)
\item Orbits of length 28 (one parameter\pmb{  n} ; stability subgroup
$\mathrm{H}^s=\mathrm{Dih_{3}}$)
\item Orbits of length 42 (one parameter\pmb{  n}; stability subgroup
$\mathrm{H}^s=\mathbb{Z}_4$) )
\item Orbits of length 56 (three parameters \pmb{ n,m,p}; stability subgroup
$\mathrm{H}^s=\mathbb{Z}_3$)
\item Orbits of length 84 (three parameters \pmb{ n,m,p}; stability subgroup
$\mathrm{H}^s=\mathbb{Z}_2$)
\item Generic orbits of length 168 (seven parameters ; stability subgroup $\mathrm{H}^s=\mathbf{1}$)
\end{enumerate}
Also in this case the above list is in some sense the
$6$-dimensional analogue of Platonic solids. It is only in some
sense, since  it is a complete classification for the group
$\mathrm{PSL(2,7)}$, yet we are not aware of a
classification of the other crystallographic subgroups of
$\mathrm{SO(6)}$, if any.
\par
\subsubsection{Synopsis of the \texorpdfstring{$\mathrm{PSL(2,7)}$}{PSL(2,7)} orbits in the weight lattice \texorpdfstring{$\Lambda_{\mathfrak{a}_6}^{\mathbf{w}}$}{La6w}} Complementing the work done
in \cite{miol168}, we obtained, also by means of  computer
calculations,  the orbits of $PSL(2,7)$  acting through its
irreducible $6$-dimensional representation on the $\mathfrak{a}_6$
weight lattice. They are briefly summarized below:
\begin{enumerate}
\item Orbits of length 7 (one parameter \textbf{  n}; stability subgroup $\mathrm{H}^s=\mathrm{O_{24A}}$)
\item Orbits of length 7 (one parameter \textbf{  n}; stability subgroup $\mathrm{H}^s=\mathrm{O_{24B}}$)
\item Orbits of length 28 (one parameter \textbf{  n} ; stability subgroup
$\mathrm{H}^s=\mathrm{Dih_{3}}$)
\item Orbits of length 21 (two parameters \textbf{ m, n}; stability subgroup
$\mathrm{H}^s=\mathrm{Dih_{3}}$)
\item Orbits of length 84 (four parameters \textbf{ n,m,p,q}; stability subgroup
$\mathrm{H}^s=\mathbb{Z}_2$)
\item Generic orbits of length 168 (six parameters ; stability subgroup $\mathrm{H}^s=\mathbf{1}$)
\end{enumerate}
Also in this case the above list is in some sense the
$6$-dimensional analogue of Platonic solids. It is only in some
sense, since  it is a complete classification for the group
$\mathrm{PSL(2,7)}$, yet we are not aware of a
classification of the other crystallographic subgroups of
$\mathrm{SO(6)}$, if any.
\section{Embedding of the group \texorpdfstring{$\mathrm{PSL(2,7)}$}{PSL(2,7)} into \texorpdfstring{$\mathrm{\mathfrak{e}_{7(7)}}$}{e77}}\label{zuccamarcia} Above we
considered the simple group $\mathrm{PSL(2,7)}$ showing
that it acts crystallographically on
$\Lambda^{\mathbf{r}}_{\mathfrak{a}_{7,6}}$ and, consequently, also
on the dual weight lattices
$\Lambda^{\mathbf{w}}_{\mathfrak{a}_{7,6}}$. In view of our goals
pursued within the context of exceptional field theory we show next
how the action of $\mathrm{PSL(2,7)}$, can be extended to the root
and weight lattices of the exceptional Lie algebra $\mathfrak{e}_7$.
\subsection{Embedding of \texorpdfstring{$\mathrm{PSL(2,7)}$}{PSL(2,7)} into \texorpdfstring{$\mathrm{Weyl}[\mathfrak{e}_7]$}{Weyl[e7]}}
Let us consider the Dynkin diagrams of the three Lie algebras
$\mathfrak{e}_7$, $\mathfrak{a}_7$ and $\mathfrak{a}_6$.
\begin{equation}
\begin{picture}(100,100)
\put (-70,35){$\mathfrak{e}_7$} \put (-20,40){\circle {10}} \put
(-23,25){$\alpha_7$} \put (-15,40){\line (1,0){20}} \put
(10,40){\circle {10}} \put (7,25){$\alpha_6$} \put (15,40){\line
(1,0){20}} \put (40,40){\circle {10}} \put (37,25){$\alpha_4$} \put
(40,70){\circle {10}} \put (48,67.8){$\alpha_5$} \put (40,45){\line
(0,1){20}} \put (45,40){\line (1,0){20}} \put (70,40){\circle {10}}
\put (67,25){$\alpha_{3}$} \put (75,40){\line (1,0){20}} \put
(100,40){\circle {10}} \put (97,25){$\alpha_{2}$} \put
(105,40){\line (1,0){20}} \put (130,40){\circle {10}} \put
(127,25){$\alpha_1$}
\end{picture}
\label{bistolfoE7}
\end{equation}
\begin{equation}
\begin{picture}(100,60)
\put (-70,45){$\mathfrak{a}_7$} \put (-20,50){\circle {10}} \put
(-23,35){$\beta_7$} \put (-15,50){\line (1,0){20}} \put
(10,50){\circle {10}} \put (7,35){$\beta_6$} \put (15,50){\line
(1,0){20}} \put (40,50){\circle {10}} \put (37,35){$\beta_5$} \put
(45,50){\line(1,0){20}}
\put (70,50){\circle {10}} \put (67,35){${\beta_4}$} \put
(75,50){\line (1,0){20}} \put (100,50){\circle {10}} \put
(97,35){$\beta_{3}$} \put (105,50){\line (1,0){20}} \put
(130,50){\circle {10}} \put (127,35){$\beta_2$}\put (135,50){\line
(1,0){20}}\put (160,50){\circle {10}}\put (157,35){$\beta_1$}
\end{picture}
\label{bistolfoA7}
\end{equation}
\begin{equation}
\begin{picture}(100,60)
\put (-70,45){$\mathfrak{a}_6$}  \put (-20,50){\circle {10}} \put
(-23,35){$\gamma_6$} \put (-15,50){\line (1,0){20}} \put
(10,50){\circle {10}} \put (7,35){$\gamma_5$} \put
(15,50){\line(1,0){20}}
\put (40,50){\circle {10}} \put (37,35){${\gamma_4}$} \put
(45,50){\line (1,0){20}} \put (70,50){\circle {10}} \put
(67,35){$\gamma_{3}$} \put (75,50){\line (1,0){20}} \put
(100,50){\circle {10}} \put (97,35){$\gamma_2$}\put (105,50){\line
(1,0){20}}\put (130,50){\circle {10}}\put (127,35){$\gamma_1$}
\end{picture}
\label{bistolfoA6}
\end{equation}
The Lie algebras $\mathfrak{a}_7$ has the same rank as
$\mathfrak{e}_7$ and the former is regularly embedded into the
latter, having the same Cartan subalgebra. Indeed given any set of
simple roots $\alpha_i$ fulfilling the relations imposed by the
Dynkin diagram (\ref{bistolfoE7}), we immediately construct a set of
simple roots $\beta_j$ fulfilling the relations imposed by the
Dinkin diagram (\ref{bistolfoA7}) by setting:
\begin{eqnarray}
  \beta_1 &=&\alpha_2\, +\, 2 \, \alpha_3 \, +\, 3 \, \alpha_4 \, +\, 2 \, \alpha_5
  \, +\, 2 \, \alpha_6\, +\, 2 \, \alpha_7 \nonumber\\
  \beta_2 &=& \alpha_1 \nonumber\\
  \beta_3 &=& \alpha_2 \nonumber\\
  \beta_4 &=& \alpha_3 \nonumber\\
  \beta_5 &=& \alpha_4 \nonumber\\
  \beta_6 &=& \alpha_6 \nonumber\\
  \beta_7 &=& \alpha_7 \label{marciofungo}
\end{eqnarray}
As one notices the $\mathfrak{a}_7$ simple roots are integer valued
linear combinations of the $\mathfrak{e}_7$ simple roots, hence they
all belong to the $\mathfrak{e}_7$ root lattice
$\Lambda^{\mathbf{r}}_{\mathfrak{e}_7}$. It follows that
$\Lambda^{\mathbf{r}}_{\mathfrak{a}_7}$ is a sublattice of the
former:
\begin{equation}\label{congrioalburro}
    \Lambda^{\mathbf{r}}_{\mathfrak{a}_7} \subset \Lambda^{\mathbf{r}}_{\mathfrak{e}_7}
\end{equation}
From Eq. (\ref{marciofungo}) we immediately read off the matrix that
performs the change of basis of $\mathrm{PSL(2,7)}$ group elements
from the basis $\beta_i$ of $\mathfrak{a}_7$ simple roots to the
basis $\alpha_i$ of $\mathfrak{e}_7$ simple roots. It is the
following one:
\begin{equation}\label{cipollacaramellata}
  \pmb{\Pi} \, = \,  \left(
\begin{array}{ccccccc}
 0 & 1 & 0 & 0 & 0 & 0 & 0 \\
 1 & 0 & 1 & 0 & 0 & 0 & 0 \\
 2 & 0 & 0 & 1 & 0 & 0 & 0 \\
 3 & 0 & 0 & 0 & 1 & 0 & 0 \\
 2 & 0 & 0 & 0 & 0 & 0 & 0 \\
 2 & 0 & 0 & 0 & 0 & 1 & 0 \\
 1 & 0 & 0 & 0 & 0 & 0 & 1 \\
\end{array}
\right)
\end{equation}
Setting:
\begin{equation}\label{franchigia}
    \mathbf{R}_{\mathfrak{e}_7}^{\mathbf{r}} \,= \, \pmb{\Pi}\, \mathcal{R}\,
    \pmb{\Pi}^{-1}\quad ; \quad \mathbf{S}_{\mathfrak{e}_7}^{\mathbf{r}} \,= \, \pmb{\Pi}\, \mathcal{S}\,
    \pmb{\Pi}^{-1}\quad ; \quad \mathbf{T}_{\mathfrak{e}_7}^{\mathbf{r}} \,= \, \pmb{\Pi}\, \mathcal{T}\,
    \pmb{\Pi}^{-1}
\end{equation}
where $\mathcal{R},\mathcal{S},\mathcal{T}$ are the generators of
the irreducible representation of $\mathrm{PSL(2,7)}$ in the
$\mathfrak{a}_7$ root basis, we obtain the generators of the same
representation in the $\mathfrak{e}_7$ root basis.
The explicit form of these $7\times 7$ matrices is given below:
\begin{eqnarray}
  \mathbf{T}_{\mathfrak{e}_7}^{\mathbf{r}}&=& \left(
\begin{array}{ccccccc}
 0 & 0 & 0 & 0 & 1 & -1 & 1 \\
 1 & 0 & 0 & 0 & 1 & -2 & 2 \\
 0 & 1 & 0 & 0 & 1 & -3 & 3 \\
 0 & 0 & 1 & 0 & 1 & -4 & 4 \\
 0 & 0 & 0 & 0 & 1 & -2 & 2 \\
 0 & 0 & 0 & 1 & 0 & -3 & 3 \\
 0 & 0 & 0 & 0 & 0 & -1 & 2 \\
\end{array}
\right)\quad ; \quad \mathbf{S}_{\mathfrak{e}_7}^{\mathbf{r}}
   \, = \,\left(
\begin{array}{ccccccc}
 0 & 0 & 0 & 0 & 1 & 0 & -1 \\
 0 & 0 & -1 & 1 & 1 & 0 & -2 \\
 0 & -1 & 0 & 1 & 1 & 0 & -3 \\
 0 & -1 & 0 & 1 & 2 & -1 & -3 \\
 0 & 0 & 0 & 0 & 1 & 0 & -2 \\
 0 & -1 & 0 & 0 & 2 & 0 & -2 \\
 -1 & 0 & 0 & 0 & 1 & 0 & -1 \\
\end{array}
\right) \\
 \mathbf{R}_{\mathfrak{e}_7}^{\mathbf{r}} &=& \left(
\begin{array}{ccccccc}
 0 & 0 & 0 & 0 & 1 & -1 & 0 \\
 0 & -1 & 1 & 0 & 1 & -1 & -1 \\
 -1 & 0 & 1 & 0 & 1 & -1 & -2 \\
 -1 & 0 & 1 & -1 & 2 & 0 & -3 \\
 0 & 0 & 0 & 0 & 1 & 0 & -2 \\
 -1 & 0 & 0 & 0 & 1 & 0 & -2 \\
 0 & 0 & 0 & 0 & 0 & 0 & -1 \\
\end{array}
\right)\label{balzamaria}
\end{eqnarray}
We can now easily verify that $\mathrm{PSL(2,7)}$ is
crystallographic with respect to the $\mathrm{\mathfrak{e}_7}$-root
lattice. It suffices to check that the above generators satisfy:
\begin{equation}\label{gagallo}
  \left(\mathbf{T}_{\mathfrak{e}_7}^{\mathbf{r}}\right)^T \, \mathfrak{C}_{\mathrm{\mathfrak{e}_7}}
  \, \mathbf{T}_{\mathfrak{e}_7}^{\mathbf{r}} \, =\, \left(\mathbf{S}_{\mathfrak{e}_7}^{\mathbf{r}}\right)^T \, \mathfrak{C}_{\mathrm{\mathfrak{e}_7}}
  \, \mathbf{S}_{\mathfrak{e}_7}^{\mathbf{r}}\, = \,  \left(\mathbf{R}_{\mathfrak{e}_7}^{\mathbf{r}}\right)^T \, \mathfrak{C}_{\mathrm{\mathfrak{e}_7}}
  \, \mathbf{R}_{\mathfrak{e}_7}^{\mathbf{r}}\, = \,  \mathfrak{C}_{\mathrm{\mathfrak{e}_7}}
\end{equation}
where:
\begin{equation}\label{cartalinoe7}
  \mathfrak{C}_{\mathrm{\mathfrak{e}_7}} \, = \, \left(
\begin{array}{ccccccc}
 2 & -1 & 0 & 0 & 0 & 0 & 0 \\
 -1 & 2 & -1 & 0 & 0 & 0 & 0 \\
 0 & -1 & 2 & -1 & 0 & 0 & 0 \\
 0 & 0 & -1 & 2 & -1 & -1 & 0 \\
 0 & 0 & 0 & -1 & 2 & 0 & 0 \\
 0 & 0 & 0 & -1 & 0 & 2 & -1 \\
 0 & 0 & 0 & 0 & 0 & -1 & 2 \\
\end{array}
\right)
\end{equation}
is the Cartan matrix of $\mathrm{\mathfrak{e}_7}$.
\par
This construction guarantees that via its $7$-dimensional
irreducible representation the group $\mathrm{PSL(2,7)}$ is embedded
into the Weyl group of $\mathfrak{e}_7$. So that we can write:
\begin{equation}
\mathrm{PSL(2,7)} \, \stackrel{\text{Irrep 7}}{\hookrightarrow} \,
\mathrm{Weyl}\left[\mathrm{\mathfrak{a}_{7}}\right] \, \subset \,
\mathrm{Weyl}\left[\mathrm{\mathfrak{e}_{7}}\right]
\end{equation}
\subsection{The second embedding of  \texorpdfstring{$\mathrm{PSL(2,7)}$}{PSL(2,7)} into \texorpdfstring{$\mathrm{Weyl}[\mathfrak{e}_7]$}{Weyl[e7]}} There is
another embedding of the $\mathrm{PSL(2,7)}$ group into
$\mathrm{Weyl}\left[\mathrm{\mathfrak{e}_{7}}\right]$ which is
governed by the crystallographic $6$-dimensional representation and
which turns out to be the relevant one to construct solutions of
Englert equations utilizing exceptional field theory:
\begin{equation}
\mathrm{PSL(2,7)} \, \stackrel{\text{Irrep 6}}{\hookrightarrow}
\,\mathrm{Weyl}\left[\mathrm{\mathfrak{a}_{6}}\right] \, \subset \,
\mathrm{Weyl}\left[\mathrm{\mathfrak{e}_{7}}\right]
\end{equation}
To understand this second embedding let us compare the Dynkin
diagram of $\mathfrak{e}_7$, in Eq. (\ref{bistolfoE7}) with that of
$\mathfrak{a}_6$ in Eq.  (\ref{bistolfoA6}). It is clear that the Lie
algebra $\mathfrak{a}_6$ is also regularly embedded in
$\mathfrak{e}_7$ since it suffices to identify the simple roots of
the former with a subset of the simple roots of the latter:
\begin{equation}\label{gammerie}
    \gamma_{1,2,3,4} \, = \, \alpha_{1,2,3,4} \quad ; \quad \gamma_5
    \, = \, \alpha_6 \quad ; \quad \gamma_6 \, = \, \alpha_7
\end{equation}
It follows that the root lattice
$\Lambda^{\mathbf{r}}_{\mathfrak{a}_6}$ of $\mathfrak{a}_6$ is a
sublattice of $\Lambda^{\mathbf{r}}_{\mathfrak{e}_7}$. Indeed we
have:
\begin{equation}\label{sublattosio}
    \mathbf{v}\in \Lambda^{\mathbf{r}}_{\mathfrak{a}_6}\subset \Lambda^{\mathbf{r}}_{\mathfrak{e}_6}
    \, \Leftrightarrow \, \mathbf{v} \, = \, v_i \, \alpha^i \quad
    \text{with} \;\;
    v_5 = 0 \;\;\text{and}\;\; v_{1,2,3,4,6,7} \in \mathbb{Z}
\end{equation}
What we need is an orthogonal decomposition of the root lattice of
$\mathfrak{e}_7$ into the root lattice $\mathfrak{a}_6$ plus its
one-dimensional complement:
\begin{equation}\label{ortofrutta}
    \Lambda^{\mathbf{r}}_{\mathfrak{e}_7} \, \supset \,
    \Lambda^{\mathbf{r}}_{\mathfrak{a}_6} \oplus \Lambda^{\mathbf{r}}_{\mathbf{1}}
\end{equation}
Orthogonality is obviously meant with respect to the Cartan matrix
$\mathfrak{C}_{\mathfrak{e}_7}$. Imposing the condition that a
vector $\mathbf{w}\in \Lambda^{\mathbf{r}}_{\mathbf{1}}$ should have
vanishing scalar product with any vector $\mathbf{v}
\in\Lambda^{\mathbf{r}}_{\mathfrak{a}_6}$:
\begin{equation}\label{carmelitus}
  0 \, = \,   \left(\mathbf{v}\, , \mathbf{w}\right)\, \equiv \, v_i\,
  w_j \, \mathfrak{C}_{\mathfrak{e}_7}^{ij}
\end{equation}
we immediately find the solution. The sublattice
$\Lambda^{\mathbf{r}}_{\mathbf{1}}$ is spanned by all vectors of the
form
\begin{equation}\label{caprifoglio}
  w^i \, = \,   \{3 m,6 m,9 m,12 m,7 m,8 m,4 m\} \quad ;
  \quad m\in \mathbb{Z}
\end{equation}
It is convenient to use a permutation and rename the simple root
$\alpha_5$ as the last one $\alpha_7$, so that the first six roots
span the $\mathfrak{a}_6$ root lattice. This is done by the matrix:
\begin{equation}\label{pmatto}
    P \, \equiv \, \left(
\begin{array}{ccccccc}
 1 & 0 & 0 & 0 & 0 & 0 & 0 \\
 0 & 1 & 0 & 0 & 0 & 0 & 0 \\
 0 & 0 & 1 & 0 & 0 & 0 & 0 \\
 0 & 0 & 0 & 1 & 0 & 0 & 0 \\
 0 & 0 & 0 & 0 & 0 & 1 & 0 \\
 0 & 0 & 0 & 0 & 0 & 0 & 1 \\
 0 & 0 & 0 & 0 & 1 & 0 & 0 \\
\end{array}
\right)
\end{equation}
In the permuted basis of simple roots the $\mathfrak{e}_7$ Cartan
matrix becomes:
\begin{equation}\label{NCarta}
    \hat{\mathfrak{C}}_{\mathfrak{e}_7}\, = \,  (P^{-1})^T \,
    \mathfrak{C}_{\mathfrak{e}_7} \, P^{-1} \, = \, \left(
\begin{array}{ccccccc}
 2 & -1 & 0 & 0 & 0 & 0 & 0 \\
 -1 & 2 & -1 & 0 & 0 & 0 & 0 \\
 0 & -1 & 2 & -1 & 0 & 0 & 0 \\
 0 & 0 & -1 & 2 & -1 & 0 & -1 \\
 0 & 0 & 0 & -1 & 2 & -1 & 0 \\
 0 & 0 & 0 & 0 & -1 & 2 & 0 \\
 0 & 0 & 0 & -1 & 0 & 0 & 2 \\
\end{array}
\right)
\end{equation}
In this basis the orthogonal decomposition (\ref{ortofrutta}) of the
$\mathfrak{e}_7$ root lattice is represented as follows:
\begin{eqnarray}
  \mathbf{v} \in  \Lambda^{\mathbf{r}}_{\mathfrak{a}_6} &\Leftrightarrow& \mathbf{v} \, = \,
  \{v_1, \,v_2,\, v_3,\, v_4, \, v_5, \, v_6, \, 0 \} \quad v_i\in \mathbb{Z}\nonumber\\
  \mathbf{w} \in  \Lambda^{\mathbf{r}}_{\mathbf{1}}  &\Leftrightarrow&
  \{3 m,\,6 m,\,9 m,\,12 m,\,8 m,\,4 m,\,7 m\} \quad m\in \mathbb{Z}\label{perrukino}
\end{eqnarray}
Using this basis we can introduce the embedding of the 6-dimensional
crystallographic representation of the $\mathrm{PSL(2,7)}$ into the
point group of the $\mathfrak{e}_7$ root lattice. We write the
following form of the three generators of the considered group:
\begin{equation}\label{cagnolinorosso}
   \mathcal{G} \, \equiv \, \left\{\rho,\,\sigma,\,\tau\right\} \, = \,
   \left\{\mathbf{R}^{\mathbf{r}}_{6+1}\, , \,
    \mathbf{S}^{\mathbf{r}}_{6+1}\, , \,
    \mathbf{T}^{\mathbf{r}}_{6+1}\right\}
\end{equation}
where
\begin{eqnarray*}
  \mathbf{R}^{\mathbf{r}}_{6+1} &=& \left(
\begin{array}{cccccc||c}
 0 & -1 & 1 & 0 & 0 & 0 & 0 \\
 -1 & 0 & 1 & 0 & 0 & 0 & 0 \\
 0 & 0 & 1 & 0 & 0 & 0 & 0 \\
 0 & 0 & 0 & 1 & 0 & 0 & 0 \\
 0 & 0 & 0 & 0 & 1 & 0 & 0 \\
 0 & 0 & 0 & 0 & 1 & -1 & 0 \\
 \hline
 \hline
 0 & 0 & 0 & 0 & 0 & 0 & 1 \\
\end{array}
\right)
\end{eqnarray*}
\begin{eqnarray*}
  \mathbf{S}^{\mathbf{r}}_{6+1} &=& \left(
\begin{array}{cccccc||c}
 -1 & 1 & 0 & 0 & 0 & 0 & 0 \\
 -1 & 1 & 0 & 0 & 0 & -1 & 1 \\
 -1 & 1 & 0 & -1 & 1 & -1 & 2 \\
 -1 & 0 & 1 & -1 & 1 & -1 & 2 \\
 -1 & 0 & 0 & 0 & 1 & -1 & 1 \\
 -1 & 0 & 0 & 0 & 0 & 0 & 1 \\
 \hline \hline
 0 & 0 & 0 & 0 & 0 & 0 & 1 \\
\end{array}
\right)
\end{eqnarray*}
\begin{eqnarray}
  \mathbf{T}^{\mathbf{r}}_{6+1} &=& \left(
\begin{array}{cccccc||c}
 0 & 0 & 0 & 0 & -1 & 1 & 1 \\
 0 & -1 & 1 & 0 & -1 & 1 & 1 \\
 0 & -1 & 0 & 1 & -1 & 1 & 1 \\
 0 & -1 & 0 & 0 & 0 & 1 & 2 \\
 1 & -1 & 0 & 0 & 0 & 1 & 1 \\
 1 & -1 & 0 & 0 & 0 & 0 & 1 \\
 \hline\hline
 0 & 0 & 0 & 0 & 0 & 0 & 1 \\
\end{array}
\right)
\end{eqnarray}
which have  the following properties:
\begin{description}
  \item[1)] The defining relations of $\mathrm{PSL(2,7)}$ displayed in
  Eq. (\ref{abstroL168Ab}) are satisfied.
  \item[2)] The generators preserve the Cartan matrix of
  $\mathfrak{e}_7$:
  \begin{equation}\label{parmigiano}
    \mathcal{G}_i^T \, \hat{\mathfrak{C}}_{\mathfrak{e}_7} \,
    \mathcal{G}_i \, = \, \hat{\mathfrak{C}}_{\mathfrak{e}_7} \quad
    \text{for}\quad i\,=\,\rho,\sigma,\tau
  \end{equation}
  \item[3)] The generators preserve the splitting (\ref{perrukino}) of the
  root lattice, namely they map any vector belonging to the
  sublattice $\Lambda^{\mathbf{r}}_{\mathfrak{a}_6}$ into a vector
   belonging to the same  sublattice and leave invariant any vector
   belonging to $\Lambda^{\mathbf{r}}_{1}$
   \begin{eqnarray}
     \mathbf{v} \in \Lambda^{\mathbf{r}}_{\mathfrak{a}_6} &\Rightarrow& \mathcal{G}_i\, \mathbf{v}
     \,\in \,\Lambda^{\mathbf{r}}_{\mathfrak{a}_6} \quad \text{for}\quad i\,=\,\rho,\sigma,\tau \nonumber\\
     \mathbf{w} \in \Lambda^{\mathbf{r}}_{1} &\Rightarrow& \mathcal{G}_i \,\mathbf{w}
     \,=\,\mathbf{w}\quad\text{for}\quad i\,=\,\rho,\sigma,\tau
   \end{eqnarray}
  \item[4)] The first $6\times 6$ blocks of the 7-dimensional
  matrices $\mathcal{G}_i$ are, respectively, the matrices $R_6$,
  $S_6$, $T_6$ displayed in Eq.s (\ref{rgen6}) and generating the
  irreducible 6-dimensional crystallographic representation of
  $\mathrm{PSL(2,7)}$ that maps the $\mathfrak{a}_6$ root lattice into itself.
\end{description}
\subsubsection{Change of basis}
Once the embedding of the 6-dimensional representation of
$\mathrm{PSL(2,7)}$ has been done in one basis it can be transformed
to any other basis. We are interested in the weight basis of the
$\mathfrak{a}_7$-lattice; hence we introduce the following product
of transformation matrices:
\begin{equation}\label{gotamatra}
    \mathfrak{M}\, = \, P \cdot \pmb{\Pi} \cdot
    \mathfrak{C}_{\mathfrak{a}_7}^{-1}
\end{equation}
where the first factor brings back to the standard labeling of
$\mathfrak{e}_7$ roots, as in Eq. (\ref{bistolfoE7}), the second
converts to the $\mathfrak{a}_7$ root lattice  and the last converts
from the root to the $\mathfrak{a}_7$ weight lattice. Setting:
\begin{eqnarray}
    \mathbf{R}^{\mathbf{w}}_{6+1} & = & \mathfrak{M}^{-1}\,
    \mathbf{R}^{\mathbf{r}}_{6+1}\,\mathfrak{M} \nonumber\\
    \mathbf{S}^{\mathbf{w}}_{6+1} & = & \mathfrak{M}^{-1}\,
    \mathbf{S}^{\mathbf{r}}_{6+1}\,\mathfrak{M} \nonumber\\
    \mathbf{T}^{\mathbf{w}}_{6+1} & = & \mathfrak{M}^{-1}\,
    \mathbf{T}^{\mathbf{r}}_{6+1}\,\mathfrak{M} \label{pigiamapalazzo}
\end{eqnarray}
we obtain:
\begin{eqnarray*}
\mathbf{R}^{\mathbf{r}}_{6+1} &=& \left(
\begin{array}{c||cccccc}
 1 & 1 & 1 & 0 & 0 & 0 & 0 \\
 \hline\hline
 0 & 0 & -1 & 0 & 0 & 0 & 0 \\
 0 & -1 & 0 & 0 & 0 & 0 & 0 \\
 0 & 1 & 1 & 1 & 0 & 0 & 0 \\
 0 & 0 & 0 & 0 & 1 & 0 & 0 \\
 0 & 0 & 0 & 0 & 0 & 1 & 1 \\
 0 & 0 & 0 & 0 & 0 & 0 & -1 \\
\end{array}
\right)
\end{eqnarray*}
\begin{eqnarray*}
  \mathbf{S}^{\mathbf{r}}_{6+1} &=& \left(
\begin{array}{c||cccccc}
 1 & 1 & 0 & 0 & 0 & 0 & 0 \\
 \hline\hline
 0 & 0 & 1 & 1 & 1 & 1 & 1 \\
 0 & 0 & 0 & 0 & 0 & -1 & -1 \\
 0 & 0 & 0 & -1 & -1 & 0 & 0 \\
 0 & 0 & 0 & 1 & 0 & 0 & 0 \\
 0 & 0 & 0 & 0 & 1 & 1 & 0 \\
 0 & -1 & -1 & -1 & -1 & -1 & 0
\end{array}
\right)
\end{eqnarray*}
\begin{eqnarray}
  \mathbf{T}^{\mathbf{w}}_{6+1} &=& \left(
\begin{array}{c||cccccc}
 1 & 1 & 1 & 1 & 1 & 1 & 0 \\
 \hline\hline
 0 & 0 & 0 & -1 & -1 & -1 & 0 \\
 0 & 0 & 0 & 1 & 0 & 0 & 0 \\
 0 & 0 & 0 & 0 & 1 & 0 & 0 \\
 0 & -1 & -1 & -1 & -1 & 0 & 0
   \\
 0 & 1 & 1 & 1 & 1 & 1 & 1 \\
 0 & 0 & -1 & -1 & -1 & -1 & -1
   \\
\end{array} \label{bugiardone}
\right)
\end{eqnarray}
Naming, respectively, $\mathbf{R}^{\mathbf{w}}_{6}$,
$\mathbf{S}^{\mathbf{w}}_{6}$ and $\mathbf{T}^{\mathbf{w}}_{6}$ the
lower $6 \times 6$ blocks of the above three matrices (they are
separated by lines in Eq.s~(\ref{bugiardone}) we obtain the three
generators of the $6$-dimensional representation of
$\mathrm{PSL(2,7)}$ which is crystallographic with respect to the
weight lattice $\Lambda^{w}_{\mathfrak{a}_6}$. In the
$\mathfrak{a}_7$ weight basis the invariant sublattice is spanned by
the vectors of the following form:
\begin{equation}\label{ocone}
   \mathfrak{w} \, = \,\mathfrak{M}^{-1} \,\{3 m,6 m,9 m,12 m,8 m,4 m,7 m\} \, = \,
   \{4 m,0,0,0,0,0,0\}
\end{equation}
and the group $\mathrm{PSL(2,7)}$ generated by the $7\times 7$
matrices (\ref{bugiardone}) leave the orthogonal complement
(\ref{ocone}) invariant.
\section{Constructing the elementary solution}
\label{baciogallone} We come next to the construction of solutions
to Englert equation utilizing, as building blocks, the minimal
solutions whose structure is governed by  Eq.  (\ref{carisma}) that
we presently retrieve.To this effect, having constructed the
explicit form of the two isomorphic groups $\mathrm{PSL(2,7)_7}$ and
$\mathrm{PSL(2,7)_{1+6}}$ we let them act on the complete
$\mathfrak{e}_7$ root system $\Delta_{126}$ containing 126 roots and
we observe how this latter splits into orbits.
\paragraph{$\mathrm{PSL(2,7)_7}$-case} We consider first the case
where $\mathrm{PSL(2,7)}$ is embedded into
$\mathrm{Weyl[\mathfrak{e}_7]}$ through its seven-dimensional
irreducible representation. Under the action of this group we find
the following four orbits:
\begin{equation}\label{cappuccionero}
    \Delta_{\mathbf{126}} = \mathcal{O}_{14A}\oplus \mathcal{O}_{14B}
    \oplus \mathcal{O}_{42} \oplus \mathcal{O}_{56}
\end{equation}
whose explicit content is displayed below:
\begin{equation}\label{panegiallo14A}
    \mathcal{O}_{14A}\, =
    \,\left\{11,33,34,40,41,47,57\right\}_{\mathrm{neg}}\bigcup
    \left\{11,33,34,40,41,47,57\right\}_{\mathrm{pos}}
\end{equation}
\begin{equation}\label{panegiallo14B}
    \mathcal{O}_{14B}\, =
    \,\left\{19,21,30,42,43,52,54\right\}_{\mathrm{neg}}\, \bigcup
    \,
    \left\{19,21,30,42,43,52,54\right\}_{\mathrm{pos}}
\end{equation}
\begin{eqnarray}
  \mathcal{O}_{42} &=&
  \left\{5,16,24,25,26,29,31,35,36,37,38,39,44,45,46,48,49,50,51,53,59\right\}_{\mathrm{neg}}
  \nonumber \\
  && \bigcup\nonumber\\
   &&\left\{5,16,24,25,26,29,31,35,36,37,38,39,44,45,46,48,49,50,51,53,59\right\}_{\mathrm{pos}}
   \label{panegiallo42}
\end{eqnarray}
\begin{eqnarray}
    \mathcal{O}_{56}& = &\left\{1,2,3,4,6,7,8,9,10,12,13,14,15,17,18,20,22,
   23,27,28,32,55,56,58,60,61,62,63\right\}_{\mathrm{neg}}\nonumber\\
    &&\bigcup\nonumber\\
   &&\left\{1,2,3,4,6,7,8,9,10,12,13,14,15,17,18,20,22,
   23,27,28,32,55,56,58,60,61,62,63\right\}_{\mathrm{pos}}\nonumber\\
   \label{panegiallo56}
\end{eqnarray}
In the above equations we have utilized the following notation: the
numbers from 1 to 63 refer to the positive roots as listed in table
\ref{racine}. The suffix pos/neg indicates whether the roots in the
brackets are the positive ones or their negatives enumerated in the
same order.
\par
The key point is that no subset of purely positive roots is left
invariant by the group $\mathrm{PSL(2,7)_7}$. This shows that this
embedding is inconvenient in order to utilize the group
$\mathrm{PSL(2,7)_7}$ as a classifier for fields $Y_{ijk}$. Indeed
in the compactification of M-theory on a $\mathrm{T^7}$ torus the
massless fields are in correspondence with the positive roots.
\begin{table}
\centering{{\small $$
\begin{array}{|r||l|}
\hline
\begin{array}{c|c}
 1 & \{1,0,0,0,0,0,0\} \\
 2 & \{0,1,0,0,0,0,0\} \\
 3 & \{0,0,1,0,0,0,0\} \\
 4 & \{0,0,0,1,0,0,0\} \\
 5 & \{0,0,0,0,1,0,0\} \\
 6 & \{0,0,0,0,0,1,0\} \\
 7 & \{0,0,0,0,0,0,1\} \\
 8 & \{1,1,0,0,0,0,0\} \\
 9 & \{0,1,1,0,0,0,0\} \\
 10 & \{0,0,1,1,0,0,0\} \\
 11 & \{0,0,0,1,1,0,0\} \\
 12 & \{0,0,0,1,0,1,0\} \\
 13 & \{0,0,0,0,0,1,1\} \\
 14 & \{1,1,1,0,0,0,0\} \\
 15 & \{0,1,1,1,0,0,0\} \\
 16 & \{0,0,1,1,1,0,0\} \\
 17 & \{0,0,1,1,0,1,0\} \\
 18 & \{0,0,0,1,0,1,1\} \\
 19 & \{0,0,0,1,1,1,0\} \\
 20 & \{1,1,1,1,0,0,0\} \\
 21 & \{0,1,1,1,1,0,0\} \\
 22 & \{0,1,1,1,0,1,0\} \\
 23 & \{0,0,1,1,0,1,1\} \\
 24 & \{0,0,1,1,1,1,0\} \\
 25 & \{0,0,0,1,1,1,1\} \\
 26 & \{1,1,1,1,1,0,0\} \\
 27 & \{1,1,1,1,0,1,0\} \\
 28 & \{0,1,1,1,0,1,1\} \\
 29 & \{0,1,1,1,1,1,0\} \\
 30 & \{0,0,1,1,1,1,1\} \\
 31 & \{0,0,1,2,1,1,0\} \\
 32 & \{1,1,1,1,0,1,1\} \\
 \end{array}
 &
 \begin{array}{c|c}
 33 & \{1,1,1,1,1,1,0\} \\
 34 & \{0,1,1,1,1,1,1\} \\
 35 & \{0,1,1,2,1,1,0\} \\
 36 & \{0,0,1,2,1,1,1\} \\
 37 & \{1,1,1,1,1,1,1\} \\
 38 & \{1,1,1,2,1,1,0\} \\
 39 & \{0,1,1,2,1,1,1\} \\
 40 & \{0,1,2,2,1,1,0\} \\
 41 & \{0,0,1,2,1,2,1\} \\
 42 & \{1,1,1,2,1,1,1\} \\
 43 & \{1,1,2,2,1,1,0\} \\
 44 & \{0,1,1,2,1,2,1\} \\
 45 & \{0,1,2,2,1,1,1\} \\
 46 & \{1,1,1,2,1,2,1\} \\
 47 & \{1,1,2,2,1,1,1\} \\
 48 & \{1,2,2,2,1,1,0\} \\
 49 & \{0,1,2,2,1,2,1\} \\
 50 & \{1,1,2,2,1,2,1\} \\
 51 & \{1,2,2,2,1,1,1\} \\
 52 & \{0,1,2,3,1,2,1\} \\
 53 & \{1,1,2,3,1,2,1\} \\
 54 & \{1,2,2,2,1,2,1\} \\
 55 & \{0,1,2,3,2,2,1\} \\
 56 & \{1,1,2,3,2,2,1\} \\
 57 & \{1,2,2,3,1,2,1\} \\
 58 & \{1,2,2,3,2,2,1\} \\
 59 & \{1,2,3,3,1,2,1\} \\
 60 & \{1,2,3,3,2,2,1\} \\
 61 & \{1,2,3,4,2,2,1\} \\
 62 & \{1,2,3,4,2,3,1\} \\
 63 & \{1,2,3,4,2,3,2\} \\
 \null& \null\\
\end{array}\\
\hline
\end{array}
$$}
} \caption{\label{racine}  Enumeration of the 63 positive roots of
$\mathfrak{e}_7$ displayed in the simple root basis. }
\end{table}
\paragraph{$\mathrm{PSL(2,7)_{1+6}}$-case} If we embed
$\mathrm{PSL(2,7)}$ into $\mathrm{Weyl[\mathfrak{e}_7]}$ through its
six-dimensional irreducible representation, the scenario of orbits
changes considerably.  Under the action of $\mathrm{PSL(2,7)_{1+6}}$
the set of 126  $\mathfrak{e}_7$ roots splits into the following
orbits:
\begin{equation}\label{crotino}
    \Delta_{\mathbf{126}}\,=\,\mathcal{O}_{7A}^+\oplus\mathcal{O}_{7A}^-
    \oplus \mathcal{O}_{7C}^+\oplus\mathcal{O}_{7C}^-\oplus
     \mathcal{O}_{28}^+\oplus\mathcal{O}_{28}^- \oplus \mathcal{O}_{42}
\end{equation}
where:
\begin{equation}\label{panenero7A}
    \mathcal{O}_{7A}^\pm\, =
    \,\left\{5,31,42,44,45,48,50\right\}_{\frac{\mathrm{pos}}{
    \mathrm{neg}}}
\end{equation}
\begin{equation}\label{panenero7C}
    \mathcal{O}_{7C}^\pm\, =
    \,\left\{55,56,58,60,61,62,63\right\}_{\frac{\mathrm{pos}}{
    \mathrm{neg}}}
\end{equation}
\begin{eqnarray}\label{panenero28}
    &&\mathcal{O}_{28}^\pm\, =\nonumber\\
    &&\left\{11,16,19,21,24,25,26,29,30,33,34,35,36,37,38,39,40,41,43,
    46,47,49,51,52,53,54,57,59\right\}_{\frac{\mathrm{pos}}{
    \mathrm{neg}}}\nonumber\\
\end{eqnarray}
\begin{eqnarray}\label{panenero42}
    \mathcal{O}_{42}& = &
    \,\left\{1,2,3,4,6,7,8,9,10,12,13,14,15,17,18,20,22,23,27,28,32\right\}_{\mathrm{pos}}\nonumber\\
    && \bigcup  \nonumber\\
    &&\,\left\{1,2,3,4,6,7,8,9,10,12,13,14,15,17,18,20,22,23,27,28,32\right\}_{\mathrm{neg}}
\end{eqnarray}
where the notation for the roots is the same as that utilized
before.
\par
The group theoretical and physical interpretation of the above
splitting is clear. The orbit of 42 roots is made by the roots of
$\mathfrak{a}_6$, that is to say of the Lie algebra of the subgroup
$\mathrm{SL(7,\mathbb{R})}\subset \mathrm{E_{7(7)}}$ parameterizing
through the coset $\frac{\mathrm{SL(7,\mathbb{R})}}{\mathrm{SO(7)}}$
the metrics on the $\mathrm{T^7}$-torus. The orbit
$\mathcal{O}_{7C}^+$ is characterized, as it can be seen from table
\ref{racine}, by the fact that all its elements have $n_5=2$, namely
their grading with respect to the root $\alpha_5$ (see
Eq. (\ref{bistolfoE7})) is $2$. Projecting these vectors onto the
fundamental weights of $\mathfrak{a}_6$ they turn out to be the
weights of the fundamental defining representation of
$\mathrm{SL(7,\mathbb{R})}$. On the other hand the roots in the two
orbits $\mathcal{O}_{7A}^+$ and $\mathcal{O}_{7A}^+$ are
characterized by the fact that their grading with respect to
$\alpha_5$ is $1$ (see table \ref{racine}). Projecting these 35
roots on the fundamental weights of $\mathfrak{a}_6$ we find the
weights of the $\mathbf{35}$-dimensional representation enumerated
in table \ref{pesotti} and there put into one--to--one
correspondence with the components of a three--time antisymmetric
tensor, namely with a triple of different integer numbers in the
range $1,2,3,4,5,6,7$. This antisymmetric tensor is $Y_{ijk}$,
namely the $3$-form defined over the $7$-torus that is supposed to
satisfy Englert equation. Summarizing we have:
\begin{equation}\label{panettonedipasqua}
    \mathbf{35}\, = \, \mathcal{O}_{7A}^+ \, \oplus \, \mathcal{O}_{28}^+
\end{equation}
which is equation (\ref{carisma}).
\begin{table}
\centering {
$$
\begin{array}{||r||l||}
\hline\hline
\begin{array}{c|c|l}
\text{Enumeration} & \text{Triple} & \text{Corresponding weight} \\
\hline
 1 & \{1,2,3\} & \{0,0,1,0,0,0\}
   \\
 2 & \{1,2,4\} &
   \{0,1,-1,1,0,0\} \\
 3 & \{1,2,5\} &
   \{0,1,0,-1,1,0\} \\
 4 & \{1,2,6\} &
   \{0,1,0,0,-1,1\} \\
 5 & \{1,2,7\} &
   \{0,1,0,0,0,-1\} \\
 6 & \{1,3,4\} &
   \{1,-1,0,1,0,0\} \\
 7 & \{1,3,5\} &
   \{1,-1,1,-1,1,0\} \\
 8 & \{1,3,6\} &
   \{1,-1,1,0,-1,1\} \\
 9 & \{1,3,7\} &
   \{1,-1,1,0,0,-1\} \\
 10 & \{1,4,5\} &
   \{1,0,-1,0,1,0\} \\
 11 & \{1,4,6\} &
   \{1,0,-1,1,-1,1\} \\
 12 & \{1,4,7\} &
   \{1,0,-1,1,0,-1\} \\
 13 & \{1,5,6\} &
   \{1,0,0,-1,0,1\} \\
 14 & \{1,5,7\} &
   \{1,0,0,-1,1,-1\} \\
 15 & \{1,6,7\} &
   \{1,0,0,0,-1,0\} \\
 16 & \{2,3,4\} &
   \{-1,0,0,1,0,0\} \\
 17 & \{2,3,5\} &
   \{-1,0,1,-1,1,0\} \\
 18 & \{2,3,6\} &
   \{-1,0,1,0,-1,1\} \\
\end{array} &
\begin{array}{c|c|l}
\text{Enumeration} & \text{Triple} & \text{Corresponding weight} \\
\hline
 19 & \{2,3,7\} &
   \{-1,0,1,0,0,-1\} \\
 20 & \{2,4,5\} &
   \{-1,1,-1,0,1,0\} \\
 21 & \{2,4,6\} &
   \{-1,1,-1,1,-1,1\} \\
 22 & \{2,4,7\} &
   \{-1,1,-1,1,0,-1\} \\
 23 & \{2,5,6\} &
   \{-1,1,0,-1,0,1\} \\
 24 & \{2,5,7\} &
   \{-1,1,0,-1,1,-1\} \\
 25 & \{2,6,7\} &
   \{-1,1,0,0,-1,0\} \\
 26 & \{3,4,5\} &
   \{0,-1,0,0,1,0\} \\
 27 & \{3,4,6\} &
   \{0,-1,0,1,-1,1\} \\
 28 & \{3,4,7\} &
   \{0,-1,0,1,0,-1\} \\
 29 & \{3,5,6\} &
   \{0,-1,1,-1,0,1\} \\
 30 & \{3,5,7\} &
   \{0,-1,1,-1,1,-1\} \\
 31 & \{3,6,7\} &
   \{0,-1,1,0,-1,0\} \\
 32 & \{4,5,6\} &
   \{0,0,-1,0,0,1\} \\
 33 & \{4,5,7\} &
   \{0,0,-1,0,1,-1\} \\
 34 & \{4,6,7\} &
   \{0,0,-1,1,-1,0\} \\
 35 & \{5,6,7\} &
   \{0,0,0,-1,0,0\} \\
\null & \null &
   \null\\
\end{array}\\
\hline\hline
\end{array}
$$}
  \caption{\label{pesotti} In this table we enumerate the 35 weights of the irreducible representation of
  $\mathfrak{a}_6$ corresponding to an antisymmetric tensor $Y_{ijk}$ in $d=7$. We associate each
  weight vector to the corresponding triple $\{i,j,k\}$ of indices.}
\end{table}
\par
As we already stressed this is the starting point in the
construction of minimal solutions

\subsection{The Minimal Solutions}\label{minsol}
Let us now illustrate step by step how to construct a set of
solutions starting from a normal form of $Y_{ijk}$ in which seven
components correspond to a Steiner triple system. The
solutions will fit orbits with respect to the $\mathrm{PSL(2,7)}$
invariance group of this septuple. To this end we introduce the
relevant notation.
\par
A septuple is conveniently characterized in terms of a distinctive
\emph{signature} $(n_0,n_1,n_2)$ which is defined as follows:
$n_\ell$, $\ell=0,1,2$, is the number of couples of triplets which
have $\ell$ indices in common. The Steiner triples have signature
$(0,21,0).$
\par The \emph{automorphism group} of a
septuple is the subgroup of the permutation group $S_7$ acting on
the internal indices $i,j,k$, which leaves the set of seven triplets
invariant, though changing the order. $\mathrm{PSL(2,7)}$ is the
automorphism group of the Fano plane and thus of the Steiner system
defining the multiplication table of the octonions $\Longrightarrow
\mathrm{PSL(2,7)}\subset \mathrm{G_2}\subset \mathrm{SO(7)}$. Below
we represent the Fano plane with identification of its vertices with
the seven triplets
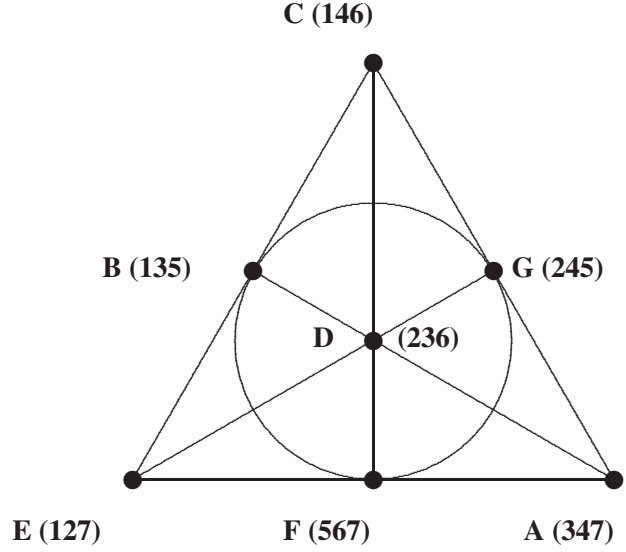
\begin{center}\label{Fanoex}
\setlength{\unitlength}{0.8cm}
\begin{picture}(8,9)(0,-1)
\put(0,0){\circle*{0.3}} \put(8,0){\circle*{0.3}}
\put(4,6.93){\circle*{0.3}} \put(4,0){\circle*{0.3}}
\put(4,2.3){\circle*{0.3}} \put(2,3.47){\circle*{0.3}}
\put(6,3.47){\circle*{0.3}} \put(0,0){\line(1,0){8}}
\put(0,0){\curve(0,0, 4,6.93)} \put(0,0){\curve(4,6.93, 8,0)}
\put(4,0){\line(0,1){6.93}} \put(0,0){\curve(2,3.47, 8,0)}
\put(0,0){\curve(6,3.47, 0,0)} \put(4,2.3){\bigcircle{4.6}} {\large
\bf \put(-2,-1){\small E (127)} \put(2.5,-1){\small F (567)}
\put(6.5,-1){\small A (347)} \put(-0.5,3.36){\small B (135)}
\put(6.3,3.36){\small G (245)} \put(3,2.2){\small D }
\put(4.4,2.2){\small (236)} \put(2.5,7.6){\small C (146)} }
\end{picture}
\end{center}
\setlength{\unitlength}{1pt}
\par
The construction of the solutions to the Englert equation proceeds as follows.
\begin{enumerate}
\item{We consider the case in which one of the two septuples, say $\vec{\sigma}$,
defines the embedding of an $\mathfrak{sl}(2)^7 $ group inside $ E_{7(7)}$ through the set of positive
roots $\alpha_{ijk}=\alpha_{\vec{\sigma}_I}$. As pointed out earlier, this system of triples is of Steiner type and defines the group ${\rm PSL}(2,7)$ with respect to which we construct the orbits of the solutions.}
\item{Then we choose the second set of 7 parameters of the minimal solution by choosing a 
septuple $\vec{\gamma}$ which is complementary to $\vec{\sigma}$. We shall classify in the sequel the independent choices of such septuples;}
\item{ Given the couple of complementary septuples $\vec\sigma$ and $\vec\gamma$, we construct
a solution $Y^{(\gamma \sigma)}$ through the formula:
{\small \begin{eqnarray}
Y^{(\gamma \sigma)}_{\vec\sigma_{P(I)}}(x^{i_I})&=&
\left(f_I\,\cos(\mu x^{i_I})+g_I\,\sin(\mu x^{i_I})\right)\,,\nonumber\\
Y^{(\gamma \sigma)}_{\vec\gamma_{P'(I)}}(x^{i_I})&=&\varepsilon_I\,\left(f_I \sin(\mu x^{i_I})
-g_I\cos(\mu x^{i_I})\right)\,\,,\,\,\,\,I=1,\dots, 7\,,\label{seed}
\end{eqnarray}}
with $\varepsilon_I=\epsilon^{i_I\,\sigma^1_{P(I)}\sigma^2_{P(I)}\sigma^3_{P(I)}
\gamma^1_{P'(I)}\gamma^2_{P'(I)}\gamma^3_{P'(I)}}$.}
\item{Being the Englert equation linear, a linear combination of solutions
$Y^{(\sigma\gamma)}$ corresponding to different choices of complementary
$\vec\sigma$ and $\vec\gamma$, is still a solution.}
\end{enumerate}
Given an elementary solution $Y^{(\gamma \sigma)}$ of the form (\ref{seed}),
we note that it sources the warp factor by a term which does not depend on the internal
coordinates $x^i$ since:
\begin{equation}
\frac{1}{6} (Y^{(\gamma \sigma)} \cdot Y^{(\gamma \sigma)})=
Y^{(\gamma \sigma)}_{\vec\sigma_{P(I)}} Y^{(\gamma \sigma)}_{\vec\sigma_{P(I)}}
+Y^{(\gamma \sigma)}_{\vec\gamma_{P'(I)}}
Y^{(\gamma \sigma)}_{\vec\gamma_{P'(I)}}=\sum_{I=1}^7 (f_I^2+g_I^2)\,,
\end{equation}
and therefore:
\begin{equation}
H=1-\frac{9}{4} e^{-2\mu U} \sum_{I=1}^7 (f_I^2+g_I^2).
\end{equation}
Notice that for this kind of solution $H=H(U)$, i.e. $H$ does not
depend on the torus coordinates $x^i$. This is reminiscent of
what happens for the 2-brane solution of seven-dimensional minimal
supergravity, studied in \cite{Fre:2015wvs}, when the internal
1-form flux, which satisfies the Arnold-Beltrami equation on a
3-torus, corresponds to the so-called \emph{ABC solution}. Combining
elementary solutions, the warp factor acquires a non-trivial
dependence on $x^i$.
\paragraph{Classifying the second septuple.}
The first septuple can be identified with the orbit ${\bf 7}_A$ in
the decomposition (\ref{carisma}) of the 35 roots $\alpha_{ijk}$ with
respect to the action of the corresponding $\mathrm{PSL(2,7)_{1+6}}$
automorphism group. The automorphism group of the second septuple
will intersect $\mathrm{PSL(2,7)_{1+6}}$ in a subgroup ${\mathrm H}$ of the
latter. We classify the second septuple by the possible choices of
${\mathrm H}$ in $\mathrm{PSL(2,7)_{1+6}}$. The condition on this subgroup is
that the decomposition of the ${\bf 28}$ orbit in (\ref{carisma}) with
respect to it should contain an order-7 orbit
${\bf 7}_B$, which is mutually non-local with respect to ${\bf
7}_A$. The septuple ${\bf 7}_B$ may also result from a combination of smaller ${\mathrm H}$-orbits. We considered the possible simple subgroups ${\mathrm H}$ classified in
section \ref{subgroupsH} and found the following results:
\begin{itemize}
\item{$\mathrm{H}={\rm O_{24A}}$ and ${\rm O_{24B}}$. The ${\bf 28}$ decomposes as:
\begin{equation}
{\bf 28}\rightarrow {\bf 12}+{\bf 12}+{\bf 4}\,.
\end{equation}
This case is not relevant to our analysis since the above decomposition contains
no order-7 orbit ${\bf 7}_B$;}
\item{$\mathrm{H}={\rm T_{12A}}$ and ${\rm T_{12B}}$. The ${\bf 28}$ decomposes as:
\begin{equation}
{\bf 28}\rightarrow {\bf 12}+{\bf 6}+{\bf 6}+{\bf 4}\,.
\end{equation}
Also this decomposition contains no orbit ${\bf 7}_B$;}
\item{$\mathrm{H}={\rm Dih_{3}}$. The ${\bf 28}$ decomposes as:
\begin{equation}
{\bf 28}\rightarrow {\bf 3}+{\bf 3}+{\bf 3}+{\bf 6}+{\bf 6}+{\bf 6}+{\bf 1}\,.
\end{equation}
In this case we checked that no-one of the combinations of orbits on the right hand side, with seven elements
realized either as ${\bf 3}+{\bf 3}+{\bf 1}$ or as ${\bf 6}+{\bf 1}$, is mutually non-local
with respect to ${\bf 7}_A$;}
\item{$\mathrm{H}={\rm G_{21}}$. The ${\bf 28}$ decomposes as:
\begin{equation}
{\bf 28}\rightarrow {\bf 7}+{\bf 21}\,.\label{G21dec}
\end{equation}
The order-7 orbit in the decomposition is mutually non-local with respect to ${\bf 7}_A$
and thus is a viable septuple ${\bf 7}_B$ for constructing a minimal solution.
Moreover this ${\bf 7}_B$ is of Steiner type;}
\item{$\mathrm{H}=\mathbb{Z}_7$. The ${\bf 28}$ decomposes as:
\begin{equation}
{\bf 28}\rightarrow {\bf 7}+{\bf 7}+{\bf 7}+{\bf 7}\,.
\end{equation}
All the four order-7 orbits in the decomposition are mutually non-local with respect
to ${\bf 7}_A$. One is of Steiner type and coincides with the one in (\ref{G21dec}),
being $\mathbb{Z}_7$ a subgroup of ${\rm G_{21}}$. The other three are not of Steiner
type and have signature $(7,7,7)$. Therefore all these four orbits are viable choices for ${\bf 7}_B$;}
\item{$\mathrm{H}=\mathbb{Z}_3$. The ${\bf 28}$ decomposes as:
\begin{equation}
{\bf 28}\rightarrow {\bf 1}+9\times {\bf 3}\,.
\end{equation}
 Also this decomposition contains septuples, realized as ${\bf 1}+{\bf 3}+{\bf 3}$, which are mutually non-local with
 respect to ${\bf 7}_A$. Some are of Steiner type and coincide with the one in (\ref{G21dec}),
 being $\mathbb{Z}_3$ a subgroup of ${\rm G_{21}}$, for isomorphic choices of ${\rm G_{21}}$
 inside ${\rm PSL(2,7)}$. The decomposition also features non-Steiner septuples
 with signatures $(3,15,3)$, $(6,9,6)$, $(0,15,6)$, $(0,18,3)$, $(3,12,6)$.
 The last three classes of ${\bf 7}_B$ are distinguished from the first two in that
 the coordinates complementary to ${\bf 7}_A$ and ${\bf 7}_B$ are not 7 but 4.
 This means that the corresponding minimal solution would only depend on 4 coordinates and the mapping $I\,\rightarrow \,i_I$ is not onto.}
\end{itemize}
Summarizing, we found viable septuples ${\bf 7}_B$ only when the
group $\mathrm{H}$ is either ${\rm G_{21}}$ or one of its subgroups.
Then we counted the possible septuples ${\bf 7}_B$ for various
isomorphic choices of $\mathrm{H}$ in ${\rm PSL(2,7)}$ and found the
multiplicities displayed in table \ref{finisterrae}.
\begin{table}
\caption{Multiplicities of the elementary solutions based on pair of
septuples $\mathbf{7}_A \oplus \mathbf{7}_B$ with fixed signature
for the second septuple. In the table we mention the invariance
group of the pair of septuples \label{finisterrae}}
\begin{center}
\begin{tabular}{|c|c|c|c|}
  \hline
  Aut & sign. & mult. & n. coord.s \\ \hline
  ${\rm G_{21}}$ & $(0,21,0)$ &  $8$ & 7 \\
  $\mathbb{Z}_7$ & $(7,7,7)$ & $24$ & 7\\
 $\mathbb{Z}_3$ & $(6,9,6)$ & $56$ & 7 \\
 $\mathbb{Z}_3$ & $(3,15,3)$ & $56$ & 7\\
 \hline
  $\mathbb{Z}_3$ & $(0,15,6)$ & $112$ & 4\\
  $\mathbb{Z}_3$ & $(0,18,3)$ & $56$ & 4\\
  $\mathbb{Z}_3$ & $(3,12,6)$ & $112$ & 4 \\
  \hline
  Total number & \null & $\mathbf{424}$&\null\\
  \hline
\end{tabular}
\end{center}
\end{table}
 The total number of independent minimal solutions is then $424$.
Only $$144=8+56+56+8$$ of these solutions depending on all the 7
coordinates. Among these latter, only 8 consists of two Steiner
systems.
\begin{table}
\centering{
$$
\begin{array}{|c||c||c||c||c||c||c||c|}
\hline \text{1st sept.} & \text{1st sept.} & \text{1st sept.}
&\text{1st sept.}
 & \text{1st sept.}& \text{1st sept.} &\text{1st sept.}& \text{1st
 sept.}\\
 \hline
\begin{array}{ccc}
 1 & 2 & 7 \\
 1 & 3 & 5 \\
 1 & 4 & 6 \\
 2 & 3 & 6 \\
 2 & 4 & 5 \\
 3 & 4 & 7 \\
 5 & 6 & 7 \\
\end{array}
 &
\begin{array}{ccc}
 1 & 2 & 7 \\
 1 & 3 & 5 \\
 1 & 4 & 6 \\
 2 & 3 & 6 \\
 2 & 4 & 5 \\
 3 & 4 & 7 \\
 5 & 6 & 7 \\
\end{array}
 &
\begin{array}{ccc}
 1 & 2 & 7 \\
 1 & 3 & 5 \\
 1 & 4 & 6 \\
 2 & 3 & 6 \\
 2 & 4 & 5 \\
 3 & 4 & 7 \\
 5 & 6 & 7 \\
\end{array}
 &
\begin{array}{ccc}
 1 & 2 & 7 \\
 1 & 3 & 5 \\
 1 & 4 & 6 \\
 2 & 3 & 6 \\
 2 & 4 & 5 \\
 3 & 4 & 7 \\
 5 & 6 & 7 \\
\end{array}
 &
\begin{array}{ccc}
 1 & 2 & 7 \\
 1 & 3 & 5 \\
 1 & 4 & 6 \\
 2 & 3 & 6 \\
 2 & 4 & 5 \\
 3 & 4 & 7 \\
 5 & 6 & 7 \\
\end{array}
 &
\begin{array}{ccc}
 1 & 2 & 7 \\
 1 & 3 & 5 \\
 1 & 4 & 6 \\
 2 & 3 & 6 \\
 2 & 4 & 5 \\
 3 & 4 & 7 \\
 5 & 6 & 7 \\
\end{array}
 &
\begin{array}{ccc}
 1 & 2 & 7 \\
 1 & 3 & 5 \\
 1 & 4 & 6 \\
 2 & 3 & 6 \\
 2 & 4 & 5 \\
 3 & 4 & 7 \\
 5 & 6 & 7 \\
\end{array}
 &
\begin{array}{ccc}
 1 & 2 & 7 \\
 1 & 3 & 5 \\
 1 & 4 & 6 \\
 2 & 3 & 6 \\
 2 & 4 & 5 \\
 3 & 4 & 7 \\
 5 & 6 & 7 \\
\end{array}
 \\
 \hline
 \hline
 \text{2nd sept.} & \text{2nd sept.} & \text{2nd sept.}& \text{2nd sept.}
 & \text{2nd sept.}& \text{2nd sept.} &\text{2nd sept.}& \text{2nd sept.}
 \\
 \hline
\begin{array}{ccc}
 1 & 2 & 5 \\
 1 & 3 & 4 \\
 1 & 6 & 7 \\
 2 & 3 & 7 \\
 2 & 4 & 6 \\
 3 & 5 & 6 \\
 4 & 5 & 7 \\
\end{array}
 &
\begin{array}{ccc}
 1 & 2 & 4 \\
 1 & 3 & 7 \\
 1 & 5 & 6 \\
 2 & 3 & 5 \\
 2 & 6 & 7 \\
 3 & 4 & 6 \\
 4 & 5 & 7 \\
\end{array}
 &
\begin{array}{ccc}
 1 & 2 & 6 \\
 1 & 3 & 4 \\
 1 & 5 & 7 \\
 2 & 3 & 5 \\
 2 & 4 & 7 \\
 3 & 6 & 7 \\
 4 & 5 & 6 \\
\end{array}
 &
\begin{array}{ccc}
 1 & 2 & 4 \\
 1 & 3 & 6 \\
 1 & 5 & 7 \\
 2 & 3 & 7 \\
 2 & 5 & 6 \\
 3 & 4 & 5 \\
 4 & 6 & 7 \\
\end{array}
&
\begin{array}{ccc}
 1 & 2 & 6 \\
 1 & 3 & 7 \\
 1 & 4 & 5 \\
 2 & 3 & 4 \\
 2 & 5 & 7 \\
 3 & 5 & 6 \\
 4 & 6 & 7 \\
\end{array}
 &
\begin{array}{ccc}
 1 & 2 & 3 \\
 1 & 4 & 5 \\
 1 & 6 & 7 \\
 2 & 4 & 7 \\
 2 & 5 & 6 \\
 3 & 4 & 6 \\
 3 & 5 & 7 \\
\end{array}
 &
\begin{array}{ccc}
 1 & 2 & 5 \\
 1 & 3 & 6 \\
 1 & 4 & 7 \\
 2 & 3 & 4 \\
 2 & 6 & 7 \\
 3 & 5 & 7 \\
 4 & 5 & 6 \\
\end{array}
 &
\begin{array}{ccc}
 1 & 2 & 3 \\
 1 & 4 & 7 \\
 1 & 5 & 6 \\
 2 & 4 & 6 \\
 2 & 5 & 7 \\
 3 & 4 & 5 \\
 3 & 6 & 7 \\
\end{array}
\\
\hline
\end{array}
$$
} \caption{\label{settupline} The eight pairs of mutually non local
Steiner septuples produced by the orbits of the 8 different
conjugate copies of subgroups $\mathrm{G_{21}}^I \subset
\mathrm{PSL(2,7)_{1+6}}$ ($I=1,\dots,8$). }
\end{table}
Notice that in general the solutions do not preserve any
supersymmetry. However, for particular choices of the parameters the
solutions can admit $\mathcal{N}=1,2,3,4,5,6$ supersymmetries. This is
quite different from the original Englert solution, which does not
preserve any supersymmetry. In the next sections we recall the
criterion for preservation of supersymmetries in the context of
these M2-brane solutions that was derived  in \cite{miol168} and we
apply it systematically to the solutions of type $(0,21,0)$
obtaining just only  from this sector a rich spectrum of
possibilities encompassing all available values of $\mathcal{N}$.
The analysis of the remaining solutions is postponed to a future
publications.
\section{The Killing spinor equation of M2-branes with Englert fluxes}
\label{kilspisection} As announced we review here the discussion of
the Killing spinor equation presented in \cite{miol168}.
\par
In order to analyze the structure of the Killing spinor equation in
the background of the M2-branes with Englert fluxes, we need a basis
of gamma matrices that is well-adapted to the splitting of the
11-dimensional manifold mentioned in Eq. (\ref{rabolatus}).
\par
Such a well adapted basis is provided by the following nested
hierarchy.
\subsection{Gamma matrices}
At the bottom of the hierarchy we have the Pauli matrices.
\paragraph{Pauli matrices.}
We use the following conventions:
\begin{equation}\label{paulisti}
\sigma_1 \,=\,\left(
\begin{array}{cc}
 0 & 1 \\
 1 & 0 \\
\end{array}
\right) \quad ; \quad \sigma_2 \,= \, \left(
\begin{array}{cc}
 0 & -i \\
 i & 0 \\
\end{array}
\right) \quad ; \quad \sigma_3 \, = \,\left(
\begin{array}{cc}
 1 & 0 \\
 0 & -1 \\
\end{array}
\right);
\end{equation}
\paragraph{Gamma matrices on the $d=3$ world-volume.} Next we construct the
 set of 2$\times $2 gamma matrices in $d=3$  in the following way
\begin{equation}\label{gammucci}
\left\{\gamma _{\underline{a}},\gamma _{\underline{b}} \right\}
=2\eta _{\underline{\text{ab}}} \quad ;\quad \gamma \, =
\,\left\{\sigma _2,\mathrm{i}\,\sigma _1,\mathrm{i}\,\sigma
_3\right\}\,\,,\,\,\,\,\underline{a},\,\underline{b}=1,2,3\,.
\end{equation}
\paragraph{Gamma matrices in $d=7$} In $d=7$ we choose gamma matrices that are real and
antisymmetric and fulfill the following Clifford algebra:
\begin{equation}\label{taumatti}
\left\{\tau _i,\tau _j\right\} = -2\delta _{{ij}}\,\,,\,\,\,\,i,j=1,\dots, 7\,.
\end{equation}
The explicit basis utilized is that one where we express the $\tau
$-matrices in terms of $\phi_{{ijk}}$, namely of the
$\mathrm{G_2}$-invariant three-tensor:
\begin{eqnarray}
\left(\tau _i\right)_{{jk}} &=& \phi_{{ijk}} \nonumber\\
\left(\tau _i\right)_{{j8}} &=& \delta _{{ij}}\quad ; \quad
\left(\tau _i\right)_{8j} = - \delta _{{ij}} \label{taubasa}
\end{eqnarray}
The explicit form of the $\phi _{{ijk}}$ tensor is given in
Eq. (\ref{gorlandus})  and it is the one well-adapted to the
immersion of the discrete group which acts crystallographically on
$\mathrm{T^7}$ into the compact $\mathrm{G_2}$ Lie group, namely
according to the canonical immersion $\mathrm{PSL(2,7)}
\longrightarrow \mathrm{G_{2(-14)}}$.
\paragraph{Gamma matrices in $d=8$}
Because of our splitting $11=3\oplus 1\oplus7 $ we need also the
gamma matrices in $d=8$ corresponding to the transverse space to the
M2-brane, namely $\mathbb{R}_+\otimes \mathrm{T^7}$. We choose the
following Clifford algebra:
\begin{equation}\label{d8gammi}
 \left\{T_I,T_J\right\} = -2 \delta _{{IJ}}\,\,,\,\,\,\,\,\,I,J=1,\dots, 8\,,
\end{equation}
and we utilize the following explicit realization:
 \begin{eqnarray}
   T_i & = & \sigma _1\otimes  \tau_i  \nonumber\\
   T_8 & = & \mathrm{i} \sigma _2\otimes  \pmb{1}_{8\times 8}  \nonumber \\
   T_9 &= & \sigma _3\otimes \pmb{1}_{8\times 8} \label{tmatre}
 \end{eqnarray}
The last matrix is the $d=8$ chirality operator which plays an
important role in the discussion of the Killing spinor equation.
\paragraph{Gamma matrices in $d=11$} At the top of the hierarchy we have the $d=11$ gamma
matrices, obeying the following Clifford algebra
\begin{equation}\label{d11gammi}
   \left\{\Gamma _a,\Gamma _b\right\}=2\eta _{\text{ab}}\,\,,\,\,\,\,\,\,a,b=0,\dots, 10\,.
\end{equation}
For them we utilize the following explicit realization:
\begin{eqnarray}
 \Gamma _{\underline{a}}& = &\gamma_{\underline{a}}\otimes T_9\nonumber\\
\Gamma _I & = &\pmb{1}_{2\times 2}\otimes  T_I
\end{eqnarray}
With these choices the charge conjugation matrix, takes the
following form:
\begin{eqnarray}
\mathcal{C}&=& \mathrm{i} \sigma _2\otimes \pmb{1}_{2\times 2} \otimes \pmb{1}_{8\times 8} \nonumber\\
\mathcal{C}\,\Gamma _a \mathcal{C}^{ -1}& = &- \Gamma _a{}^T
\end{eqnarray}
Equipped with this set of properly chosen gamma matrices we can turn
to the investigation of the Killing spinor equation.
\subsection{The tensor structure of the Killing spinor equation}
The rheonomic solution of the $d=11$ Bianchi identities (see
Eq. (\ref{rheoFDA})) allows us to write the Killing spinor equation
in the following  general form:
\begin{equation}\label{spikillusequo}
\mathcal{D} \xi -\frac{\mathrm{i}}{3}\Gamma
^{\text{abc}}V^dF_{\text{abcd} }\xi -\frac{\mathrm{i}}{24}\Gamma
_{\text{abcdf}} F^{\text{abcd}}V^f\xi  = 0
\end{equation}
where
\begin{equation}\label{lorenzoderivo}
    \mathcal{D} \xi  \equiv \text{d$\xi $} - \frac{1}{4}\omega ^{\text{ab}}\Gamma _{\text{ab}} \xi
\end{equation}
is the Lorentz covariant differential in $d=11$.
\par
 Equation (\ref{spikillusequo}) can be usefully rewritten as follows:
 \begin{equation}\label{genCovar}
 \nabla \xi
\equiv \text{d$\xi $}+\Omega
 \xi  =0
 \end{equation}
where $\Omega $ is a generalized connection in the 32--dimensional
spinor space, defined as follows:
\begin{equation}\label{genOmme}
    \Omega  \equiv \Theta _L+\Theta _1^{[F]}+\Theta _2^{[F]}
\end{equation}
In the above equation we have  introduced the following definitions:
\begin{eqnarray}
 \Theta _L &\equiv& - \frac{1}{4}\omega ^{\text{ab}}\Gamma_{\text{ab}}\nonumber\\
 \Theta _1^{[F]} &\equiv& -\frac{\mathrm{i}}{3}\Gamma ^{\text{abc}}V^dF_{\text{abcd} }\nonumber \\
 \Theta _2^{[F]} &\equiv& -\frac{\mathrm{i}}{24}\Gamma _{\text{abcdf}} F^{\text{abcd}}V^f
\end{eqnarray}
Next let us make another splitting of the overall generalized
connection:
\begin{equation}\label{omegasplittus}
   \Omega  = \Omega _H+ \Omega _Y
\end{equation}
where $\Omega _H$ depends only on the (inhomogeneous)-harmonic
function H and it is obtained from $\Omega $ by setting $Y_{{ijk}}
\,\rightarrow \, 0$. Instead, the other part $\Omega _Y$,  is just
the difference and it depends linearly on $Y_{ijk}$
\subsubsection{M2-branes without Englert fluxes: tensor structure of \texorpdfstring{$\Omega_H$}{OH}}
As shown in \cite{miol168}, by introducing the following operators:
\begin{eqnarray}
V\circ \gamma
 &= & V^{\underline{a}}\gamma _{\underline{a}}\label{oppo1}\\
 \mathbb{P}_{\pm } &=& \frac{1}{2}\left(\pmb{1}_{16} \pm  T_9\right) \label{oppo2}\\
 \partial H\circ T &=&\frac{1}{3}H ^{-\frac{7}{6}}\partial _IH T^I \label{oppo3}\\
 V\diamond \partial H\circ T & = &- \frac{1}{12} H ^{-\frac{7}{6}}
 V_{[I}\partial _{J]}H T^{\text{IJ}} \label{oppo4}\\
\text{$\mathbf{d}$H} & = &\frac{1}{6}H ^{-\frac{7}{6}} \sum _{I=1}^8
\partial _IH V^I\label{oppo5}
\end{eqnarray}
we get that the H-part of the generalized connection has the
following tensor structure:
\begin{equation}\label{omegaH}
\Omega _H = V\circ \gamma \otimes  \partial H\circ T \,\mathbb{P}_-
+ \pmb{\pmb{1}_2}\pmb{ }\otimes \text{ }V\diamond \partial H\circ T
\, \mathbb{P}_- +  \pmb{1}_2 \otimes\text{$\mathbf{d}$H}\,T_9
\end{equation}
From equation (\ref{omegaH}) one readily derives the form of the
Killing spinors for pure M2-brane solutions. Writing the 32
component Killing spinor as a tensor product:
\begin{equation}\label{stensorokillo}
   \xi  = \epsilon \otimes \chi
\end{equation}
we find that, in the absence of Y-fields, the Killing spinor
equation is satisfied provided:
\begin{eqnarray}
 T_9\chi & = & \chi \quad \Rightarrow \quad \mathbb{P}_{-}\chi \, = \, 0\label{chiraltus}\\
 \chi  &=& H^{ -\frac{1}{6}} \, \chi_0
\end{eqnarray}
where $H$ is the (inhomogeneous)-harmonic function appearing in the
metric (\ref{colabrodo}) and $\chi _0$ is a constant spinor with
commuting components. Indeed, in view of our 2-brane interpretation
of these backgrounds, we assume that the two--component spinors
$\epsilon $ are the anticommuting objects.
\par
Using the tensor structure of the $d=8$ T-matrices we set:
\begin{equation}\label{splittochi}
    \chi \, = \, \kappa \otimes \lambda
\end{equation}
where $\kappa $ is a two component spinor:
\begin{equation}\label{cappus}
\kappa =\left(
\begin{array}{c}
 \kappa _1 \\
 \kappa _2 \\
\end{array}
\right)
\end{equation}
with commuting components, while $\lambda $ is an eight--component
spinor:
\begin{equation}\label{lambdus}
\lambda \, = \,\left\{
 \lambda _1 , \,
 \lambda _2 , \,
 \lambda _3 , \,
 \lambda _4 , \,
 \lambda _5 , \,
 \lambda _6 , \,
 \lambda _7 , \,
 \lambda _8 \right\}
\end{equation}
also with commuting components.
\par
In this language the most general 32--component spinor has the form:
\begin{equation}\label{tritensori}
    \xi  = \epsilon \otimes \kappa \otimes\lambda
\end{equation}
and  the general solution for the Killing spinor at $Y_{{ijk}}\,= \,
0$ is obtained by setting:
\begin{equation}\label{crisaltus}
    \kappa _2 = 0 \quad ; \quad \kappa _1 = H^{ -\frac{1}{6}}
\end{equation}
This shows that the M2-branes without Englert-fluxes preserve 16
supersymmetries, namely $\frac{1}{2}$ of the total SUSY.
\subsubsection{M2-branes with Englert fluxes: tensor structure of \texorpdfstring{$\Omega _Y$}{OY}}
We come next to analyze the structure of the Y-part of the
connection $\Omega _Y$.
\par
We begin by introducing two  $d=7$ operators constructed with the
Englert field $Y_{{ijk}}$, the flat 8-dimensional  vielbein
$\hat{V}^I\equiv dy^I$ and the $\tau $-matrices:
\begin{equation}\label{operati}
    \mathcal{B} \equiv \tau_{{ijk}} \, {Y}_{{ijk}}\quad ; \quad\mathcal{T} = \hat{V}^i\tau _i
\end{equation}
in \cite{miol168} it was shown that $\Omega _Y$ can be written as
follows:
\begin{eqnarray}
   \Omega _Y &=&
\mathrm{i}\frac{1}{12}\mu e^{-U \mu }H^{-2/3}\times \nonumber\\
&& \left[V\circ \gamma
 \left(
\begin{array}{cc}
2 \mathcal{B} & 0
\\ 0 & 0 \\
\end{array}
\right) +\pmb{1}\otimes \left(
\begin{array}{cc}
\hat{V}^0\mathcal{B} &
0 \\ 0 & 0 \\
\end{array}
\right) +\frac{1}{2}\pmb{1}\otimes \left(
\begin{array}{cc}
0 & 3 \mathcal{B}
\mathcal{T} \\
-\mathcal{T} \mathcal{B} & 0
\\ \end{array}
\right) \right]\nonumber\\
\label{gomorra}
\end{eqnarray}
Eq.~(\ref{gomorra}) reveals the mechanism behind the preservation of
supersymmetry by M2-branes with Englert fluxes. Writing the
candidate Killing spinor in the tensor product form
(\ref{tritensori}) we see that the connection $\Omega _Y$
annihilates it if $\kappa = \left(\begin{array}{c}
H^{-\ft 16} \\
0
\end{array}\right)$ as we already established from consideration of the H-part of the connection and
if the 8-component $\lambda $ is a null-vector of $\mathcal{B}$:
\begin{equation}
   \mathcal{B} \, \lambda = 0\label{Binullo}
\end{equation}
This is the only possibility to integrate the Killing spinor
equation. Indeed the term with V$\circ \gamma $ which mixes the
internal coordinateswith the world volume ones has to vanish since
it cannot be compensated in any other way. This implies
Eq.~(\ref{Binullo}). The magic thing is that  the precise values of
the coefficients provided by the rheonomic solution of Bianchi
identities in $d=11$, produce  the structure in Eq. (\ref{gomorra}). In
this way the condition (\ref{Binullo}) suffices to annihilate also
the action of the other terms in the connection.
\par
In conclusion M2-branes with Englert fluxes preserve part of the
Killing spinors existing in the case of $Y=0$ if and only if the
operator $\mathcal{B}$ has a non trivial Null-Space, namely if the
Rank of $\mathcal{B}$ is $<$ 8. Every $\lambda $ satisfying
(\ref{Binullo}) corresponds to a preserved supersymmetry.\par
In Appendix \ref{travertinus} the above conditions on the matrix $\mathcal{B}$ for the solution to preserve an amount of supersymmetry are shown to be a special case of the general supersymmetry conditions worked out in the literature on M2-branes with self-dual fluxes.
\section{Supersymmetry of the solutions of type \texorpdfstring{$(0,21,0)$}{(0,21,0)}}
\label{0210susy} In this section we present the results we have
obtained for the supersymmetry of solutions of type $(0,21,0)$,
mentioned in table \ref{finisterrae}.
\par
According to the previously explained rules for the construction of minimal
solutions we have derived each of the eight $14$-parameter solutions
for the three-form $\mathbf{Y}$ obtained by pairing the standard
septuple $\mathbf{7}_A$ with one of the eight different septuples
$\mathbf{7}_B^I$ displayed in table \ref{settupline}. Let us name
them $\mathbf{Y}^{14}_I$, $I=1,\dots,8$. Since Englert equation is
linear, the sum of these solutions is also a solution:
\begin{equation}\label{romilato}
    \hat{\mathbf{Y}} \, = \, \sum_{I=1}^8 \, \mathbf{Y}^{14}_I
\end{equation}
which apparently depends on $8\times 14 = 112$ parameters. Actually
the independent combinations of differentials $dx^i\wedge dx^j
\wedge dx^k$ with the trigonometric functions $\cos(\mu x^\ell)$ and
$\sin(\mu x^\ell)$ that is produced in this sum are not 112 but rather $56$, since each combination appears twice. Renaming
$\delta_\alpha$, $\alpha=1,\dots,56$ the coefficients of the
independent combinations $\mathbf{B}^\alpha$ (they are listed in
table \ref{baciucco}), we have obtained a general solution of the
following form:
\begin{equation}\label{caramelladierbe}
    \mathbf{Y}^{56}\left(\pmb{x}|\pmb{\delta}\right) \, = \, \sum_{\alpha = 1}^{56} \, \delta_\alpha \,
\mathbf{B}^\alpha
\end{equation}
\begin{table}[H]
$$\begin{array}{|c||c|}
\hline \hline
\begin{array}{c|c}
\mathbf{B}^{1} & \cos (\mu   {x^4}) d {x^1}\wedge d {x^2}\wedge
   d {x^3}\\
\mathbf{B}^{ 2} & \sin (\mu   {x^4}) d {x^1}\wedge d {x^2}\wedge
   d {x^3}\\
 \mathbf{B}^{3} & \cos (\mu   {x^3}) d {x^1}\wedge d {x^2}\wedge
   d {x^4}\\
 \mathbf{B}^{4} & \sin (\mu   {x^3}) d {x^1}\wedge d {x^2}\wedge
   d {x^4}\\
 \mathbf{B}^{5} & \cos (\mu   {x^6}) d {x^1}\wedge d {x^2}\wedge
   d {x^5}\\
 \mathbf{B}^{6} & \sin (\mu   {x^6}) d {x^1}\wedge d {x^2}\wedge
   d {x^5}\\
 \mathbf{B}^{7} & \cos (\mu   {x^5}) d {x^1}\wedge d {x^2}\wedge
   d {x^6}\\
 \mathbf{B}^{8} & \sin (\mu   {x^5}) d {x^1}\wedge d {x^2}\wedge
   d {x^6}\\
 \mathbf{B}^{9} & \cos (\mu   {x^3}) d {x^1}\wedge d {x^2}\wedge
   d {x^7}\\
 \mathbf{B}^{10} & \cos (\mu   {x^4}) d {x^1}\wedge d {x^2}\wedge
   d {x^7}\\
 \mathbf{B}^{11} & \cos (\mu   {x^5}) d {x^1}\wedge d {x^2}\wedge
   d {x^7}\\
 \mathbf{B}^{12} & \cos (\mu   {x^6}) d {x^1}\wedge d {x^2}\wedge
   d {x^7}\\
 \mathbf{B}^{13} & \sin (\mu   {x^3}) d {x^1}\wedge d {x^2}\wedge
   d {x^7}\\
 \mathbf{B}^{14} & \sin (\mu   {x^4}) d {x^1}\wedge d {x^2}\wedge
   d {x^7}\\
 \mathbf{B}^{15} & \sin (\mu   {x^5}) d {x^1}\wedge d {x^2}\wedge
   d {x^7}\\
 \mathbf{B}^{16} & \sin (\mu   {x^6}) d {x^1}\wedge d {x^2}\wedge
   d {x^7}\\
 \mathbf{B}^{17} & \cos (\mu   {x^2}) d {x^1}\wedge d {x^3}\wedge
   d {x^4}\\
 \mathbf{B}^{18} & \sin (\mu   {x^2}) d {x^1}\wedge d {x^3}\wedge
   d {x^4}\\
 \mathbf{B}^{19} & \cos (\mu   {x^2}) d {x^1}\wedge d {x^3}\wedge
   d {x^5}\\
 \mathbf{B}^{20} & \cos (\mu   {x^4}) d {x^1}\wedge d {x^3}\wedge
   d {x^5}\\
 \mathbf{B}^{21} & \cos (\mu   {x^6}) d {x^1}\wedge d {x^3}\wedge
   d {x^5}\\
 \mathbf{B}^{22} & \cos (\mu   {x^7}) d {x^1}\wedge d {x^3}\wedge
   d {x^5}\\
 \mathbf{B}^{23} & \sin (\mu   {x^2}) d {x^1}\wedge d {x^3}\wedge
   d {x^5}\\
 \mathbf{B}^{24} & \sin (\mu   {x^4}) d {x^1}\wedge d {x^3}\wedge
   d {x^5}\\
 \mathbf{B}^{25} & \sin (\mu   {x^6}) d {x^1}\wedge d {x^3}\wedge
   d {x^5}\\
 \mathbf{B}^{26} & \sin (\mu   {x^7}) d {x^1}\wedge d {x^3}\wedge
   d {x^5}\\
 \mathbf{B}^{27} & \cos (\mu   {x^7}) d {x^1}\wedge d {x^3}\wedge
   d {x^6}\\
 \mathbf{B}^{28} & \sin (\mu   {x^7}) d {x^1}\wedge d {x^3}\wedge
   d {x^6}\\
   \end{array}&
\begin{array}{c|c}
 \mathbf{B}^{29} & \cos (\mu   {x^6}) d {x^1}\wedge d {x^3}\wedge
   d {x^7}\\
 \mathbf{B}^{30} & \sin (\mu   {x^6}) d {x^1}\wedge d {x^3}\wedge
   d {x^7}\\
 \mathbf{B}^{31} & \cos (\mu   {x^7}) d {x^1}\wedge d {x^4}\wedge
   d {x^5}\\
 \mathbf{B}^{32} & \sin (\mu   {x^7}) d {x^1}\wedge d {x^4}\wedge
   d {x^5}\\
 \mathbf{B}^{33} & \cos (\mu   {x^2}) d {x^1}\wedge d {x^4}\wedge
   d {x^6}\\
 \mathbf{B}^{34} & \cos (\mu   {x^3}) d {x^1}\wedge d {x^4}\wedge
   d {x^6}\\
 \mathbf{B}^{35} & \cos (\mu   {x^5}) d {x^1}\wedge d {x^4}\wedge
   d {x^6}\\
 \mathbf{B}^{36} & \cos (\mu   {x^7}) d {x^1}\wedge d {x^4}\wedge
   d {x^6}\\
 \mathbf{B}^{37} & \sin (\mu   {x^2}) d {x^1}\wedge d {x^4}\wedge
   d {x^6}\\
 \mathbf{B}^{38} & \sin (\mu   {x^3}) d {x^1}\wedge d {x^4}\wedge
   d {x^6}\\
 \mathbf{B}^{39} & \sin (\mu   {x^5}) d {x^1}\wedge d {x^4}\wedge
   d {x^6}\\
 \mathbf{B}^{40} & \sin (\mu   {x^7}) d {x^1}\wedge d {x^4}\wedge
   d {x^6}\\
 \mathbf{B}^{41} & \cos (\mu   {x^5}) d {x^1}\wedge d {x^4}\wedge
   d {x^7}\\
 \mathbf{B}^{42} & \sin (\mu   {x^5}) d {x^1}\wedge d {x^4}\wedge
   d {x^7}\\
 \mathbf{B}^{43} & \cos (\mu   {x^2}) d {x^1}\wedge d {x^5}\wedge
   d {x^6}\\
 \mathbf{B}^{44} & \sin (\mu   {x^2}) d {x^1}\wedge d {x^5}\wedge
   d {x^6}\\
 \mathbf{B}^{45} & \cos (\mu   {x^4}) d {x^1}\wedge d {x^5}\wedge
   d {x^7}\\
 \mathbf{B}^{46} & \sin (\mu   {x^4}) d {x^1}\wedge d {x^5}\wedge
   d {x^7}\\
 \mathbf{B}^{47} & \cos (\mu   {x^3}) d {x^1}\wedge d {x^6}\wedge
   d {x^7}\\
 \mathbf{B}^{48} & \sin (\mu   {x^3}) d {x^1}\wedge d {x^6}\wedge
   d {x^7}\\
 \mathbf{B}^{49} & \cos (\mu   {x^1}) d {x^2}\wedge d {x^3}\wedge
   d {x^4}\\
 \mathbf{B}^{50} & \sin (\mu   {x^1}) d {x^2}\wedge d {x^3}\wedge
   d {x^4}\\
 \mathbf{B}^{51} & \cos (\mu   {x^7}) d {x^2}\wedge d {x^3}\wedge
   d {x^5}\\
 \mathbf{B}^{52} & \sin (\mu   {x^7}) d {x^2}\wedge d {x^3}\wedge
   d {x^5}\\
 \mathbf{B}^{53} & \cos (\mu   {x^1}) d {x^2}\wedge d {x^3}\wedge
   d {x^6}\\
 \mathbf{B}^{54} & \cos (\mu   {x^4}) d {x^2}\wedge d {x^3}\wedge
   d {x^6}\\
 \mathbf{B}^{55} & \cos (\mu   {x^5}) d {x^2}\wedge d {x^3}\wedge
   d {x^6}\\
 \mathbf{B}^{56} & \cos (\mu   {x^7}) d {x^2}\wedge d {x^3}\wedge
   d {x^6}\\
\end{array}\\
\hline
\end{array}
$$
\caption{\label{baciucco} List of the addends $\mathbf{B}_\alpha$ in
the general solution of Englert equation corresponding to septuples
of signature $(0,21,0)$.}
\end{table}
The action of the group $\mathrm{PSL(2,7)}$ on the Englert form
$\mathbf{Y}^{56}(\pmb{x}|\pmb{\delta})$ is generated by the action
of the group on the seven coordinates $x^i$ which is only by means
of permutations. The explicit form of this action which is
consistent with the action on the $\mathbf{35}$ representation
considered as weights of the $\mathfrak{a}_6$ Lie algebra, according
with the conversion rule of table \ref{pesotti}, is that provided by
the following identification of the three generators:
\begin{equation}\label{idepermute}
    \begin{array}{ccc}
      R & \twoheadrightarrow & \{{x^1}\to {x^3},\,\,\,{x^2}\to {x^2},\,\,\,{x^3}\to
   {x^1},\,\,\,{x^4}\to {x^4},\,\,\,{x^5}\to {x^5},\,\,\,{x^6}\to
   {x^7},\,\,\,{x^7}\to {x^6}\} \\
      S & \twoheadrightarrow & \{{x^1}\to {x^7},\,\,\,{x^2}\to {x^1},\,\,\,{x^3}\to
   {x^4},\,\,\,{x^4}\to {x^5},\,\,\,{x^5}\to {x^3},\,\,\,{x^6}\to
   {x^6},\,\,\,{x^7}\to {x^2}\}\\
      T & \twoheadrightarrow & \{{x^1}\to {x^5},\,\,\,{x^2}\to {x^7},\,\,\,{x^3}\to
   {x^2},\,\,\,{x^4}\to {x^3},\,\,\,{x^5}\to {x^4},\,\,\,{x^6}\to
   {x^1},\,\,\,{x^7}\to {x^6}\}
    \end{array}
\end{equation}
Let us name $\mathfrak{g}_7 \in \mathrm{PSL(2,7)}$ any element of
the group in the 7-dimensional representation generated by the
transformations (\ref{idepermute}). Since the basis forms
$\mathbf{B}^\alpha$ are permuted among themselves by this action it
follows that $\mathfrak{g}_7 \in \mathrm{PSL(2,7)}$ induces a
corresponding linear transformation $\mathfrak{g}_{56}$ on the $56$
parameters $\delta_\alpha$ according with:
\begin{equation}\label{minnesota}
    \mathbf{Y}^{56}\left(\mathfrak{g}_7 \pmb{x}|\pmb{\delta}\right)\, = \,
    \mathbf{Y}^{56}\left(\pmb{x}|\mathfrak{g}_{56} \pmb{\delta}\right)
\end{equation}
In this way we obtain a $56$-dimensional representation of the group
$\mathrm{PSL(2,7)}$ group of which we can consider the decomposition
into irreducible representations. We obtain:
\begin{equation}\label{decompazzopsl27}
 \mathbf{56}\,   \stackrel{\mathrm{PSL(2,7)}}{\Longrightarrow}4 \,
 \mathrm{D}_7+2 \, \mathrm{D}_8+2 \, \text{DA}_3+2 \,\text{DB}_3
\end{equation}
This clearly means that there are in this sector no Englert fields
that are invariant under the full $\mathrm{PSL(2,7)}$ group, since
no singlets do appear in the above decomposition. Calculating
instead the decomposition of the same representation under the
maximal subgroup $\mathrm{G_{21}} \subset \mathrm{PSL(2,7)}$ we
obtain the following decomposition:
\begin{equation}\label{decompazzog21}
\mathbf{56}\, \stackrel{\mathrm{G_{21}}}{\Longrightarrow}\,  4 \,
\mathrm{D_1}+8\, \text{DA}_3+8 \text{DB}_3+2\, \text{DX}_1+2\,
\text{DY}_1
\end{equation}
This means that there exists a $4$-parameter solution of Englert
equation that is invariant with respect to the full group
$\mathrm{G_{21}}$. As we are going to see a $2$-parameter
subspace of this solution preserves also $\mathcal{N}=1$
supersymmetry.
\par
In order to study residual supersymmetry of the considered solutions
we have proceeded as follows. Naming
$Y^{56}_{ijk}\left(\pmb{x}|\pmb{\delta}\right)$ the components of
the form (\ref{caramelladierbe}) we have constructed the
corresponding symmetric $8\times 8$ matrix $\mathcal{B}$:
\begin{equation}\label{Bione}
    \mathcal{B}\left[\pmb{\delta},\pmb{x}\right] \, = \, \tau^{ijk} Y^{56}_{ijk}
    \left(\pmb{x}|\pmb{\delta}\right)
\end{equation}
The condition of $\mathcal{N}=1$ supersymmetry is provided by
requiring that, independently from the point $\pmb{x}$, one should
have:
\begin{equation}\label{condazio}
    \mathcal{B}\left[\pmb{\delta},\pmb{x}\right]_{I,8} \, = \, 0
    \quad ; \quad I=1,\dots,8
\end{equation}
This yields 14 linear conditions on the 56 parameters
$\pmb{\delta}$. We can view this as an orthogonal splitting of the
56-dimensional parameter space $\mathcal{M}^{56}$ of the following
type:
\begin{eqnarray}\label{zompino}
   \mathcal{M}^{56} & = & \mathcal{M}_{N=1} \oplus
   \mathcal{M}_{N=1}^\perp\\
   \mbox{dim}\mathcal{M}_{N=1} & = & 42\\
   \mbox{dim}\mathcal{M}_{N=1}^\perp & = & 14
\end{eqnarray}
The 42-dimensional subspace $\mathcal{M}_{N=1}$ is the space of
$\mathcal{N}=1$ supersymmetric Englert solutions. We can inquire
what is the subgroup $\mathrm{G}\subset \mathrm{PSL(2,7)}$ that
preserves the splitting (\ref{zompino}), namely:
\begin{equation}\label{crisculetto}
    \mathrm{G} \quad : \quad \mathcal{M}_{N=1}\longrightarrow \mathcal{M}_{N=1} \quad ; \quad
    \mathrm{G} \quad : \quad
    \mathcal{M}_{N=1}^\perp\longrightarrow \mathcal{M}_{N=1}^\perp
\end{equation}
By explicit calculation we find that $\mathrm{G}\, \sim \,
\mathrm{G_{21}}$, namely it is one of the eight conjugate copies of
$\mathrm{G_{21}}$ contained in $PSL(2,7)$. We already know from
Eq. (\ref{decompazzog21}) that with respect to this group there are
invariant Englert solutions and indeed we find that the invariant
subspace:
\begin{equation}\label{quidazzo}
    \mathcal{M}_{N=1}^{inv} \, \subset \, \mathcal{M}_{N=1}
\end{equation}
of those Englert fields that preserve $\mathcal{N}=1$ supersymmetry
and are invariant under the full group $\mathrm{G_{21}}$ stabilizing
the space $\mathcal{M}_{N=1}$ has dimension:
\begin{equation}\label{cragiletto}
    \mbox{dim} \,\mathcal{M}_{N=1}^{inv}\, = \, 2
\end{equation}
In other words there is a $2$-parameter $\mathrm{G_{21}}$-invariant
solution of Englert equation that preserves $\mathcal{N}=1$
supersymmetry.
\par
The scan of various supersymmetries was performed along these same
lines defining:
\begin{equation}\label{cretinus}
  \pmb{\delta}\, \in\,  \mathcal{M}_{N} \quad \Leftrightarrow \quad
  \mathcal{B}\left[\pmb{\delta},\pmb{x}\right]_{I,9-K} \, = \, 0
    \quad ; \quad I=1,\dots,8 \quad ; \quad K=1,\dots, N
\end{equation}
The result of this scan are summarized in the table here below:
\begin{equation}
\begin{array}{|c|c|c|c|c|c|c|c|} \hline
 \text{SUSY} & \text{Stability subgroup} & \text{Order} & \text{dim of} & \text{dim of} &
\text{dim of}  &  \text{Max inv.} & \text{Order} \\
\null & \text{of }\mathcal{M}_N &  \text{of G} &\mathcal{M}_N^{inv}
&  \mathcal{M}_N &
\mathcal{M}_N^\perp &  \text{of N sol} & \text{of $\Gamma$} \\
\hline
 N & \mathrm{G}\subset \text{PSL}(2,7) & |G| & n^{\text{inv}}_N & n_N & n_N^\perp &
 \Gamma \subset \mathrm{G} &
 |\Gamma | \\
\hline
 1 & \mathrm{G_{21}} & 21 & 2 & 42 & 14 & \mathrm{G_{21}} & 21 \\
\hline
 2 & \text{Dih}_3 & 6 & 2 & 30 & 26 & \text{Dih}_3 & 6 \\
\hline
 3 & \mathbb{Z}_3 & 3 & 8 & 20 & 36 & \mathbb{Z}_3 & 3 \\
\hline
 4 & \mathrm{T_{12}} & 12 & 4 & 12 & 44 & \mathbb{Z}_3\subset \mathrm{T_{12}} & 3 \\
\hline
 5 & \mathbb{Z}_3 & 3 & 2 & 6 & 50 & \mathbb{Z}_3 & 3 \\
\hline
 6 & \text{Dih}_3 & 6 & 2 & 2 & 54 & \mathbb{Z}_3\subset \text{Dih}_3 & 3 \\
\hline
 7 & \text{PSL}(2,7) & 168 & 0 & 0 & 56 & \text{PSL}(2,7) & 168 \\
\hline
\end{array}\label{barimondo}
\end{equation}
Let us comment on the notation. The names of the subgroups
$\mathrm{G}$ are those used in the previous sections and need no
explanation. By definition we name $\Gamma\subset \mathrm{G}$ the
subgroup with respect to which the space $\mathcal{M}_N$ contains
singlets. Except for the cases $N=4,6$ the subgroup $\Gamma$
coincides with the full group $\mathrm{G}$.
\par
Table (\ref{barimondo}) suffices to show that we have a rich
collection of solutions to Englert equation solutions leading to
exact M2-brane solutions of $d=11$ supergravity endowed with
prescribed $\mathcal{N}=N=1,2,3,4,5,6$ supersymmetries and
possessing also a non trivial group $\Gamma$ of discrete symmetries.
The complete analysis of all the $424$ solutions classified in
previous sections is postponed to a future publication.
\section{Conclusions and outlook}
\label{zumildo} In this paper we have achieved an exhaustive
classification of all $M2$-brane solutions of $d=11$ supergravity of
the type described in equations
(\ref{colabrodo}),(\ref{coffilasio}),
(\ref{3harmoniusca3}-\ref{englertaF}). The key item  in this
classification of M2-branes is the exhaustive classification of
solutions to Englert equation on a 7-torus which is precisely what
we have obtained utilizing the properties of the discrete group
$\mathrm{PSL(2,7)}$. We have also shown that this rich collection of
solutions possesses equally rich subclasses with three-dimensional
supersymmetries of all types from $\mathcal{N}=1$ to
$\mathcal{N}=6$. These exact solutions are of a genuinely new type,
so far never considered in M-theory.
\par
The open problem is that of the possible interpretation of our new
solutions in the following  contexts:
\begin{enumerate}
  \item The conformal gauge/gravity correspondence in the case  a suitable
  change of coordinates revealed an asymptotic
  factorization of the $d=11$ space of the following form:
  \begin{equation}\label{carneoldo}
    \mathcal{M}_{11} \, \stackrel{asymptotically}\longrightarrow \,
    \mathrm{AdS_4} \, \times \, \mathrm{SE_7}
  \end{equation}
$\mathrm{SE_7}$ denoting some Sasaki-Einstein $7$-manifold.
  \item The domain-wall/quantum field theory correspondence if by means
  of some other suitable change of coordinates we succeeded in
  achieving domain wall configurations.
  \item Effective four-dimensional gauged supergravity description if
suitable conditions on the parameters were revealed for which our solutions 
admit a well-defined $d=4$ limit.

\end{enumerate}
Independently of the above listed possibilities a mandatory
analysis of the physical content of our new class of M-theory
solutions is the systematic derivation of their Kaluza-Klein
spectrum. Indeed seven of the eleven dimensions are chosen to be
those corresponding to a  compact $7$-torus and an expansion in the corresponding 
normal modes is well-defined and natural. We plan to perform such
analysis in a forthcoming future publication.
\par
Last but not least let us remark that one key algebraic item of our
constructions is the discovery that not only the 7-dimensional
irreducible representation of $\mathrm{PSL(2,7)}$ is
crystallographic with respect to the $\mathfrak{a}_7$ and
$\mathfrak{e}_7$ root lattices, but also the 6-dimensional one is
crystallographic with respect to the $\mathfrak{a}_6$ lattice. An
appealing conjecture is that also the 8-dimensional irreducible
representation might be crystallographic with respect to the
$\mathfrak{a}_8$ and $\mathfrak{e}_8$ root lattices. This might lead
to interesting consequences for $\mathrm{E_{(8,8)}/SO(16)}$ sigma
model representing supergravity degrees of freedom in three
dimensions.
\section*{Aknowledgements}
We acknowledge important clarifying discussions with Massimo Bianchi,
Gianfranco Pradisi, Ugo Bruzzo, Pietro Antonio Grassi, Riccardo
D'Auria and Laura Andrianopoli.
\appendix
\section{Main Formulas of \texorpdfstring{$d=11$}{d=11} Supergravity and Conventions}
\label{peritoniteA} In this Appendix we recall the main formulas
of $d=11$ supergravity \cite{Cremmer:1978km} and give the
dictionary relating the relevant quantities in the formalism of the
original paper to those of the rheonomic Free-Differential Algebra
formulation \cite{FreDauriaHidden} that were utilized in
\cite{miol168} as well as in the present
paper. The former will be distinguished from the latter by a tilde,
when different.
 The $d=11$ supergravity
bosonic fields consist in the metric $\hat{g}_{\hat{\mu}\hat{\nu}}$
and the 3-form field $\tilde{A}_{\hat{\mu}\hat{\nu}\hat{\rho}}$ and the
bosonic action reads
\begin{equation}
\hat{e}^{-1}\mathcal{L}=
-\frac{1}{4}\,\tilde{R}-\frac{1}{48}\,\tilde{F}_{\hat{\mu}\hat{\nu}\hat{\rho}\hat{\sigma}}
\tilde{F}^{\hat{\mu}\hat{\nu}\hat{\rho}\hat{\sigma}}+
\frac{2}{\hat{e}\,(12)^4}\,\epsilon^{\hat{\mu}_1\dots
\hat{\mu}_{11}}\tilde{F}_{\hat{\mu}_1\dots\hat{\mu}_4}
\tilde{F}_{\hat{\mu}_5\dots\hat{\mu}_8}\tilde{A}_{\hat{\mu}_9\hat{\mu}_{10}\hat{\mu}_{11}}\,,
\end{equation}
where $\hat{e}\equiv \sqrt{|{\rm
det}(\hat{g}_{\hat{\mu}\hat{\nu}})|}$,
$\hat{\mu},\hat{\nu},\dots=0,\dots, 10$. We use the ``mostly minus''
notation and $\epsilon_{01\dots\,10}=\epsilon^{01\dots\,10}=+1$.\par
The Einstein equation and the field equation for the 3-form read:
\begin{align}
\tilde{R}_{\hat{\mu}\hat{\nu}}&=-\frac{1}{3}\,\tilde{F}_{\hat{\mu}\hat{\mu}_1\hat{\mu}_2\hat{\mu}_3}
\tilde{F}_{\hat{\nu}}{}^{\hat{\mu}_1\hat{\mu}_2\hat{\mu}_3}+\frac{1}{36}\,
\hat{g}_{\hat{\mu}\hat{\nu}}\,\tilde{F}_{\hat{\mu}_1\hat{\mu}_2\hat{\mu}_3\hat{\mu}_4}
\tilde{F}^{\hat{\mu}_1\hat{\mu}_2\hat{\mu}_3\hat{\mu}_4}\,,\nonumber\\
\partial_{\hat{\mu}}\left(\hat{e}\,\tilde{F}^{\hat{\mu}\hat{\nu}\hat{\rho}\hat{\sigma}}\right)&
=-\frac{3}{(12)^3}\epsilon^{\hat{\nu}\hat{\rho}\hat{\sigma}
\hat{\mu}_1\dots
\hat{\mu}_{8}}\,\tilde{F}_{\hat{\mu}_1\dots\hat{\mu}_4}\tilde{F}_{\hat{\mu}_5\dots\hat{\mu}_8}\,.
\end{align}
Below we give the dictionary between this notation and that of
\cite{miol168}, in which the relevant quantities are denoted by untilded symbols:
\begin{align}
\tilde{R}_{\hat{\mu}\hat{\nu}}&=- 2\,{R}_{\hat{\mu}\hat{\nu}}\,,\nonumber\\
\tilde{F}_{\hat{\mu}\hat{\nu}\hat{\rho}\hat{\sigma}}&=4\,\partial_{[\hat{\mu}}\tilde{A}_{\hat{\nu}\hat{\rho}\hat{\sigma}]}
=6\, {F}_{\hat{\mu}\hat{\nu}\hat{\rho}\hat{\sigma}}
=6\,\partial_{[\hat{\mu}}{A}_{\hat{\nu}\hat{\rho}\hat{\sigma}]}\,\nonumber\\
\tilde{A}_{\hat{\nu}\hat{\rho}\hat{\sigma}}&=\frac{3}{2}\,{A}_{\hat{\nu}\hat{\rho}\hat{\sigma}}\,\nonumber\\
\tilde{{\bf F}}^{[4]}&=d\tilde{{\bf A}}^{[3]}=\frac{1}{4}\,{F}^{[4]}=\frac{1}{4}\,d{A}^{[3]}\,,\label{dic}
\end{align}
where we have defined:
\begin{align}
\tilde{{\bf F}}^{[4]}&\equiv
\frac{1}{4!}\,\tilde{F}_{\hat{\mu}\hat{\nu}\hat{\rho}\hat{\sigma}}\,
dx^{\hat{\mu}}\wedge \dots \wedge dx^{\hat{\sigma}}\,,\nonumber\\
\tilde{{\bf A}}^{[3]}&\equiv
\frac{1}{3!}\,\tilde{A}_{\hat{\mu}\hat{\nu}\hat{\rho}}\,dx^{\hat{\mu}}\wedge
dx^{\hat{\nu}} \wedge dx^{\hat{\rho}}\,,\nonumber\\
{F}^{[4]}&\equiv
{F}_{\hat{\mu}\hat{\nu}\hat{\rho}\hat{\sigma}}
\,dx^{\hat{\mu}}\wedge \dots \wedge dx^{\hat{\sigma}}\,,\nonumber\\
{A}^{[3]}&\equiv {A}_{\hat{\mu}\hat{\nu}\hat{\rho}}\,
dx^{\hat{\mu}}\wedge dx^{\hat{\nu}} \wedge dx^{\hat{\rho}}\,.
\end{align}
We also introduce the 6-form $\tilde{{\bf A}}^{[6]}$ dual to $\tilde{{\bf A}}^{[3]}$ by Legendre
transforming the $d=11$ action. Its 7-form field strength reads:
\begin{equation}
\tilde{{\bf F}}^{[7]}=d\tilde{{\bf A}}^{[6]}+\tilde{{\bf F}}^{[4]}\wedge \tilde{{\bf A}}^{[3]}={}^*\tilde{{\bf F}}^{[4]}\,.
\end{equation}
In components:
\begin{align}
\tilde{{\bf F}}^{[7]}&=\frac{1}{7!}\tilde{F}_{\hat{\mu}_1\dots
\hat{\mu}_{7}}\,dx^{\hat{\mu}_1}\wedge\dots\wedge
dx^{\hat{\mu}_7}\,,
\nonumber\\
\tilde{F}_{\hat{\mu}_1\dots
\hat{\mu}_{7}}&=7\,\partial_{[\hat{\mu}_1}\tilde{A}_{\hat{\mu}_2\dots
\hat{\mu}_7]}+35\,\tilde{F}_{[\hat{\mu}_1\dots
\hat{\mu}_{4}}\tilde{A}_{\hat{\mu}_5\dots
\hat{\mu}_7]}=\frac{e}{4!}\epsilon_{\hat{\mu}_1 \dots
\hat{\mu}_{7}\hat{\mu}_8\dots \hat{\mu}_{11}}\,\tilde{F}^{\hat{\mu}_8\dots
\hat{\mu}_{11}}\,.\label{F7}
\end{align}
As for the fermionic sector, the gamma matrices $\tilde{\Gamma}^{\hat{\mu}}$ in \cite{Cremmer:1978km}
differ by an overall sign from those in the present paper ${\Gamma}^{\hat{\mu}}$:
\begin{equation}
\tilde{\Gamma}^{\hat{\mu}}=-{\Gamma}^{\hat{\mu}}\,,
\end{equation}
while the gravitino field is the same in the two notations. The supersymmetry variation of the latter field therefore reads:
\begin{align}
\delta\Psi_{\hat{\mu}}&=\mathcal{D}_{\hat{\mu}}\epsilon+\frac{i}{144}
\left(\tilde{\Gamma}^{\hat{\mu}_1\hat{\mu}_2\hat{\mu}_3\hat{\mu}_4}{}_{\hat{\mu}}-8\,
\tilde{\Gamma}^{\hat{\mu}_1\hat{\mu}_2\hat{\mu}_3}\delta^{\hat{\mu}_4}_{\hat{\mu}}\right)
\Psi\,\tilde{F}_{\hat{\mu}_1\hat{\mu}_2\hat{\mu}_3\hat{\mu}_4}=\nonumber\\
&=\mathcal{D}_{\hat{\mu}}\epsilon-\frac{i}{24}
\left({\Gamma}^{\hat{\mu}_1\hat{\mu}_2\hat{\mu}_3\hat{\mu}_4}{}_{\hat{\mu}}-8\,
{\Gamma}^{\hat{\mu}_1\hat{\mu}_2\hat{\mu}_3}\delta^{\hat{\mu}_4}_{\hat{\mu}}\right)\Psi\,
{F}_{\hat{\mu}_1\hat{\mu}_2\hat{\mu}_3\hat{\mu}_4}\,.
\end{align}
\section{M2-Brane Solutions with Transverse Flux}
\label{travertinus} The $M2$-brane solutions considered in the
present work are part of a general class of solutions characterized
by the presence of a self-dual 4-form flux along the transverse
eight-dimensional space
\cite{Duff:1997xja,Hawking:1997bg,Cvetic:2000mh,Cvetic:2000db}. The
Ansatz for the $d=11$ metric is the one given in Eq.
(\ref{colabrodone}) while the 3-form field has the following general
expression:
\begin{equation}
{\bf A}^{[3]}=\frac{2}{H(y)}\,\Omega^{[3]}+\mathring{{\bf A}}^{[3]}(y)\,,\label{A3sd}
\end{equation}
where $H(y)$ is a function of the eight transverse coordinates $y^I$ and $\mathring{{\bf A}}^{[3]}(y)$
is a 3-form in the transverse space.
The 4-form field strength reads:
\begin{equation}
{\bf F}^{[4]}=d{\bf A}^{[3]}=-\frac{2}{H(y)^2}\,\partial_I H\,dy^I \wedge \Omega^{[3]}+\mathring{{\bf F}}^{[4]}(y)\,,
\end{equation}
where $$\mathring{{\bf F}}^{[4]}=d\mathring{{\bf A}}^{[3]}=\mathring{F}(y)_{IJKL}\,dy^I\wedge dy^J\wedge dy^K\wedge dy^L\,.$$
We require $\mathring{{\bf F}}^{[4]}$ to be self-dual in the transverse Euclidean space:
\footnote{Had we chosen the $M2$-brane with the opposite charge with respect to ${\bf A}^{[3]}$,
we should have taken $\mathring{{\bf F}}^{[4]}$ to be anti-self-dual.}
\begin{equation}
\star_8\mathring{{\bf F}}^{[4]}=\mathring{{\bf F}}^{[4]}\,.\label{selfdual}
\end{equation}
Plugging the above Ansatz in the field equations we find for $H(y)$
\begin{equation}
\Box_8 H\,=\, -3\, \mathring{F}_{IJKL}\mathring{F}^{IJKL}\,,
\end{equation}
where $\Box_8$ is the d'Alembertian in the flat transverse space: $\Box_8 H\equiv \partial_I \partial_I H$.\par
\paragraph{Supersymmetry.}
Substituting the above Ansatz in the Killing spinor equation:
\begin{align}
\mathcal{D}_{\hat{\mu}}\xi-\frac{i}{24}\left({\Gamma}^{\hat{\mu}_1\hat{\mu}_2\hat{\mu}_3\hat{\mu}_4}{}_{\hat{\mu}}-8\,
{\Gamma}^{\hat{\mu}_1\hat{\mu}_2\hat{\mu}_3}\delta^{\hat{\mu}_4}_{\hat{\mu}}\right)\,
\xi\,{F}_{\hat{\mu}_1\hat{\mu}_2\hat{\mu}_3\hat{\mu}_4}=0\,,
\end{align}
and writing $\xi=\epsilon \otimes \chi$, as in (\ref{stensorokillo}), after some algebra, one finds the following conditions:
\begin{align}
\mathbb{P}_-\,\chi=0\,\,,\,\,\,\,\mathring{F}_{IJKL}\,T^{IJKL}\,\chi=0\,\,,\,\,\,\,
\mathring{F}_{I_1\dots I_4}T^{I_1\dots I_4}\,T_I\,\chi=0\,,\label{susyeqs1}
\end{align}
where $\mathbb{P}_-\equiv\frac{1}{2}\,({\bf 1}_{16}-T_9)$.
One can show, following \cite{Duff:1997xja}, that the above conditions can be recast in the following equivalent form:
\begin{equation}
\mathbb{P}_-\,\chi=0\,,\,\,\,\mathring{F}_{IJKL}\,T^{JKL}\,\chi=0\,.
\end{equation}
The self-duality condition of $\mathring{F}$ further simplifies equations (\ref{susyeqs1}) since
\begin{equation}
\mathring{F}_{IJKL}\,T^{IJKL}=\mathring{F}_{IJKL}\,T^{IJKL}\,\mathbb{P}_+\,.
\end{equation}
Therefore, choosing $\chi$ so that $\mathbb{P}_-\chi=0$, the last of Eq.s~(\ref{susyeqs1}) is automatically satisfied, and
the supersymmetry conditions reduce to:
\begin{equation}
\mathbb{P}_-\,\chi=0\,,\,\,\,\mathring{F}_{IJKL}\,T^{IJKL}\,\chi=0\,.\label{lastsusy}
\end{equation}
The existence of solutions to the above equations depends on the detailed structure of $\mathring{F}_{IJKL}$.
As we have shown, the form of the self-dual flux in the class of solutions considered here does allow for solutions
with different degrees of supersymmetry. Let us show below how the Englert equation implements the self-duality condition
 (\ref{selfdual}) for the class of solutions discussed in the present work.
\paragraph{$M2$-branes with Englert fluxes.} These solutions are obtained by choosing the transverse space of the
form $\mathbb{R}_+\times T^7$, splitting $(y^I)=(x^i,\,U)$ and further specializing the Ansatz (\ref{A3sd})
by choosing the inner components of the 3-form as follows:
\begin{equation}
\mathring{{\bf A}}^{[3]}= e^{-\mu\, U}\,Y_{ijk}(x)\,dx^i\wedge dx^j\wedge dx^k\,,
\end{equation}
where $i,j,k=1,\dots,7 $. The self-duality condition (\ref{A3sd}) then reduces to the Englert equation in $Y_{ijk}$.
Formally this amounts to a Scherk-Schwarz reduction \cite{Scherk:1979zr} from the Euclidean eight-dimensional transverse space to the seven-torus
and the original self-duality condition reduces to the ``self-duality'' in odd-dimensions of \cite{Townsend:1983xs}.\par
As far as the supersymmetry conditions (\ref{lastsusy}) are concerned, if we further split $\chi=\kappa\otimes \lambda$,
as in Eq.~(\ref{splittochi}), where now $\sigma_3 \kappa=\kappa$, condition $\mathbb{P}_-\chi=0$ is satisfied,
being $T_9=\sigma_3\otimes {\bf 1}_8$. The last of equations (\ref{lastsusy}) now boils down to:
\begin{equation}
0=\mathring{F}_{0ijk}\,T^{0ijk}\,\chi\propto \kappa\otimes \mathcal{B}\lambda\,\,\Leftrightarrow\,\,\,\,\,\mathcal{B}\lambda=0\,,
\end{equation}
where $\mathcal{B}\equiv \tau^{ijk}\,Y_{ijk}$. We then retrieve the
equation (\ref{Binullo}), whose solutions have been studied in the
present work.
\newpage
  \bibliography{petrusbiblio11X18}

\begin{thebibliography}{10}

\bibitem{Maldacena:1997re}
J.~Maldacena, ``The large-{N} limit of superconformal field theories and
  supergravity,'' {\em International Journal of Theoretical Physics}, vol.~4,
  no.~38, pp.~1113--1133, 1999.
\newblock doi:10.1023/A:1026654312961 [hep-th/9711200].

\bibitem{Kallosh:1998ph}
R.~Kallosh and A.~Van~Proeyen, ``Conformal symmetry of supergravities in {AdS}
  spaces,'' {\em Physical Review D}, vol.~60, no.~2, p.~026001, 1999.
\newblock doi:10.1103/PhysRevD.60.026001 [hep- th/9804099].

\bibitem{Ferrara:1998jm}
S.~Ferrara and C.~Fronsdal, ``Gauge fields as composite boundary excitations,''
  {\em Physics Letters B}, vol.~433, no.~1, pp.~19--28, 1998.
\newblock doi:10.1016/S0370-2693(98)00664-9 [hep-th/9802126].

\bibitem{Ferrara:1998ej}
S.~Ferrara, C.~Fronsdal, and A.~Zaffaroni, ``On {$\mathcal{N}=8$} supergravity
  in {AdS}$_5$ and {$\mathcal{N}=4$} superconformal {Y}ang-{M}ills theory,''
  {\em Nuclear Physics B}, vol.~532, no.~1-2, pp.~153--162, 1998.
\newblock doi:10.1016/S0550-3213(98)00444-1 [hep-th/9802203].

\bibitem{sergiotorino}
A.~Ceresole, G.~Dall'Agata, R.~D'Auria, and S.~Ferrara, ``Spectrum of type
  {IIB} supergravity on {$AdS_5\times T^{11}$}: predictions on
  {$\mathcal{N}=1$} {SCFT}'s,'' {\em Physical Review D}, vol.~61, no.~6,
  p.~066001, 2000.
\newblock [hep-th/9905226].

\bibitem{witkleb}
I.~R. Klebanov and E.~Witten, ``{Superconformal field theory on three-branes at
  a Calabi-Yau singularity},'' {\em Nucl. Phys.}, vol.~B536, pp.~199--218,
  1998.

\bibitem{Fabbri:1999hw}
D.~Fabbri, P.~Fr{\'e}, L.~Gualtieri, C.~Reina, A.~Tomasiello, A.~Zaffaroni, and
  A.~Zampa, ``3{D} superconformal theories from {S}asakian seven-manifolds: new
  non-trivial evidences for {$\mathrm{AdS}_4 / \mathrm{CFT}_3$},'' {\em Nuclear
  Physics B}, vol.~577, no.~3, pp.~547--608, 2000.
\newblock [hep- th/9907219].

\bibitem{Fabbri:1999ay}
D.~Fabbri, P.~Fr\'e, L.~Gualtieri, and P.~Termonia, ``{$\mathrm{Osp(N|4)}$}
  supermultiplets as conformal superfields on {$\partial\mathrm{AdS}_4$} and
  the generic form of {$\mathcal{N}=2$}, {D=3} gauge theories,'' {\em Classical
  and Quantum Gravity}, vol.~17, no.~1, p.~55, 2000.
\newblock [hep-th/9905134].

\bibitem{Aharony:2008ug}
O.~Aharony, O.~Bergman, D.~L. Jafferis, and J.~Maldacena, ``{N=6}
  superconformal {C}hern-{S}imons-matter theories, {M2}-branes and their
  gravity duals,'' {\em Journal of High Energy Physics}, vol.~2008, no.~10,
  p.~091, 2008.
\newblock doi:10.1088/1126-6708/2008/10/091 [arXiv:0806.1218 [hep-th]].

\bibitem{Gaiotto:2007qi}
D.~Gaiotto and X.~Yin, ``Notes on superconformal {C}hern-{S}imons-{M}atter
  theories,'' {\em Journal of High Energy Physics}, vol.~2007, no.~08, p.~056,
  2007.
\newblock doi:10.1088/1126-6708/2007/08/056 [arXiv:0704.3740 [hep-th]].

\bibitem{Gaiotto:2009tk}
D.~Gaiotto and D.~L. Jafferis, ``{Notes on adding D6 branes wrapping RP**3 in
  AdS(4) x CP**3},'' {\em JHEP}, vol.~11, p.~015, 2012.

\bibitem{Duff:1997xja}
M.~J. Duff, J.~M. Evans, R.~R. Khuri, J.~X. Lu, and R.~Minasian, ``{The
  Octonionic membrane},'' {\em Phys. Lett.}, vol.~B412, pp.~281--287, 1997.
\newblock [Nucl. Phys. Proc. Suppl.68,295(1998)].

\bibitem{Hawking:1997bg}
S.~W. Hawking and M.~Taylor, ``{Bulk charges in eleven-dimensions},'' {\em
  Phys. Rev.}, vol.~D58, p.~025006, 1998.

\bibitem{Cvetic:2000mh}
M.~Cveti\v{c}, H.~Lu, and C.~N. Pope, ``{Brane resolution through
  transgression},'' {\em Nucl. Phys.}, vol.~B600, pp.~103--132, 2001.

\bibitem{Cvetic:2000db}
M.~Cveti\v{c}, G.~W. Gibbons, H.~Lu, and C.~N. Pope, ``{Ricci flat metrics,
  harmonic forms and brane resolutions},'' {\em Commun. Math. Phys.}, vol.~232,
  pp.~457--500, 2003.

\bibitem{miol168}
P.~Fr\'e, ``{Supersymmetric M2-branes with Englert fluxes, and the simple group
  PSL(2, 7)},'' {\em Fortsch. Phys.}, vol.~64, no.~6-7, pp.~425--462, 2016.

\bibitem{pietroantoniosorin}
P.~Fr\'e, P.~A. Grassi, and A.~S. Sorin, ``{Hyperinstantons, the Beltrami
  Equation, and Triholomorphic Maps},'' {\em Fortsch. Phys.}, vol.~64, no.~2-3,
  pp.~151--175, 2016.

\bibitem{Fre:2015wvs}
P.~Fr{\'e}, P.~A. Grassi, L.~Ravera, and M.~Trigiante, ``{Minimal $D=7$
  Supergravity and the supersymmetry of Arnold-Beltrami Flux branes},'' {\em
  JHEP}, vol.~06, p.~018, 2016.

\bibitem{englert}
F.~Englert, ``Spontaneous compactification of eleven-dimensional
  supergravity,'' {\em Physics Letters B}, vol.~119, no.~4-6, pp.~339--342,
  1982.

\bibitem{Fre:2015mla}
P.~Fr\'e and A.~S. Sorin, ``Classification of {A}rnold-{B}eltrami flows and
  their hidden symmetries,'' {\em Phys. Part. Nucl}, vol.~46, no.~4,
  pp.~497--632, 2015.

\bibitem{Cerchiai:2010tk}
B.~L. Cerchiai and B.~van Geemen, ``{From qubits to E7},'' {\em J. Math.
  Phys.}, vol.~51, p.~122203, 2010.

\bibitem{FreDauriaHidden}
R.~D'Auria and P.~Fr{\'e}, ``{Geometric Supergravity in d = 11 and Its Hidden
  Supergroup},'' {\em Nucl. Phys.}, vol.~B201, pp.~101--140, 1982.
\newblock [Erratum: Nucl. Phys.B206,496(1982)].

\bibitem{FDAgauge}
P.~Fr{\'e}, ``{Comments on the Six Index Photon in $D=11$ Supergravity and the
  Gauging of Free Differential Algebras},'' {\em Class. Quant. Grav.}, vol.~1,
  p.~L81, 1984.

\bibitem{maiobuk}
P.~G. Fr{\'e}, {\em Gravity, a Geometrical Course}, vol.~1,2.
\newblock Springer Science \& Business Media, 2012.

\bibitem{Cremmer:1978km}
E.~Cremmer, B.~Julia, and J.~Scherk, ``{Supergravity Theory in
  Eleven-Dimensions},'' {\em Phys. Lett.}, vol.~B76, pp.~409--412, 1978.
\newblock [,25(1978)].

\bibitem{Cremmer:1979up}
E.~Cremmer and B.~Julia, ``{The SO(8) Supergravity},'' {\em Nucl. Phys.},
  vol.~B159, pp.~141--212, 1979.

\bibitem{kingus}
R.~C. King, F.~Toumazet, and B.~G. Wybourne, ``A finite subgroup of the
  exceptional {L}ie group {G2},'' {\em Journal of Physics A: Mathematical and
  General}, vol.~32, no.~48, pp.~8527 -- 8537, 1999.

\bibitem{ramonus}
C.~Luhn, S.~Nasri, and P.~Ramond, ``Simple finite non-{A}belian flavor
  groups,'' {\em Journal of Mathematical Physics}, vol.~48, no.~12, p.~123519,
  2007.
\newblock ArXiv:0709.1447 [hep-th].

\bibitem{Bruzzo:2017fwj}
U.~Bruzzo, A.~Fino, and P.~Fr\'e, ``{The K\"ahler Quotient Resolution of
  $\mathbb{C}^3/\Gamma$ singularities, the McKay correspondence and D=3
  $\mathcal{N}=2$ Chern-Simons gauge theories}.'' {\tt arXiv:1710.01046}, 2017.

\bibitem{marcovaldo}
D.~Markushevich, ``Resolution of {${\bf C}^3/H_{168}$},'' {\em Math. Ann.},
  vol.~308, no.~2, pp.~279--289, 1997.

\bibitem{Scherk:1979zr}
J.~Scherk and J.~H. Schwarz, ``{How to Get Masses from Extra Dimensions},''
  {\em Nucl. Phys.}, vol.~B153, pp.~61--88, 1979.
\newblock [,79(1979)].

\bibitem{Townsend:1983xs}
P.~{T}ownsend, K.~{P}ilch, and P.~van {N}ieuwenhuizen, ``Selfduality in odd
  dimensions,'' {\em Phys. Lett.}, vol.~136B, no.~1, p.~38, 1984.

\end{thebibliography}
  \bibliographystyle{ieeetr}
\end{document}